\documentclass[referee,sn-basic]{sn-jnl}% referee option is meant for double line spacing

\graphicspath{ {Figures/} }
\usepackage{adjustbox}
\usepackage{array}

\newcolumntype{R}[2]{%
    >{\adjustbox{angle=#1,lap=\width-(#2)}\bgroup}%
    l%
    <{\egroup}%
}
\newcommand*\rot{\multicolumn{1}{R{90}{1em}}}% no optional argument here, please!
\usepackage{lipsum}
\usepackage{mathtools}
\usepackage{cuted}

\jyear{2022}%

\theoremstyle{thmstyleone}%
\theoremstyle{thmstyletwo}%

\theoremstyle{thmstylethree}%

\begin{document}

\title[ ]{Land Use Trade-offs in Decarbonization of Electricity Generation in the American West}

\author*[1]{\fnm{Neha} \sur{Patankar}}\email{npatankar@binghamton.edu}
\author[2,3]{\fnm{Xiili} \sur{Sarkela-Basset}}
\author[4]{\fnm{Greg} \sur{Schivley}}
\author[5]{\fnm{Emily} \sur{Leslie}}
\author[2,3]{\fnm{Jesse} \sur{Jenkins}}

\affil*[1]{\orgdiv{Systems Science and Industrial Engineering}, \orgname{Binghamton University}, \orgaddress{\city{Binghamton}, \state{NY}, \postcode{13902}, \country{USA}}}

\affil[2]{\orgdiv{Andlinger Center for Energy and the Environment}, \orgname{Princeton University}, \orgaddress{\city{Princeton}, \state{NJ}, \postcode{08544}, \country{USA}}}

\affil[3]{\orgdiv{Mechanical \& Aerospace Engineering}, \orgname{Princeton University}, \orgaddress{\city{Princeton}, \state{NJ}, \postcode{08544},  \country{USA}}}

\affil[4]{\orgname{Carbon Impact Consulting}, \orgaddress{\city{New Providence}, \state{NJ}, \postcode{07974}, \country{USA}}}

\affil[5]{\orgname{Montara Mountain Energy}, \orgaddress{\city{Pacifica}, \state{CA}, \postcode{94044}, \country{USA}}}


\abstract{Land-use conflicts may constrain the unprecedented rates of renewable energy deployment required to meet the decarbonization goals of the Inflation Reduction Act (IRA). This paper employs geospatially resolved data and a detailed electricity system capacity expansion model to generate 160 affordable, zero-carbon electricity supply portfolios for the American west and evaluates the land use impacts of each portfolio. Less than 4\% of all sites suitable for solar development and 17\% of all wind sites appear in this set of portfolios. Of these sites, 53\% of solar and 85\% of wind sites exhibit higher development risk and potential for land use conflict. We thus find that clean electricity goals cannot be achieved in an affordable manner without substantial renewable development on sites with potential for land use conflict. However, this paper identifies significant flexibility across western U.S. states to site renewable energy or alter the composition of the electricity supply portfolio to ameliorate potential conflicts.}




\maketitle

\section{Introduction}\label{sec1}

Decarbonizing the electric power sector is a priority in the recently-passed Inflation Reduction Act (IRA) \citep{IRA}. Ambitious emissions reduction targets and Clean Electricity Standards (CES) have also recently been mandated in several states in the western U.S., including California, Washington, Nevada, New Mexico, and Oregon, which have all committed to 100\% carbon-free electricity by mid-century or sooner \citep{CES}. Reaching IRA's goal of 40\% emission reduction below 2005 levels by 2030 may increase the annual addition of wind and solar in 2025-26 to 39 GW/year ($\sim$2X the 2020 pace) and 49 GW/year ($\sim$5X the 2020 pace) in the United States, respectively \citep{REPEAT}. To further decarbonize the electricity sector and reach a net-zero energy system by 2050, a mix of increasingly affordable and mature VRE technologies, mainly solar photovoltaic (PV) and onshore wind turbines will need to be deployed in the United States at historically unprecedented rates \citep{larsonnetzero2021, baik2021different,jenkins2021mission}. The land intensive nature of solar and wind facilities may lead to land-use conflicts hindering the pace of electricity decarbonization. For the United States to achieve decarbonization goals at the necessary pace and scale, power sector models should provide decision support to inform cost-effective pathways and policies to achieve zero-carbon electricity while mitigating potential land use conflicts. 

Power sector planning models, also known as capacity expansion models (CEMs), solve optimization problems that attempt to meet electricity demand in a given planning year while minimizing the cost of electricity supply and maximizing utility of consumption within operational, physical, engineering, and policy constraints. The resulting least-cost technology portfolios typically: 1) do not produce scenarios diverse enough to capture the full range of possible future outcomes, instead relying on a few sensitivities performed on preliminary assumptions \citep{klein2015comparing}; 2) do not capture uncertain, varied, and/or conflicting priorities of electricity system stakeholders and actors such as electric utilities, power sector investors, impacted communities, non-governmental and public interest organizations, and policymakers  \citep{wilson2013best, hobbs1995optimization, swisher1997tools}; and 3) do not quantitatively address non-engineering priorities or constraints, such as state-specific political feasibility or land conservation constraints \citep{decarolis2011using}.  

Our study builds on previous literature \citep{hernandez2015solar, nock2019holistic, klein2015comparing, mai2021interactions, wu2020low} and addresses the above limitations by, first, employing the modeling to generate alternatives (MGA) technique \citep{brill1982modeling, decarolis2011using} in concert with a detailed, open-source CEM, GenX \citep{jenkins2017enhanced, GenX}. We create a model of the U.S. portion of the Western Interconnection (aka WECC) to explore the technology options, key trade-offs, and policy considerations associated with achieving 100\% carbon-free electricity supply by 2045, consistent with the timelines required by clean electricity standard (CES) policies enacted in multiple western states. Using the MGA technique, we create 160 possible alternative generation portfolios with no more than 10\% greater cost than the least-cost portfolio. Second, we quantify renewable siting trade-offs across alternative portfolios by producing a spatially-explicit map and evaluating the land use impact of each portfolio. It provides a methodology for states to collaboratively identify potential areas of land-use conflict and strategies to resolve conflicts without significantly increasing costs of a carbon-free system.  See Section \ref{section:Method} and Appendices \ref{SI:method}-\ref{SI:Data} for further details. 
 
Existing CEMs, which do not consider multiple stakeholder preferences nor include highly resolved spatial and temporal detail, often solve for renewables development in land areas with both high renewable resource potential and high value for conservation, agriculture, recreation or other alternative land uses. This type of analysis has the potential to misinform decision makers by overestimating the precision of the model and broadcasting an inappropriate sense of certainty regarding the suggested system transformation \citep{decarolis2017formalizing}. A rapid transition to zero-carbon electricity systems would be better informed by robust and spatially resolved models that can quantitatively assess multiple salient land-use outcomes, illuminate trade-offs and flexibility between alternative technology portfolios, and account for potential non-engineering constraints. The modeling approach demonstrated in this study provides a wide range of alternative technology portfolios and identifies siting flexibility for achieving zero-carbon goals. While this paper focuses on land-use impacts, the methods demonstrated in this study can be readily extended to quantify a wide range of other non-cost related trade-offs and multiple system-wide objectives salient to the social, environmental, economic and political priorities of various stakeholders in regional energy transitions such as air pollution and public health, water consumption, implications for landowner and government revenues, and the distribution and composition of energy-related employment.

\section{Results}\label{section:Results}

This study uses spatially-resolved wind and solar development potential estimates based on a geo-spatial site suitability analysis for the continental United States performed using the MapRE tool \cite{ranjit_deshmukh_grace_wu_multi-criteria_2019, wu2020low}. Results from the land suitability analysis show that nearly 45\% of the total land area of the American West is potentially available for onshore wind and/or utility-scale solar PV development (SI \ref{SI:LandAvail}). That is, more than 1.4 million km\textsuperscript{2} of land area has no constraints such as existing infrastructure, military bases, water bodies, strenuous terrain, administrative and conservation protections such as parks or conservation trusts, or other constraints described in SI Table \ref{table:SiteSuitabilityDatasets} for building new onshore wind turbines or utility-scale solar PV. However, monetary, ecological, and social priorities not captured in geo-spatial datasets readily available for site suitability analysis can impose previously unrecognised challenges or trade-offs for wind and solar development. 

To understand these potential challenges and trade-offs, we first downscale (Section \ref{section:down-scaling}) the capacity expansion results from the MGA analysis to a spatially explicit map (Section \ref{section:RMap}) showing the total available area with cost-effective sites in each state. Second, we create three categories of land types that could act as proxy metrics for unmodeled land constraints and quantify the extent of wind and solar development on riskier land areas in Section \ref{section:RRisk}. Third, we study the impact of ten potential stakeholder/policymaker objectives on the siting of wind and solar and the composition of the electricity supply portfolio across five categories of riskier land types in Section \ref{section:RExtrLand}. Lastly, we show the potential for flexibility in the wind and solar site distribution across the states in the American West in Section \ref{section:RFlex} and identify strategies for states to collaboratively reduce riskier land usage. This approach illuminates the state-specific land-use, land cover and technology trade-offs for renewable development in the WECC. This analysis offers insights for creating policies to decrease overall land impact required to achieve the pace and scale of renewable development required to decarbonize the WECC while considering conflicting social, political and technological objectives. These contributions can help enable a more rapid and politically durable zero-carbon electricity transition. 

\subsection{Cost-effective variable renewable energy development} \label{section:RMap}

\begin{figure}
  \centering
  \includegraphics[width=\textwidth]{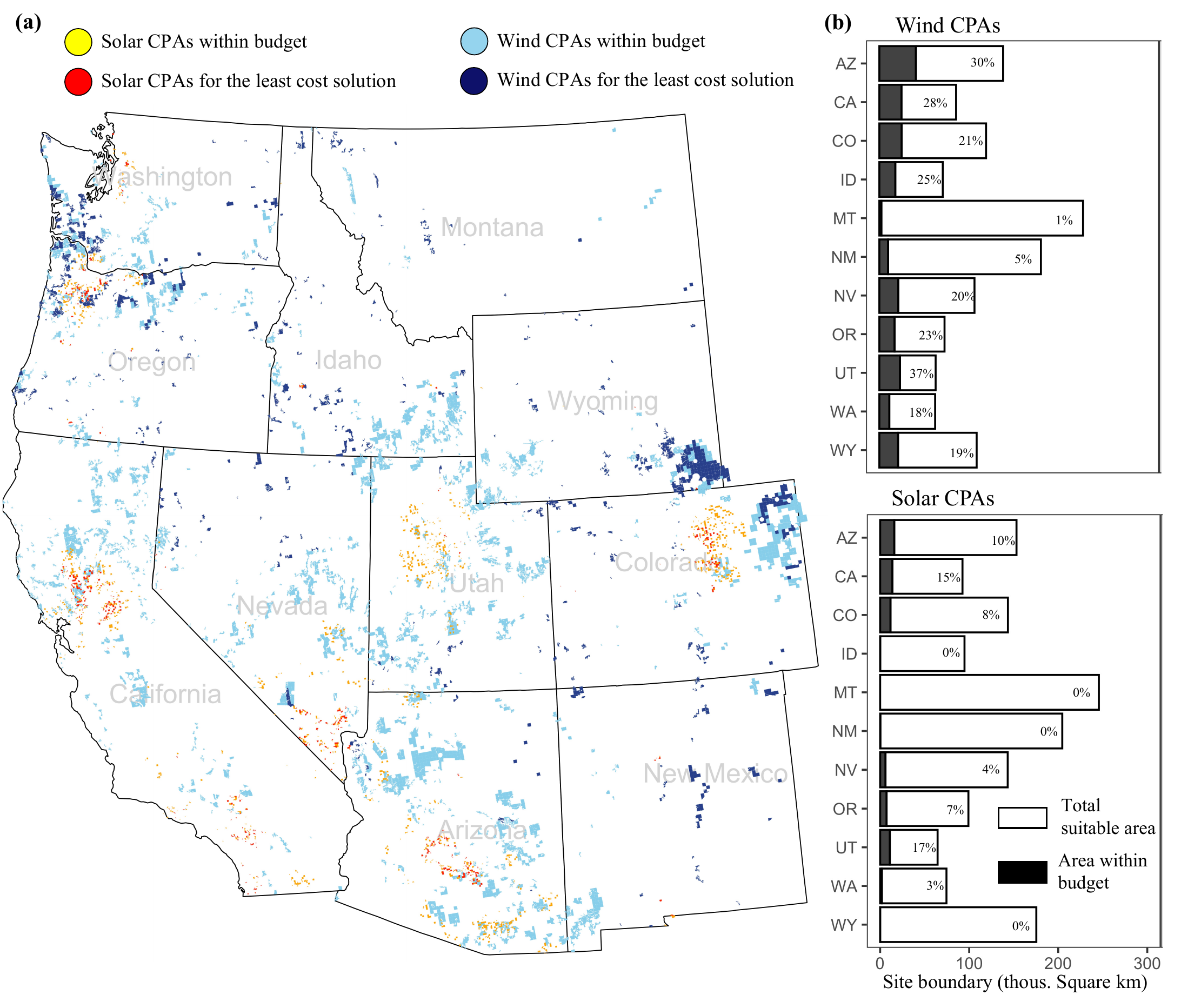}
  \caption{(a) Down-scaling of least-cost onshore wind and utility PV capacity results from the power system capacity expansion model (CEM) for all 160 MGA iterations and the least-cost solution. This figure illustrates selection of individual candidate project areas consistent with CEM results using a least-cost siting procedure described in Section \ref{section:down-scaling}. (b) Total suitable area for wind and solar development in each state and total area in wind and solar candidate project areas selected in one or more portfolios with total system cost within 10\% of the least-cost system. For example, ~102,000 km\textsuperscript{2} area in California has no site suitability constraints for solar development. However, only 15\% of that area can be used without increasing the total system cost by more than 10\%.}
  \label{fig:Downscaling}
\end{figure}

In this section, we ``downscale'' the capacity expansion results by using a least-cost algorithm to site and visualize individual wind and solar facilities using the procedure outlined in Section \ref{section:down-scaling}. Figure \ref{fig:Downscaling}(a) shows the `cost-effective' wind and solar sites, i.e., sites chosen in the least-cost solution or in one or more MGA iterations. These sites are present in at least one portfolio that achieves 100\% carbon-free electricity supply at a cost no greater than 10\% above the least-cost portfolio. Conversely, sites that are not present in at least one MGA iteration are unlikely to be present in an affordable portfolio. 

Figure \ref{fig:Downscaling}(b) computes the total land area of CPAs selected in one or more MGA iterates. We find that the total cost-effective area available for wind and solar development is less than 17\% (215,000 km\textsuperscript{2}) and 4\% (68,000 km\textsuperscript{2}), respectively, of all suitable wind and solar sites in the American West. This implies that good-quality, reasonable-cost sites are substantially more limited than the total technical potential for wind and solar development that passes basic site suitability screens. However, the total area of cost-effective wind and solar sites is 3-4 times higher than the area required for achieving zero-carbon electricity supply in the WECC at a least possible cost.  Site flexibility analysis can thus identify high priority areas where proactive efforts to mitigate land-use conflicts will have high value and other areas where siting may be flexibly configured so as to minimize conflict.

Figure \ref{fig:Downscaling}(a) and (b) show that the good-quality, reasonable-cost sites for wind and solar development are asymmetrically distributed across WECC. More than 60\% of the cost-effective area for solar development is in California, Arizona and Colorado. On the other hand, wind sites are more evenly distributed across the states with Wyoming, Utah, Nevada, Colorado, Arizona, and California, each accounting for more than 10\% of all cost-effective sites. Understanding the asymmetrical distribution of cost-effective wind and solar sites can assist multi-state collaboration to achieve zero-carbon electricity supply across the American West.

\subsection{Risk assessment of the cost-effective wind and solar sites} \label{section:RRisk}
To perform the analysis in Section \ref{section:RMap}, we exclude the sites based on site suitability constraints described in \ref{section:CPA}. However, the potential cost-effective wind or solar sites can experience other non-modeled constraints, such as not-in-my-backyard (NIMBY) opposition for wind development from local residents. Similarly, potential sites on a previously undisturbed land can experience ecological constraints due to wildlife habitat or bird migration patterns, making the site infeasible for development. In this section, we perform risk assessment of cost-effective wind and solar sites generated from the MGA analysis. 

\begin{figure}
  \centering
  \includegraphics[width=\textwidth]{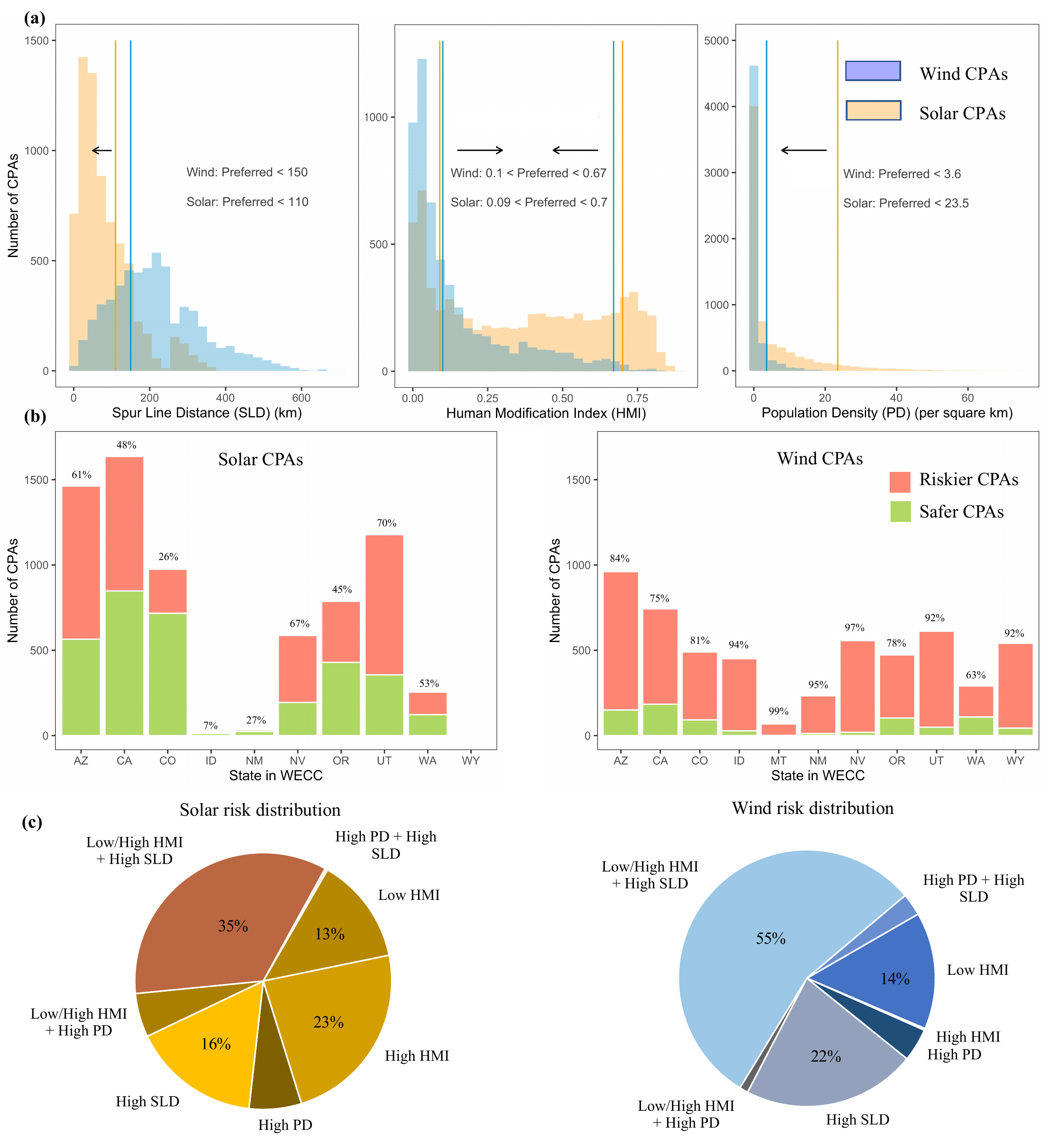}
  \caption{(a) Distribution of cost-effective solar and wind CPAs in 3 categories of riskier land use based on distance of site from the nearest metro area ($>$750,000 population), population density at the site and human modification index (HMI) of the site. (b) Total number of cost-effective wind and solar sites that fall under one or more riskier categories given in 5(a). (c) Distribution of riskier sites by risk category for wind and solar CPAs.}
  \label{fig:RiskLand}
\end{figure}

Limitations on the computational capabilities and the scarcity of the input data restricts us from explicitly considering all relevant social, geographical and ecological constraints. To overcome this limitation, we quantify three types of riskier land-use based on 1) the distance of transmission lines needed to interconnect a candidate project area (CPA) (`spur line' distance (SLD)), 2) the population density (PD) in the area of the CPA, and 3) the human modification index (HMI) of candidate solar and wind sites. HMI is a measure of land modification by human activity, with a measure of 0 being completely undisturbed and a measure of 1 being completely modified for human use (such as an urban area or roadway). We analyze the empirical distribution of each of these risk indicator metrics for existing solar and wind sites to determine the threshold for preferred sites. SI Figure \ref{fig:solar_wind_stats} provides 10\textsuperscript{th} and 90\textsuperscript{th} percentile values for the distribution of spur line distance, population density and human modification index for existing wind and solar sites. We assume that solar and wind sites far from a large metro area, i.e. sites in top 10\% of spur line distribution, can impose unforeseen constraints while building the site-to-metro transmission lines. Moreover, we assume that sites with high population density or high HMI are populated enough to generate conflict over alternative land uses or concerns about visual impact on residents. Conversely, sites with low HMI, i.e. sites in bottom 10\% of HMI distribution, are assumed to be undeveloped lands with potentially high conservation value \cite{theobald_david_et_al_detailed_2020}. Figure \ref{fig:RiskLand}(a) shows the distribution of CPAs in each category and threshold for preferred sites for cost-effective wind and solar sites. We categorize the CPAs as `riskier' if they satisfy one or more of the criteria given in Figure \ref{fig:RiskLand}(a). Figure \ref{fig:RiskLand}(b) shows the percentage of riskier, cost-effective solar and wind CPAs by state. Figure \ref{fig:RiskLand}(c) shows the distribution of riskier land types by risk category. The sites with non-preferred HMI are split into low and high HMI sites. These metrics are only rough proxies for potential drivers of land use conflict, and they are unlikely to be exhaustive. Additionally, \textit{potential} for land use or siting conflict does not mean such sites are impossible to develop. Nevertheless, this approach illustrates the substantial portion of wind and solar candidate project areas that may be more challenging to develop than initial site suitability screens indicate. 

Figure \ref{fig:RiskLand}(b) shows that of the cost-effective sites, nearly 53\% of solar sites and more than 85\% of wind sites have riskier characteristics suggesting that wind sites may experience significantly higher social or ecological constraints than solar sites in the American West. Figure \ref{fig:RiskLand}(b) also shows that a higher number of cost-effective sites does not necessarily lead to a high potential for land-related conflict resolution. For example, a higher number of solar CPAs are cost-effective in Arizona, given their proximity to the population centres in California. However, 61\% of those solar CPAs fall into one or more of the riskier categories. In contrast, solar sites in Colorado, although lower in total number, have a higher number of safer sites. As a result, land-related conflicts can be partially resolved by favouring solar development in Colorado over Arizona. 

Land-use conflicts, however, may not be fully resolved by only using safer sites for wind and solar development. Achieving a clean-electricity future may require renewable development on riskier sites. Therefore, understanding the type of risk for cost-effective sites is crucial for future renewable development. Figure \ref{fig:RiskLand}(c) shows that a majority of the riskier wind and solar sites are in areas with unfavorable HMI, indicate strong trade-off between ecological and social challenges. In other words, reducing renewable development in sites with low HMI would drive the development to the sites with high HMI or high population density, increasing potential social challenges. Notably, population centres tend to be closer to good-quality solar sites and farther away from good-quality wind sites with high wind speed. As a result, the potential wind sites are pushed away from the population centres, increasing the spur line distance and decreasing the HMI of the sites. Knowing these trade-offs in advance will help decision makers prepare to (1) develop sites that achieve goals of ecological preservation, (2) lower the length of transmission lines to connect sites to the metro area, and (3) avoid new development in densely populated areas. Note that only 1\% of the total land area within a wind site boundary is directly impacted by development, compared to 91\% of land within a solar PV facility. As such, wind farms may represent a lower ecological impact for a given project area than solar. Nonetheless, the high spur line distance and relatively untouched nature of lands suitable for cost-effective wind sites significantly increase the potential for land-related conflicts.

\subsection{Impact of potential objectives on the riskier land use} \label{section:RExtrLand}

\begin{figure}
  \centering
  \includegraphics[width=\textwidth]{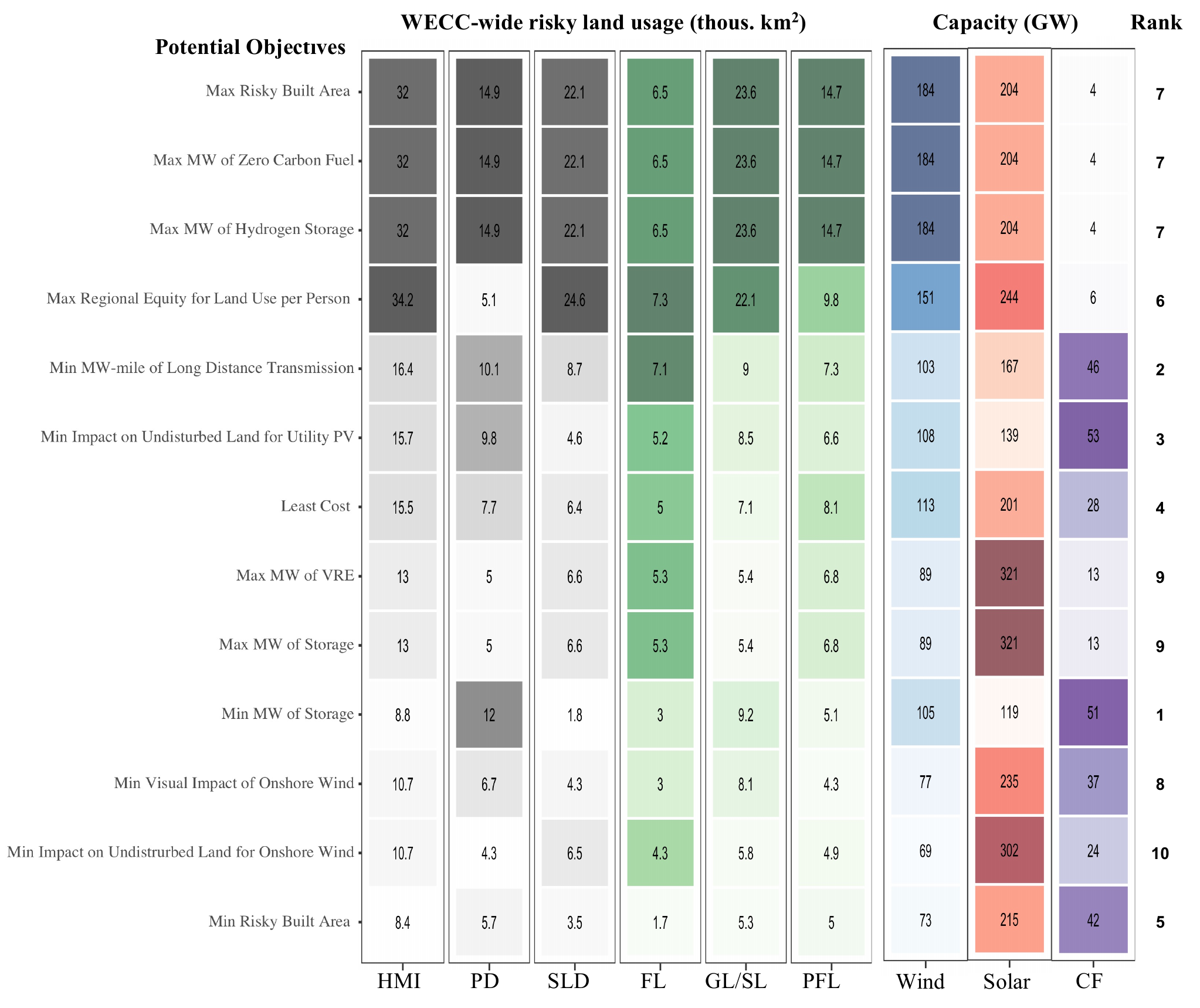}
  \caption{Impacts of ten potential WECC-wide objectives on riskier land use for solar and wind development and technology capacity. Potential objectives are in descending order of total renewable site boundary. Footprint of solar is 91\% and footprint of wind is 1\% of total site boundary. Ranking of a policy objective is based on direct renewable footprint in the riskier area (1-Lowest; 10-Highest). Non-preferred attributes: HMI - Human modification index (0.1 $>$ wind HMI $>$ 0.67, 0.09 $>$ solar HMI $>$ 0.7); PD - Population density ( wind PD $>$ 3.5, solar PD $>$ 23.5); SLD - Spur-line distance (wind SLD $>$ 150, solar SLD $>$ 110); FL - Forest land; GL/SL - Grass land or Shrub land; PFL - Prime farm land; CF - Clean firm capacity (i.e. nuclear, biomass, geothermal and natural gas power plants with carbon capture and sequestration (CCS))}
  \label{fig:LandExtr}
\end{figure}

The vast number of riskier sites from Section \ref{section:RRisk} indicates the necessity to explicitly consider potential for land use conflict while planning for solar and wind development. It also raises the question of whether less-risky, cost-effective sites are enough to attain the level of solar and wind development required to achieve clean electricity goals. This section answers the above question by analyzing the aggregated effects of potential policy objectives on riskier land use and technology capacity in the WECC. We create ten potential objectives apart from the original least-cost criterion as well as criteria for minimum or maximum total renewable site boundary in riskier areas as shown in Figure \ref{fig:LandExtr}. We assess the impact of each objective on renewable development at sites with non-preferred HMI, non-preferred population density, non-preferred spur line distance, forest land, grass-land or shrub-land, and prime farmland. We thus move from a spatially explicit map of the entire trade-off space to a quantitative assessment of various land use impacts across WECC associated with achieving a clean-electricity objective. We do so by 1) quantifying total wind and solar development area across riskier lands for each of the 160 MGA iterations, 2) selecting the iteration that satisfies each of the 10 potential objectives and 3) plotting the WECC-wide land usage within that iteration. For example, in Figure \ref{fig:LandExtr}, ‘Max MW of Clean Firm’ objective selects the iteration with maximum clean firm capacity (i.e. nuclear, biomass, geothermal and natural gas power plants with carbon capture and sequestration (CCS) \citep{baik2021different, sepulveda2018role}) from the 160 total iterations and then visualizes the total developed area in riskier land types and the capacity of wind, solar and clean firm technology category across WECC for that criterion (See SI \ref{SI:capacity} for other technology capacity). We order the potential objectives in a descending order based on the total site boundary of wind and solar and also rank them based on direct renewable footprint in riskier land areas. 

Figure \ref{fig:LandExtr} shows that policy, planning, and technology choices play a vital role in mitigating or exacerbating potential land related conflicts. With appropriate policy drivers, renewable footprint on the riskier sites can be varied by $-36\%$ to $+46\%$ compared to the least-cost solution. However, `Min Risky Built Area' criteria  shows that nearly 14,000 km\textsuperscript{2} of solar and wind sites with riskier characteristics must be developed at minimum in order to reach 100\% carbon-free supply across WECC at a cost no more than 10\% greater than the least-cost portfolio, meaning that development of riskier sites cannot be entirely eliminated if clean electricity goals are to be achieved at a reasonable cost. 

The high variability in different types of riskier land use exhibited in Figure \ref{fig:LandExtr} also indicates that decision-makers and stakeholders can manage the trade-offs and impacts of wind and solar development by pursuing different policy and planning objectives or technology priorities. For example, decarbonization policies or strategies that prioritize clean firm generation (equivalent to the results of the 'Min MW of Storage' objective) or seek to minimize development of solar on previously undisturbed lands can significantly reduce overall wind and solar development on riskier land areas. In contrast, decarbonization policies that eschew or minimize clean firm capacity (equivalent to the results of 'Max MW of Zero Carbon Fuel/Hydrogen Storage') causes 90\% increase in renewable footprint on riskier sites, with 1) 24\% increase in sites with high population density, 2) 156\% increase in development on grass/shrub land, 3) 188\% increase in development on prime farmland, and 4) 1127\% increase in short-distance site-to-metro transmission relative to the best available, i.e., 'Min MW of Storage' scenario.

Figure \ref{fig:LandExtr} also indicates trade-offs across different risk proxy metrics. Policies intended to limit the visual impact of wind farm development, for example, may increase overall development risk by increasing the total distance of transmission interconnection required to connect more remote wind sites (SLD) and increasing development of land-intensive solar PV. An objective of minimizing long-distance transmission lines also results in increased forest land usage by 42\%, and increased dependence on clean firm capacity by 64\% relative to the least-cost scenario. 

In this analysis, we also considered a policy objective of ensuring equity in land area impacted by wind and solar development per person across Western states. Though the equitable distribution of land use across states might not be a primary objective of policymakers, land use equity can be used as a proxy metric that can assist in revealing stakeholder preferences such as equity of perceived fairness in employment outcomes, the visual impact of wind development, development of riskier land types across states, or other outcomes that correlate with wind and solar area. 

These examples of upcoming trade-offs for decision-makers demonstrate urgent need for proactive land planning to accommodate growing share of renewables and careful consideration of the potential unintended impacts of narrowing the portfolio of low-carbon technologies, such as excluding or limiting clean firm generation. 

Note that the degree of variation in land use or technology capacity is dependent on other variables in the system. Also, the extent of land use and technology flexibility depends on the technology portfolio, inter-regional transmission network, resource quality, electricity demand and renewable cost uncertainty. Though dependent on underlying uncertainty, these insights still allow policymakers to steer away from policies that may have negative impacts on their objective.

\subsection{State-wise trade-off in the riskier land use} \label{section:RFlex}
\begin{figure}
  \centering
  \includegraphics[width=\textwidth]{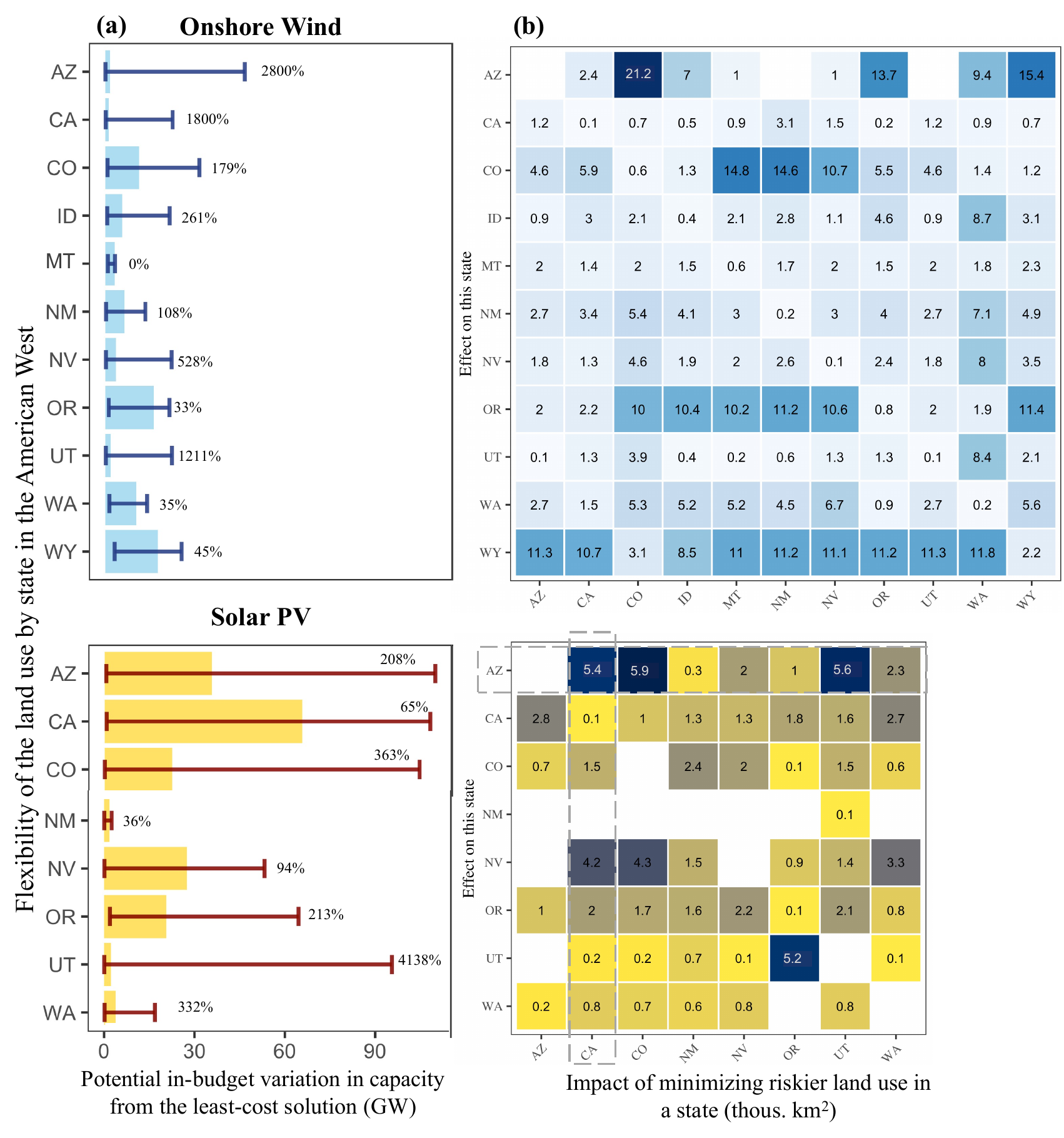}
  \caption{(a) Solar and wind capacity in the least-cost solution for each state in the WECC. The error bars show potential variation in state-wise capacity without increasing the total system cost by more than 10\% as compared to the least-cost solution, reflecting the minimum and maximum installed capacity across 160 MGA iterations. Percent values show the maximum possible percentage increase in the technology capacity as compared to the least-cost solution. (b) Each row quantifies the impact of minimizing riskier land use in one state on the rest of the states in the WECC, depicted across each column in that row. The diagonal entries from top-left to bottom-right show that riskier land use can be eliminated in nearly all states, but with increase in risky land development in various neighboring states.}
  \label{fig:FlexLand}
\end{figure}

The analysis in Section \ref{section:RRisk} show that a significant number of wind and solar CPAs may experience financial, ecological or social conflicts. Section \ref{section:RExtrLand} shows that the wind and solar development on the riskier land types cannot be completely eliminated. In this section, we explore if states can collaboratively manage renewable development on riskier sites. We learn from Section \ref{section:RMap} that the total area for cost-effective solar and wind development is 3-4 times larger than required to reach a zero-carbon electricity system in the American West by 2045 at the least cost, allowing for flexibility in site selection to resolve potential land-related conflicts. This section provides insights on the degree of state-specific site flexibility, potential ways for states to collaboratively benefit from site flexibility, and state-wise trade-offs of riskier land use for renewable development in the American West. The bars in Figure \ref{fig:FlexLand}(a) show the solar and wind capacity in the least-cost solution for each state in the WECC and the error bars show the variation in capacity across 160 MGA iterations.  In Figure \ref{fig:FlexLand}(b), each row quantifies the impact of minimizing risky land use in a given state on each other state in the WECC. To do so, we, first, identify the MGA iteration with minimum riskier wind and solar site development for each state. We then plot the riskier wind and solar site development in each state in WECC for that iteration.

This analysis shows that total land use as well as riskier land use for solar and wind development in each state in the WECC is exceedingly flexible. The diagonal going from top-left to bottom-right in Figure \ref{fig:FlexLand}(b) suggests that each state has an ability to substantially reduce if not entirely avoid renewable development on riskier sites within their borders without significantly increasing the total system cost. However, minimizing impact in one state involves increase in risky land development in one or several other states. For example, in Figure \ref{fig:FlexLand}(b), the highlighted row and column illustrates that if California seeks to minimize risky solar development within its borders, doing so forces solar development on 5400 km\textsuperscript{2} of riskier sites in Arizona, 116\% more than the least-cost scenario. As this example indicates, state planning and policy strategies that attempt to minimize risky land development in one state may significantly increase risks of siting conflict elsewhere in the WECC. If care is not taken to manage these state-wise trade-offs, unilateral state efforts to minimize development conflicts in their borders may inadvertently imperil the overall WECC-wide transition to 100\% carbon-free electricity by exacerbating conflict in other states or increasing overall system cost. As such, our analysis identifies two sets of states that act as strong substitutes for wind and solar development where more/less development on riskier lands in one state results in less/more risky development in the remaining states in the grouping. First, Arizona, California, Nevada, and Utah can act as substitutes for riskier solar development. Second, Arizona, Colorado, Oregon, and Wyoming can act as substitutes for riskier wind development. The degree of substitution between states may vary based on the types of riskier land, input data for cost and performance parameters of technologies, criteria for categorizing the riskier and safer land type, or other non-modeled geographical and social attributes. However, this analysis indicates that policy-makers in these groupings of states should take care to coordinate planning and policy for wind and solar siting to help collaboratively manage and mitigate overall siting risk and conflicts. State-wise trade-offs in total land use and trade-offs by land cover type are given in SI \ref{SI:MinState}, SI \ref{SI:MinCover}, and SI \ref{SI:state_landcover}.

\section{Discussion}\label{sec12}

Utility-scale solar PV and onshore wind farms are expected to play a key role in the transition to a zero-carbon electricity systems due to their rapidly declining costs and subsidies from the Inflation Reduction Act. Their land-intensive nature entails potential siting and land-use related conflicts. Proactive consideration and mitigation of these potential conflicts is likely to be critical to maintain the pace and scale required to meet the goals of ambitious clean energy policies. This study demonstrates the utility of a high-resolution power system capacity expansion model combined with a spatially explicit resource siting process for analyzing the land-use impacts and trade-offs associated with renewable energy development in the American West. These methods equip decision-makers and stakeholders with multiple possible siting distributions and technology portfolios to help manage land-related conflicts without substantially increasing the total system cost.

We consider a wide range of siting constraints associated with existing infrastructure, setback limits from restricted areas, water bodies, wetlands, military bases, and land conservation constraints. This initial site suitability screening suggests that roughly half of the land area of U.S. states served by the Western Interconnection (WECC) is potentially suitable for solar and wind development. Despite this vast resource potential, Modeling to Generate Alternatives (MGA) analysis conducted in this paper finds good quality, cost-effective sites represent less than 10\% of all suitable sites in the WECC. Furthermore, out of these cost-effective sites, we find that 53\% of solar sites and 85\% of wind sites exhibit characteristics such as high surrounding population density, low prior human modification, or long transmission interconnection distances that indicate higher development risk and potential for land use conflict. As a result, some development must take place on riskier wind and solar candidate project areas in order to achieve WECC-wide clean electricity goals without substantial increase in total electricity costs (e.g. $>10\%$ relative to the least-cost portfolio). Potential development risks for wind and solar sites across WECC arise from the vast acreage of previously undisturbed ecologically important landscapes, concentrated population centres in the western part of the region, the remoteness of good-quality and cost-effective wind resources, and long transmission distances required to inter-connect the regions. 

The limited area of cost-effective, low-risk sites suitable for wind and solar development in the WECC could easily lead to land-use conflicts and bottlenecks that would slow the development of renewable energy. However, this study shows that the pool of cost-effective wind and solar sites is still 3-4 times larger than required to reach a zero-carbon electricity system. As a result, western U.S. states retain significant flexibility to site wind and solar or alter the configuration of the carbon-free electricity portfolio to ameliorate (if not eliminate) potential land use conflicts. To exploit this flexibility and unlock the potential to achieve a socially acceptable distribution of impacts and benefits, decision support tools are needed that can quantify and explain trade-offs across multiple salient outcomes associated with alternative portfolios and spatial distributions of energy infrastructure. 

This paper contributes to this need by demonstrating a novel approach to map siting flexibility across cost-effective wind and solar sites and quantify the impacts and trade-offs associated with pursuing different potential objectives, such as reducing visual impacts of wind projects, land area impacted by solar, or the role of clean firm resources such as nuclear power. This work also illustrates how attempts to minimizing potential siting conflicts in one state can negatively impact land use and development risks in other states in the region. In particular, we identify two groupings of states for which siting decisions in one state have strong implications for outcomes in other states in the grouping, indicating the importance of collaborative planning and policy making across these states.

Our study constitutes a preliminary effort to combine a detailed electricity system capacity expansion model, modeling to generate alternatives, and a spatially-explicit downscaling method for wind and solar siting to improve decision support for low-carbon energy transitions. However, we outline three limitations to be addressed in future work. First, our study considers land-use restrictions and siting criteria in the post-processing steps. These criteria should ideally be modeled as constraints in the power system model to fully assess these interactions since increasing wind and solar development would affect the social acceptability and increase the land-use conflicts. Second, our capacity expansion model, GenX, assumes perfect competition, perfect foresight, perfect access to information for all market participants, equal risk preferences for investors in each technology category, as well as efficient price formation and market rules. As a result, CEMs like GenX are best employed not as predictive models but rather as decision support tools to explore possible futures using a consistent analytical framework incorporating critical engineering, economic, and policy constraints and incentives. Finally, future work involves the development of better underlying geo-spatial data and metrics to a) represent brown-field sites and other suitable areas for development that reduce strain on ecologically important areas, b) include other layers to improve the selection of underlying CPAs (e.g., fire risk), c) analyze trade-offs at a county level, and d) quantify a wider range of localized impacts and benefits associated with siting wind, solar, and other electricity system infrastructure, such as air pollution, water consumption, local property tax or lease payments, and employment-related impacts.

In the real world, siting decisions associated with renewable development ultimately take place at the local level and at a higher spatial resolution that would be more computationally intensive and thus challenging to perform. Communities will need to develop their plans to identify least-regrets policies that incorporate their community values. The framework developed in this study can make it easier for communities to develop such plans, help identify the areas where this kind of effort may be most valuable, and identify preferred spatial patterns of clean energy infrastructure deployment. As the most affordable decarbonization strategies depend on achieving unprecedented wind and solar development levels in the United States, this paper provides critical guidance for managing land use risk in future power system planning.

\section{Methods}\label{section:Method}

\begin{figure}
  \centering
  \includegraphics[width=\textwidth]{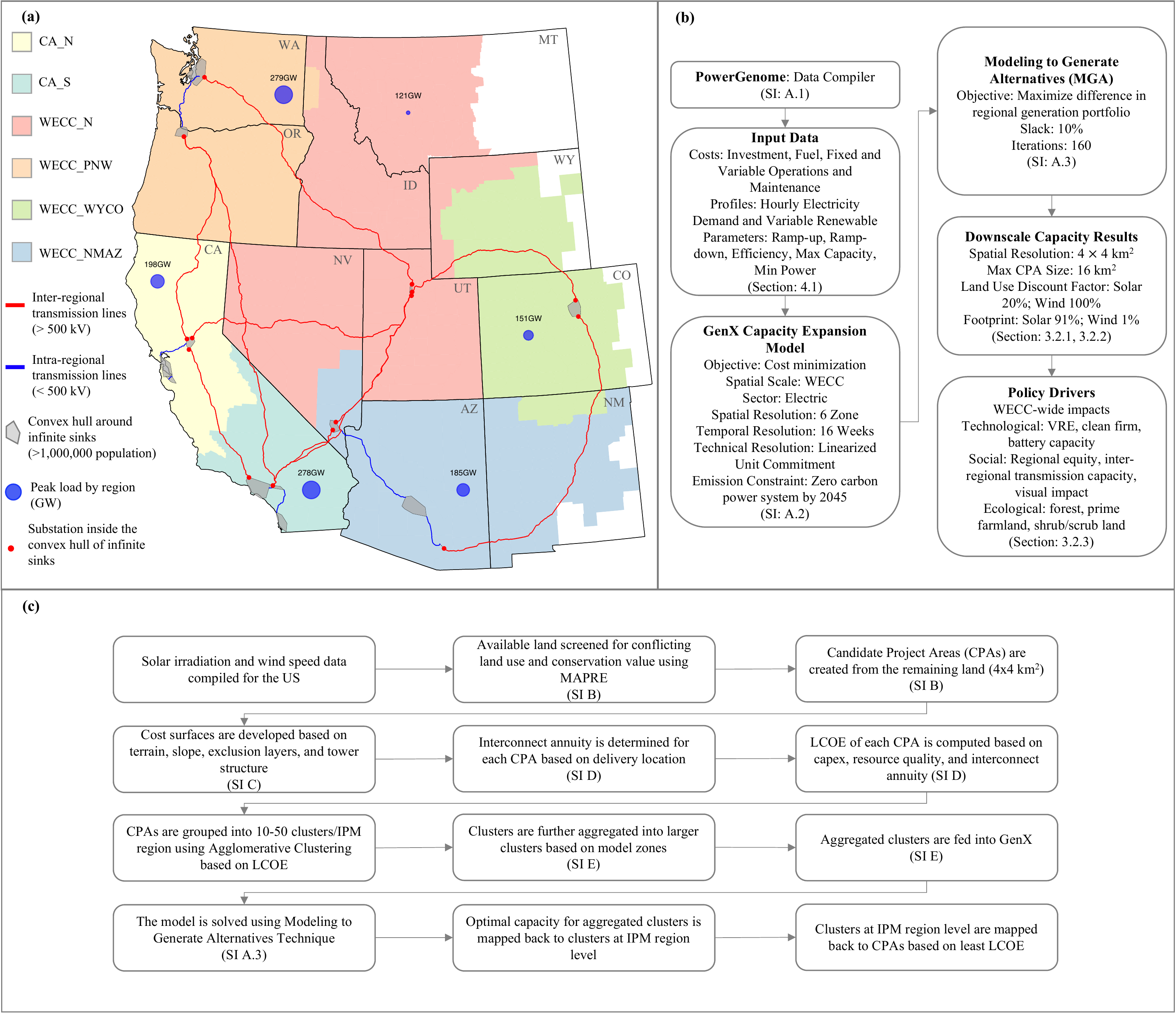}
  \caption{(a) The six model regions with inter-regional transmission lines (red lines), intra-regional transmission backbone (blue lines) and total electricity demand by region (circles) for a high electrification scenario for the year 2045. (b) Methodology for exploring land-use trade-off in the American West. (c) Overview of the methodology for creating Candidate Project Areas (CPAs) for wind and solar sites, aggregating them into larger clusters for a power system model, and mapping the optimal results from capacity expansion model back to the CPAs}
  \label{fig:flowchart_1}
\end{figure}

Our approach for quantifying land-use trade-offs involves building an electricity system planning model for the American West, implementing modeling to generate alternatives (MGA) techniques, exploring the effects of various criteria on the land-use for renewable development and understanding the ecological trade-offs in land cover types across the states associated with achieving 100\% carbon-free electricity supply by 2045. Figure \ref{fig:flowchart_1} outlines the flow of information through the analysis. 

First, Figure \ref{fig:flowchart_1}(a) depicts a 6-zone representation of the U.S. portion of the Western Electricity Coordinating Council (WECC) electricity system, also known as the Western Interconnection (see Section \ref{section:Inputs} for details). 

Second, beginning in the upper left hand corner of the flowchart in Figure \ref{fig:flowchart_1}(b), we assemble data from public sources using an open-source data compilation tool called PowerGenome (SI \ref{SI:PG}). It includes spatially-resolved land-use data to restrict wind and solar development in urban and developed areas, flood zones, areas of critical environmental concern, legally and administratively protected areas with high ecological or cultural value, and a variety of other areas in which renewable energy development is challenging or prohibited (Section \ref{section:CPA}).

Third, we feed the data into an open-source electricity system planning model called GenX (SI \ref{SI:GenX}). GenX solves the cost minimization model with linearized unit commitment constraints assuming high electrification level, low cost of mature variable renewable energy (VRE) technologies (e.g. wind, solar photovoltaic (PV), and battery storage), and low cost of zero-carbon fuel (e.g. hydrogen or another equivalent fuel with zero emissions at a cost of \$15 per million British thermal units (MMBtu)) (SI \ref{SI:Data}).

Fourth, we apply the MGA technique (SI \ref{SI:MGA}) to generate 160 near-least-cost electricity resource portfolios, each with costs less than 110\% of the least-cost solution. Note that all 160 MGA portfolios have the same input parameters but different technology compositions that all lead to similar total system cost. SI \ref{SI:MGA} demonstrates that 160 MGA iterations are adequate for exploring the near-optimal feasible space in this study. 

Fifth, we downscale the capacity expansion results (Section \ref{section:down-scaling}) from the MGA iterations for utility PV and onshore wind to individual Candidate Project Areas (CPAs) at a 4$\times$4 km\textsuperscript{2} resolution. Figure \ref{fig:flowchart_1}(c) provides details for CPA identification and downscaling methodology.

Sixth, we define a set of potential objectives (Section \ref{section:ExtrScen}) to explore the near-optimal feasible space obtained from the MGA iterations. These portfolios quantitatively assess MGA iterations using a number of key metrics, including capacity deployment requirements, regional equity in wind and solar capacity outcomes across the states in the American West, impacts on specific land cover types, land area impacts in each state, transmission expansion, and other outcomes of interest. We find that these criteria offer additional insight and alternative portfolios more salient to decision-makers than the conventional 'maximally-different' iterates selected as per the Harmonic Mean of Squared Euclidean Distance (HMSED) methodology proposed in \cite{berntsen2017ensuring}. 

As a result of this analysis, we first identify all solar and wind CPAs in each state that appear in at least one MGA iterate and define these as `cost-effective' sites, since they can be developed in at least one feasible portfolio with costs no greater than 110\% of the least-cost portfolio (Section \ref{section:RMap}). Second, we use several criteria as proxies (population density, human modification index, and spur line distance) to identify solar and wind sites that may experience additional ecological or social constraints or siting conflicts and categorize them as riskier sites as discussed in Section \ref{section:RRisk}. Third, we quantify the effects of several potential objectives on land use and land cover trade-offs across the states in WECC as described in Section \ref{section:RExtrLand}. Fourth, in Section \ref{section:RFlex}, we quantify the overall variation in wind and solar capacity deployed in each state across all MGA iterates as well as all state-wise trade-offs in risky land development. Lastly, we identify two groupings of states that can act together to resolve potential wind and solar siting conflict by shifting sites with higher potential for conflict to more suitable sites in another state. 

We run the model using a computing cluster containing two Multi-Core IntelXeon E5-2600 series processors, representing a total of 24 compute cores. The resulting linear model is solved using CPLEX. The computational time to solve the least cost model is 250-300 minutes while solving each MGA iteration take 180-250 minutes for a given budget constraint.

\subsection{WECC case study}  \label{section:Inputs}

This paper analyzes the U.S. portion of the Western Electricity Coordinating Council (WECC) power system, one of three synchronized electricity grids serving the United States. We aggregate the states based on the U.S. Environmental Protection Agency's IPM regions into six model zones to build the GenX compatible power system database, henceforth called the WECC database. 

The zones reflect major transmission path constraints between regions in the WECC, as shown in Figure \ref{fig:flowchart_1}(a). The long-distance transmission network is divided into two types: inter-regional transmission and intra-regional transmission. The inter-regional transmission represents the transmission lines that connect the model regions defined in the GenX model (with transmission paths represented by the red routes in the Figure \ref{fig:flowchart_1}(a)). The intra-region transmission network represents transmission 'backbone' lines within each zone connecting major metropolitan areas with population $>$1 million if more than one is present within the zone (blue routes in the figure).  Transmission power flows, constraints and capacity expansion decisions are explicitly modeled between the six zones. Power flows within each region are assumed to be unconstrained and are not modeled explicitly, but for each MW increase in the inter-regional transmission capacity, the model needs to build an equivalent capacity of the intra-region network to ensure power may flow between major population centers within each zone, and this cost is added to the cost of transmission expansion between zones (SI \ref{SI:inter-regional transmission}). Additionally, the cost of ‘spur lines’ or shorter-distance transmission lines to connect from specific generating resource locations to demand centres within each zone are explicitly modeled as part of the location-specific cost of each generating resource in each region (SI \ref{SI:LCOE}).

We draw the WECC power system database for each IPM region from various publicly available sources (SI Table \ref{Tab:datasource}) and aggregate into model zones using PowerGenome \citep{PowerGenome} for 2045 (see SI \ref{SI:PG}). We then cluster the hourly load and renewable profile data into 16 representative weeks using a k-means clustering algorithm \citep{likas2003global, mallapragada2018impact} to reduce the computational complexity of the model. 

To understand the land cover impacts of our various generation portfolios, we use the National Land Cover Database (NLCD) \citep{jin2019national} to map the selected CPAs to their respective land cover types located in seven distinct land cover types out of the 20 land cover types described in NLCD. We aggregate the selected land cover types into three types: forest (evergreen deciduous, and mixed forest), grass/shrub land (grass, lichen, moss, sedge, dwarf scrub and scrub land), and prime farmland. Developed or barren lands experience low density of renewable development and hence, are not reported in this analysis. Detailed assumptions for the input data are provided in SI \ref{SI:Data}. All the input datasets and the model used for this analysis are publicly available for testing and verification on a Zenodo repository \citep{patankar_neha_2022_6897346}.

\subsection{Experimental design for exploring land-related trade-offs} \label{section:ExpDesign}

Given all input assumptions in the WECC database, we determine the least-cost portfolio consistent with a 100\% carbon-free electricity supply across the WECC, and then perform MGA analysis to identify all near-optimal feasible solutions with costs no greater than 110\% of the least-cost solution. Region-specific capacity results from this MGA analysis uncover potential technology trade-offs in the WECC. However, aggregate regional capacity results fail to uncover potential land use conflicts, ecological impacts, or regional conflict reduction strategies on a more granular level. We address this limitation by down-scaling the region-specific capacity results to individual wind and solar candidate project areas (CPAs) at 4 km x 4 km resolution using the methodology described in Sections \ref{section:CPA} and \ref{section:down-scaling}. We downscale the capacity expansion results for all 160 MGA iterations to fully explore the near-cost-optimal feasible space. We then filter MGA iterations that satisfy ten potential objectives for uncovering trade-off between five types of risky land usage and prime farmland usage in Section \ref{section:ExtrScen}.

\subsubsection{Identification of candidate project areas (CPAs)} \label{section:CPA}

To downscale the region-specific capacity expansion results, we first use sixty GIS layers in the Multi-Criteria Analysis for Planning Renewable Energy (MAPRE) tool \citep{ranjit_deshmukh_grace_wu_multi-criteria_2019, wu2020low} to create siting exclusions based on well-defined techno-economic and environmental criteria. Second, we construct candidate project areas (CPAs) for wind and solar sites in a 4 km x 4 km grid and identify developable areas within each grid cell that pass all sixty siting exclusions. Third, we estimate the maximum wind or solar capacity that can be deployed in each CPA based on developable area in each CPA and an average power density of 2.7 MW/km\textsuperscript{2} for onshore wind and 45 MW/km\textsuperscript{2} for solar based on empirical analysis of existing wind/solar farms performed in \cite{larsonnetzero2021} Annex D. We further derate the CPA capacity to reflect spacing between individual solar panels, spacing and wake effect of wind farms, and effects of population density as detailed in SI \ref{SI:CPA_postprocessing}. Lastly, we assign CPA attributes such as capital cost, transmission cost, and population density using the input datasets summarized in SI Tables \ref{table:rawResource} and \ref{table:attributeDatasets}. SI Table \ref{table:SiteSuitabilityDatasets} provides an overview of site exclusions based on site slope, population density, urban area buffers, water bodies, military installations, active mines, airports and railroads. Methodologies for determining CPAs, creating routing cost surface, computation of levelized cost of electricity (LCOE), and clustering CPAs are provided in SI \ref{SI:CPA selection},\ref{SI:cost surface},\ref{SI:LCOE} and \ref{SI:renewable clusters}.

\subsubsection{Down-scaling of the capacity expansion results} \label{section:down-scaling}
Large-scale capacity expansion models have a limited computational capacity to consider renewable development at the CPA resolution level. To maintain computational tractability, we first use the agglomerative clustering methodology \citep{ward1963application} to aggregate the CPAs into larger clusters using the method shown in Figure \ref{fig:flowchart_1}. Second, aggregated clusters are fed into the power system model, GenX, and the model is solved with capacity decisions at the cluster level using the MGA technique. Through the aggregation process, we retain the link of each aggregated cluster to individual CPAs. Lastly, we develop a site selection algorithm that selects individual CPAs that provide the total capacity selected by GenX at the cluster level by minimizing the levelized cost of energy, taking into account resource quality (location-specific wind speed and solar irradiance), capital cost, and transmission cost of connecting CPAs to the nearest population centre. We use this methodology to downscale the capacity expansion results for each MGA iteration.

While down-scaling the capacity expansion results, we focus on two primary definitions of land use impacts: the total site boundary of new-build wind farms and utility-scale PV, and the direct footprint (e.g. the area of a project impacted by the placement of wind turbine pads, solar arrays, roads, substations, and other equipment). Total site boundary acts as a proxy matrix for understanding visual impacts of new onshore wind turbines and the total developable area used by a wind or solar facility. Direct renewable footprint is a measure of the land area unavailable for other uses. Given that wind farms require significant spacing between turbines to avoid wake effects, the footprint (directly impacted land) of wind farms is only 1\% of the total wind farm area, which allows land within the total site boundary of a wind farm to be used for multiple purposes, such as cultivated crops or grazing lands \citep{rand2017thirty, cohen2014re}. Utility scale PV projects are much more land-intensive within a given site boundary, with a 91\% footprint. Note that we focus here on the siting of wind and solar facilities, and we do not account for the upstream, transmission, and downstream effects of a given technology. 

\subsubsection{Potential objectives} \label{section:ExtrScen}

The methodology in Section \ref{section:down-scaling} lets us analyze the state-specific land use for all the MGA iterations. To highlight this range of impacts, we select MGA iterates that reflect variation in ten potential criteria across three types of objectives: technology capacity, social objectives, and ecological objectives. Technology capacity criteria are defined as the minimum and maximum installed capacity (WECC-wide) for, separately: clean firm, variable renewable energy (VRE), storage, and zero-carbon fuel power plants (ZCF). For ecological criteria, we focus on minimizing the impact of wind and solar development on previously undisturbed land. For social criteria, we include minimizing the capacity of high-voltage inter-regional transmission lines, minimizing the visual impact of new wind turbines, and maximizing equity for land use per person across the states in the WECC. Additionally, we identify two portfolios with the minimum and maximum overall renewable site boundary on riskier land sites of onshore wind and utility-scale solar PV as described in Section \ref{section:RRisk}. Table \ref{Tab:extr} shows the technology eligibility for each criteria explored in this analysis.

\begin{table*}
  \footnotesize
	\centering
	\caption{Qualified technologies for quantifying the impact of ten potential criteria} \label{Tab:extr}
\resizebox{\textwidth}{!}{%
\begin{tabular}{c|c|c|ccccccccccccc}
  \toprule
  Types&	Sub-Types &	Criteria & \rot{Utility PV} & \rot{Onshore Wind} & \rot{Offshore Wind} & \rot{Nuclear} & \rot{NGCC-ZCF} & \rot{NGCT-ZCF} & \rot{NGCCS} & \rot{Biomass} & \rot{Geothermal}  & \rot{Hydroelectric} & \rot{Li-ion Battery} & \rot{Pumped-hydro} & \rot{H\textsubscript{2} Storage}\\
	\hline
	Technology &	VRE &	Max &  \checkmark & \checkmark & \checkmark &\\
	& Zero Carbon Fuel & Max & & & & & \checkmark & \checkmark\\
	& Hydrogen Storage & Max & & & & & & & & & & & &  & \checkmark\\
	& Storage & Max & & & & & & & & & & & \checkmark & \checkmark & \\
	& Storage & Min & & & & & & & & & & & \checkmark & \checkmark & \\
	\hline
	Social & Inter-Regional Transmission & Min & & & \multicolumn{9}{c}{MW-mile of long distance transmission}\\
	& Visual Impact of Wind & Min & & \checkmark\\
	& Regional Equity & Max & \checkmark & \checkmark \\
	\hline
	Ecological & Previously undisturbed land & Min & \checkmark &  &\\
	Ecological & Previously undisturbed land & Min &  & \checkmark &\\

	\bottomrule
\end{tabular}
}
\begin{tablenotes}
\scriptsize
\item
VRE - Variable Renewable Energy; ZCF - zero-carbon fuel, CCS - Carbon Capture and Sequestration
\end{tablenotes}
\end{table*}

For minimizing the impact on previously undisturbed land, we minimize the selection of CPAs with human modification index (HMI) less than 0.1. Similarly, for minimizing the visual impact of wind sites, we minimize the CPA selection in areas with population density greater than 3.5 people per km\textsuperscript{2} for wind. For quantifying the uniformity of the distribution of land use per person (as a proxy measure of equity across states), we use a statistical measure based on the population-weighted Gini index \cite{gini1912memorie} as described in \cite{sasse2019distributional}. Statistical measure equivalent to 100\% denotes perfectly equitable and 0\% denotes perfectly inequitable distributions. We computes equity indices separately for onshore wind and utility PV and add to find the MGA iterate with highest regional equity. Equity index for land-use can be considered as a proxy metric for equitable distribution of both siting-related impacts and localized benefits associated with renewable energy deployment, such as employment, air quality, and water usage. Future work could assess these outcomes of interest explicitly.

\backmatter

\section{Declarations}

\begin{itemize}
\item Availability of data and materials - All the data and material used for the analysis in this paper is available at \cite{patankar_neha_2022_6897346}.
\item Code availability - All the code used for the analysis in this paper is available at \cite{patankar_neha_2022_6897346} or \cite{GenX}.
\item Authors' contributions - Conceptualization, N.P., J.J.; methodology, J.J, N.P., E.L, G.S.; investigation, N.P, X.B.; visualization, N.P., X.B., G.S., E.L.; supervision, J.J.; writing - original draft, N.P., X.B.; writing- review \& editing, J.J., E.L., G.S.
\end{itemize}

\begin{appendices}

\section{Modeling tools} \label{SI:method}

\subsection{Data compilation tool: PowerGenome} \label{SI:PG}

This study requires the data for existing generation units, transmission constraints between model regions, hourly load profiles, hourly generation profiles for variable renewable energy (VRE) resources, fuel price projections, and cost estimates for generating units for the WECC power system for 2045. We build the input dataset required for this study using a python-based open-source data compilation tool called PowerGenome \citep{PowerGenome}.  PowerGenome uses electric utility data from Public Utility Data Liberation (PUDL) database \citep{PUDL}, which collates a relational database using public data from the U.S. Energy Information Administration, Federal Energy Regulatory Commission, and Environmental Protection Agency. PowerGenome also uses cost estimates for new generation and storage resources from the NREL Annual Technology Baseline (ATB) report \citep{akar20202020}, wind and solar availability profiles (at 13-km resolution) from Vibrant Clean Energy \citep{clack2016demonstrating, clack2017modeling} using the NOAA RUC assimilation model data, and distributed generation profiles from Renewable Ninja web platform \citep{pfenninger2016long}. PowerGenome acts as a pre-processing step for creating inputs for power system CEMs by cleaning, standardizing, and cross-linking the data for user-specified aggregation of EPA's Integrated Planning Model (IPM) regions. Moreover, PowerGenome uses a k-means clustering algorithm to aggregate the hourly VRE and load profiles into a user-specified number of representative time slots for the given model year. The aggregation of time series data allows the user to adjust the computational complexity of the model. The data generated by PowerGenome is fed into a power system CEM called GenX, described in the following section.

\subsection{Power system capacity expansion model: GenX} \label{SI:GenX}

Since this paper uses a power system capacity expansion model (CEM), we provide a brief description of a GenX model \citep{jenkins2017enhanced, GenX} used to optimize generation portfolios that achieve deep decarbonization with high renewable energy penetration. GenX is a highly configurable optimization modeling tool that solves for system-wide cost-optimal electricity generation, storage, demand-side resources, and transmission infrastructure in a future planning year subject to grid operational, engineering constraints and policy constraints, and with a user-defined level of spatial and temporal granularity. The model incorporates large numbers of individual renewable resource sites to reflect the increasing shares of variable renewable energy resources in the electricity generation portfolio and represent heterogeneous temporal profiles, resource quality, and transmission grid connection costs of different wind and solar sites. Variable renewables require high-resolution temporal and spatial data to ensure reliable operations and address spatial trade-offs in generator siting and network expansion decisions, increasing the size of CEM optimization problems \citep{mallapragada2018impact, jenkins2018electricity}. In this study, GenX was made more computationally tractable while still capturing the constraints and decision variables of increasingly renewable energy systems by using the k-means time domain reduction technique based on \cite{mallapragada2018impact} that temporally aggregates hourly time series data (wind, solar, hydro and demand profiles, including a key intra-annual extreme point in the time series) into a smaller set of representative time periods without averaging out peak load periods \citep{gabrielli2018optimal, kotzur2018impact, pfenninger2017dealing, schutz2018comparison}. The k-means algorithm clustered year-long hourly time series into 16 representative weeks with 168 consecutive hourly time steps. For the given set of technologies $g \in G$, model zones $z \in Z$, time steps $t \in T$, transmission lines $l \in L$ and demand segments $d \in D$, the objective function of GenX after weighted k-means clustering is given as follows.

  \begin{equation}\label{eq:GenXObj}
  \begin{split}
  &\text{min} \enskip f(\Omega,\Delta,\theta,\pi,\Lambda,\chi,\gamma) = \text{min} \enskip \Bigg\{\\& \sum_{g \in G}\sum_{z \in Z} \left ( C^{I}_{gz}U_{gz}\Omega_{gz} +  C^{F}_{gz}U_{gz} \left (\Omega^{0}_{gz}+\Omega_{gz}-\Delta_{gz} \right ) \right )\\& 
  + \sum_{g \in G}\sum_{z \in Z}\sum_{t' \in T'} \left ( w_{t'}(C^{V}_{gz}+ C^{f}_{gz}) \sum_{h \in t'}(\theta_{gzh}) + w_{t'}C^{V} \sum_{h \in t'}(\pi_{g,z,h}) \right )\\& 
  + \sum_{z \in Z}\sum_{t' \in T'}\sum_{d \in D} (C^{N}_{s}\sum_{h \in t'}(\Lambda_{zhs}))\\&
  + \sum_{g \in G}\sum_{z \in Z}\sum_{t' \in T'} (C^{S}\sum_{h \in t'}(\chi_{gzh}))+ \sum_{l \in L}(C^{T}_{l}\gamma_{l}) \Bigg\}\\
  \end{split} 
  \end{equation}

where, $C^{I}, C^{F}, C^{V}, C^{f}, C^{N}, C^{S}$ and $C^{T}$ are annually amortized investment cost, fixed cost, variable cost, fuel cost, cost of non-served energy, start up cost and transmission power flow reinforcement cost, respectively. The decision variables in the above problem are new capacity $\Omega$, retiring capacity $\Delta$, electricity generation $\theta$, electricity consumption $\pi$, non-served energy $\Lambda$, generator start-up event $\chi$ and transmission capacity expansion $\gamma$. The parameters $U$ and $\Omega^{0}$ are unit size and existing capacity of the technology, respectively. The above objective function is subject to various constraints detailed in \citep{GenX}. The constraints include resource investment and retirement constraints, transmission network flow, demand balance, transmission network loss, operational constraints, linearized unit commitment constraints, storage, flexible demand management and emission limit constraints. 

The parameter $w_{t}$ are hourly weight parameters that reflect the number of time periods in the full year time series represented by each of the model’s sample time periods (e.g. 16 7-day sample periods used herein) as per the k-means clustering based time series aggregation method described above. The parameter $t'$ is defined as the sample time period (day/week) and h is defined as the hours of that typical period. 

\subsection{Modeling to generate alternatives (MGA)} \label{SI:MGA}

For the analysis in this paper, we modify the cost-minimization objective function of GenX given in Equation \ref{eq:GenXObj} to employ the modeling to generate alternatives (MGA) technique. MGA sequentially solves a series of optimization problems in order to automatically generate near-optimal electricity resource portfolios. MGA was introduced by \cite{brill1982modeling} as hop-skip-jump algorithm that minimizes the weighted sum of decision variables from the previous run. MGA has since been modified and applied to various energy system optimization models \citep{brill1990mga, decarolis2011using, decarolis2016modelling, li2017investment, jing2019exploring, price2017modelling, trutnevyte2012context, lombardi2022redundant}. Exploring the near-optimal solution space of these optimization problems is achieved by introducing an additional constraint to the system limiting total system cost as a percent increase relative to the least-cost portfolio’s total cost, and then replacing the primary objective of minimizing the total system cost with a secondary objective, which in this study is to maximize the difference in electricity generation by resource type and region. For this study we thus use \cite{berntsen2017ensuring} heuristic algorithm that sidesteps the computational intensity of mixed integer MGA formulation proposed by \cite{price2017modelling} and instead computes harmonic mean of squared euclidean distance (HMSED) to iteratively explore the near optimal solution space and identify maximally different generation portfolios from hundreds of options. Our linearized approach aggregates the power generation of each generator $g \in G$ by resource type $r \in R$ for each model region $z \in Z$ using the variable $P_{zr}$ and a slack constraint representing the maximum permitted increase in total annualized system costs relative to the least-cost portfolio. This study’s MGA formulation is thus: 

  \begin{align}
  \text{max/min} \quad \label{eq:MGA}
  &\sum_{z \in Z}\sum_{r \in R} \beta_{zr}^{k}P_{zr}\\
  \text{s.t.} \quad
  &P_{zr} = \sum_{g \in G}\sum_{t' \in T'} w_{t} \theta_{gzt'r} \label{eq: 9} \\ 
  & f \leq f^* + \delta \label{eq: 10}\\ 
  &Ax = b \label{eq: 11}
  \end{align}
  
The parameter $\theta_{gzt'r}$ is electricity generation from a resource of technology $g$ in resource category $r$ in zone $z$ in a typical time period $t'$ within resource type $r$. $\delta$ is defined as the maximum permitted increase in cost versus the least-cost solution (this study uses an increase of 10\%), and $\beta_{z,r}$ is a random objective function coefficient between (0, 1) for MGA iteration $k$. We solve the above MGA formulation with minimization and maximization objective functions and for 120 iterations to explore the near optimal solution space of the optimization problem given in Equation \ref{eq:GenXObj}. 

\begin{figure}
  \centering
  \includegraphics[scale=0.4]{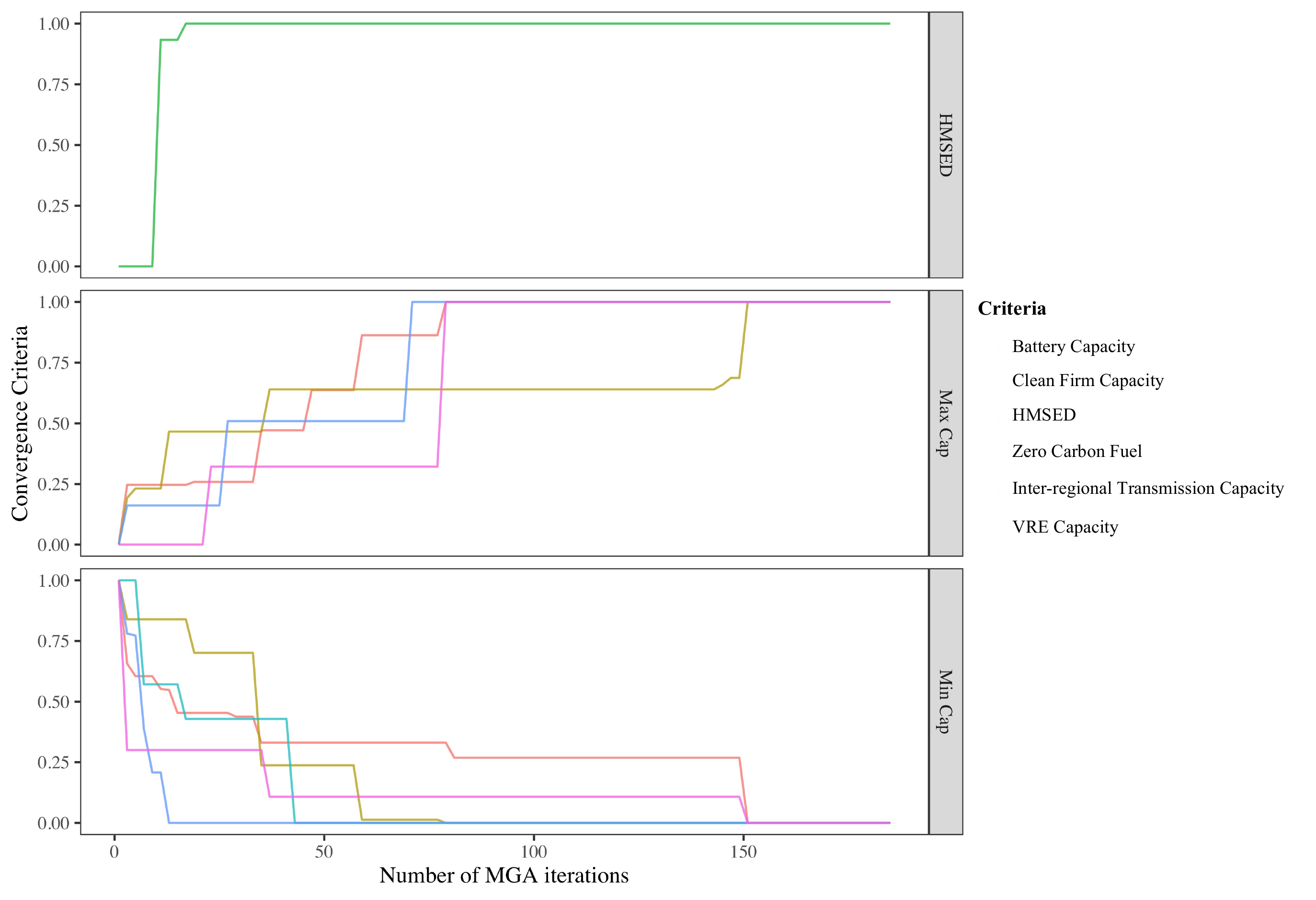}
  \caption{MGA iterations needed for convergence}
  \label{fig:Conv}
\end{figure}

We employ the methodology developed by \cite{berntsen2017ensuring}  and \cite{trutnevyte2012context} to find the iteration with maximum harmonic mean of squared euclidean distance (HMSED) from the least-cost solution. In other words, we find the technology portfolio that is maximally different than the least-cost solution. The following SI Figure \ref{fig:Conv} demonstrates that we have generated an adequate number of MGA iterations to ensure convergence towards complete coverage of the near-optimal feasible space for all interested criteria. To determine if we have an adequate number of iterations, we plot each criteria -- the harmonic mean of squared euclidean distance (HMSED) from the least-cost solution as well as maximum and minimum capacities of various resources as defined by our extreme portfolios -- on the y-axis, with iterations on the x-axis. We select $N$ MGA iterations at a time, find the iteration with min/max capacity and save that capacity value. We increase $N$ iterations gradually and find the iterations with more extreme max/min capacities of a given resource such that further exploration of the near-optimal feasible space is a process of discovering more extreme portfolios.  

\section{Solar and wind candidate project area (CPA) selection} \label{SI:CPA selection}
We begin this section by identifying solar and wind candidate project areas (CPAs) that are suitable for new development. We then assign hourly generation profiles for each CPA. Finally, we adjust the attributes of CPAs to allow for more realistic development patterns.

\subsection{Candidate project area identification} \label{SI:CPA identification}
We identify utility-scale solar photovoltaic (PV), onshore wind, and offshore wind Candidate Project Areas (CPAs) in a 4$\times$4km\textsuperscript{2} grid, through a site suitability analysis conducted using the  Multi-Criteria Analysis for Planning Renewable Energy (MAPRE) \cite{ranjitdeshmukhgracewumulticriteria2019} tool. To identify initial CPAs, resource potential and site suitability datasets were gathered and pre-processed as inputs to the MAPRE tool. We select site suitability screens based on the applicable renewable resource type (solar, onshore wind, offshore wind). CPA data sets are produced as the output of the site suitability analysis run using MAPRE script tool B based on the applicable site suitability screens.\footnote{Note: this analysis customizes the estimation of LCOE in a post-processing step rather than use the base LCOE calculator in the MAPRE Script B tool.} For a detailed listing of the data sets included in site suitability analyses, see Table \ref{table:SiteSuitabilityDatasets}. 

\begin{table*}
\caption{Site suitability datasets (percent excluded)} \label{table:SiteSuitabilityDatasets}
\resizebox{\textwidth}{!}{%
\begin{tabular}{llllll}
\toprule
Dataset & Type & Source & Solar & Onshore Wind & Offshore wind \\
\hline
Slope (lower 48) & techo-econ & USGS & exclude $> 10^{\circ}$ & exclude $> 19^{\circ}$  & na  \\
Water bodies and rivers & techo-econ & EZMT & exclude $<250$m &  & na \\
Urban Areas & techo-econ & US Census Bureau & exclude $<500$m  &exclude $<1000$m  & na \\
Population Density (person/km\textsuperscript{2})& techo-econ & Landscan & exclude areas $> 100$  & exclude areas $> 100$ \\
Military Installation Areas & techo-econ & EZMT & exclude $<1000$m & exclude $<3000$m  & na \\
Active mines & techo-econ & USGS & exclude $<1000$m from sites & na\\
Airports arrivals and departures $>1000$/yr & techo-econ & EZMT & exclude $<1000$m & exclude $<5$mi & na \\
Airport runways & techo-econ & EZMT & exclude $<1000$m & exclude $<5$mi & na \\
Railways & techo-econ & EZMT & exclude $<250$m  & exclude $<250$m & exclude $<250$m \\
Capacity factor & techo-econ & VCE & na & $<20\%$ & na \\
Flood Zones - FEMA 1\% annual & techo-econ & FEMA & 100\% & 100\%  & na \\
Marine Protected Area - except BOEM lease areas & enviro & PADv2 & 100\% & 100\% & 100\% \\
BOEM offshore lease areas for renewable energy & techno-econ & BOEM & 100\% & 100\%  & 0\% \\
Area of Critical Environmental Concern & enviro & PADv2 & 100\% & 100\%  & na \\
Conservation easements & enviro & PADv2 & 100\% & 100\% &  na \\
Fish and Wildlife Service Areas & enviro & EZMT & 100\% & 100\%  & na \\
Historic or Cultural Area & enviro & PADv2 & 100\% & 100\%  & na \\
Inventoried Roadless Areas & enviro & PADv2 & 100\% & 100\%  & na \\
Landscape intactness (Theobald human modification index (HMI)) & techno-econ & paper/data & 0\% & 0\% & na \\
Local Conservation Area & enviro & PADv2 & 100\% & 100\%  & na \\
Local Historic or Cultural Area & enviro & PADv2 & 100\% & 100\%  & na \\
Military Land & techo-econ & PADv2 & 100\% & exclude $<5$mi except for ridge crests  & na \\
Military installations, ranges and training areas & techo-econ & DoD & exclude $<5$mi & exclude $<5$mi & na \\
Mitigation Land or Bank & enviro & PADv2 & 100\% & 100\%  & na \\
National Conservation area & enviro & PADv2 & 100\% & 100\% & na \\
National Forest & enviro & PADv2 & 100\% & except for ridge crests & na \\
National Historic or Scenic Trail & enviro & PADv2 & 100\% & 100\%  & na \\
National Lakeshore or Seashore & enviro & PADv2 & 100\% & 100\% & na \\
National Park & enviro & PADv2 & 100\% & 100\%  & na \\
National Recreation area & enviro & PADv2 & 100\% & 100\%  & na \\
National Scenic, Botanical or Volcanic Area & enviro & PADv2 & 100\% & 100\%  & na \\
National Wildlife Refuge & enviro & PADv2 & 100\% & 100\% & na \\
Native American Land Area & enviro & PADv2 & 100\% & 100\% & na \\
NEXRAD radar & techno-economic & NOAA & na & $<3000$m & na \\
Prime Farmland (USA Soils Farmland Class, ESRI Living Atlas) & techno-econ & ESRI & 0\% & 0\%  &  na \\
Private Conservation (includes National Natural Landmarks) & enviro & PADv2 & 100\% & 100\%  & na \\
Private Forest Stewardship & enviro & PADv2 & 100\% & except for ridge crests & na \\
Private Forest Stewardship Easement & enviro & PADv2 & 100\% & 100\%  & na \\
Research Natural Area & enviro & PADv2 & 100\% & 100\%  & na \\
Special Designation Area & enviro & PADv2 & 100\% & 100\%  & na \\
State Conservation area  (includes National Natural Landmarks) & enviro & PADv2 & 100\% & 100\%  & na \\
State Historic or Cultural Area & enviro & PADv2 & 100\% & 100\%  & na \\
State Forests (in multiple PAD designations) & enviro & PADv2 & 100\% & 100\%  & na \\
State Park & enviro & PADv2 & 100\% & 100\% & na \\
State Wilderness & enviro & PADv2 & 100\% & 100\%  & na \\
Watershed Protection Area & enviro & PADv2 & 100\% & 100\%  & na \\
Weather radar stations & techno-econ & HIFLD & na & exclude $<5$mi & na \\
Wild and Scenic Rivers & enviro & PADv2 & 100\% & 100\%  & na \\
Wilderness area & enviro & PADv2 & 100\% & 100\%  & na \\
Wilderness study area & enviro & PADv2 & 100\% & 100\%  & na \\
National Wetlands Inventory & enviro & FWS & 100\% & 100\%  & na \\
BLM-WIND: "High" and "Moderate" Siting Considerations & enviro & BLM - WWWMP & 0\% & 100\%  & na \\
BLM-Solar Exclusions & enviro & BLM - SEP, WSP & 100\% & 0\%  & na \\
Marine Protected Areas & enviro & EZMT &  na & na  & 100\% \\
Shipping Lanes & techno-econ & EZMT & na & na  & 100\% \\
Military Installation Areas & techno-econ & EZMT & na & na & 100\% \\
Offshore Military Danger Zones & techno-econ & EZMT & na & na  & 100\% \\
Marine Restriction - Sanctuary & enviro & EZMT & na & na & 100\% \\
Areas Outside Leasing, Planning Areas & techno-econ & BOEM &  na & na  & 0\%\\
\bottomrule
\end{tabular}%
}
\begin{tablenotes}
\scriptsize
\item Abbreviations : USGS - U.S. Geological Survey; EZMT - Energy Zone Mapping Tool; LandScan - LandScan 2017 high-resolution global population data set; VCE - Vibrant Clean Energy; FEMA - Federal Emergency Management Agency; PADv2 - Protected Areas Database Version 2; DoD - Department of Defence; ESRI - Environmental Systems Reserach Institute; NOAA - National Oceanic and Atmospheric Administration; BLM WWWMP - Bureau of Land Management West-wide Wind Mapping Project; BLM SEP WSP - Bureau of Land Management Solar Exclusion Plan; BOEM - Bureau of Ocean Energy Management.
\end{tablenotes}
\end{table*}

\begin{table*}
\caption{Input datasets for CPA attribute calculations} \label{table:attributeDatasets}
\resizebox{\textwidth}{!}{%
\begin{tabular}{lll}
\toprule
Data & Source & URL \\
\hline
Population density & ORNL Landscan & \cite{bright2018landscan} \\
Electric transmission lines and substations & US DHS Homeland Infrastructure Foundation-Level Data (HIFLD) & \cite{hifldelectric2020}\\
Generation capital cost & NREL ATB & \cite{nrel2020} \\
Transmission capital cost & Regional Energy Deployment System (ReEDS) Model & \cite{cohen2019regional}\\
Existing Solar Facilities & USGS National Solar Arrays & \cite{SolarUSGS}\\
Existing Wind Facilities & USWTDB & \cite{WindUSGS}\\
Planned Wind and Solar Facilities & US EIA 860 & \cite{EIA2019}\\
\bottomrule
\end{tabular}%
}
\end{table*}

For each CPA, nameplate capacity of the applicable renewable resource is estimated based on the CPA shape area and a standard assumed power density set forth in Table \ref{table:powerDensity}. For solar power, we empirically estimate the national mean power density from the USGS National Solar Array dataset \citep{carr2016}, based on the subset remaining after the following facilities are removed: 1) plants that do not have an AC nameplate rating; or that have an AC nameplate rating, but an DC:AC ratio of less than 1.1;\footnote{Because the time frame of this analysis is 2045, and because industry trends indicate increasing inverter loading ratios over time, the lower DC:AC ratios of the historic dataset were deemed irrelevant.  Higher DC:AC ratios are anticipated going forward.} 2) power density $\leq10$ or $\geq100$ MW/ km\textsuperscript{2} (ac);\footnote{These cutoffs are arbitrary and result in the removal of the bottom 1 percent and top 7 percent of projects in the USGS database.} and 3) where power densities based on nameplate capacities reported in the USGS dataset \citep{carr2016} and EIA Form 860 \citep{EIA8602020} differ by $>5$ MW/km\textsuperscript{2}.\footnote{We chose to compare power density rather than reported total capacities as it allowed a little more flexibility in capacity discrepancies in larger systems - of which there are fewer in the data set, while not allowing relatively large reported discrepancies between small systems. This $>5$ MW/km\textsuperscript{2} cutoff is arbitrary. Some discrepancy between the USGS and EIA estimated power densities	of larger systems is allowed as the nameplate AC rating relies completely on the de-rating factor used by the company/person reporting the system. Allowed ac/dc de-rating factor range from between 1.1 to 2.4.}

For wind power, we calculate national weighted average power density from the USGS US Wind Turbine Data Base \citep{hoen2018}. Existing facilities with the following characteristics are included in the sample: 1) commercial operation date in 2017 or later; 2) nameplate capacity $\geq$20 MW (ac); 3) power density $>1$ MW/km\textsuperscript{2} (to remove outliers).

\begin{table*}
\caption{Power densities for solar PV, onshore wind and offshore wind } \label{table:powerDensity}
\centering
\begin{tabular}{ c|c } 
\toprule
Technology & Power Density (MW/km\textsuperscript{2}) \\
\hline
Solar  & 45.0 \\
Onshore wind & 2.7 \\
Offshore wind & 5.0 (fixed); 8.0 (floating) \\
\bottomrule
\end{tabular}
\end{table*}

The mean power densities calculated for this study were compared to the values reported in prior studies. Prior studies have estimated solar power densities ranging from 6.6 MW/km\textsuperscript{2} \citep{johnvanzalk2018}, to 27.8 MW/km\textsuperscript{2} \citep{Ong2013}, and 48 MW/km\textsuperscript{2} \citep{lopez2012}. Our estimate is on the high end of the range, but considered reasonable due to recent industry trends such as  increasing nameplate power rating for photovoltaic panels \citep{markbolingerutilityscale2020}. Previous studies have estimated wind power density to be 3.0 $\pm$ 1.7 MW/km\textsuperscript{2} \citep{denholm2009}.  Our wind power density is comparable, if on the low end. This is consistent with the industry trend of  declining specific power in American wind facilities \citep{bolinger2020}, as manufacturers seek to maximize annual energy production on low wind-speed sites, thereby increasing rotor diameter and decreasing nameplate specific power. For offshore wind, fixed turbine power density is assumed to be 5.0 MW/km\textsuperscript{2}  and floating turbine power density is 8.0 MW/km\textsuperscript{2}, as prior studies indicate a range of 3.1 to 18.7  \citep{rasmus2018,beiter2016spatial,energiewende2020making}.

\subsection{Methodology for creating renewable resource profiles} \label{SI:resource_profiles}
This section translates the nameplate capacity of CPAs to annual generation using hourly generation profiles for solar photovoltaic (PV), onshore wind, and offshore wind from the weather year 2012. Each CPA is assigned an hourly generation profile based on data from Vibrant Clean Energy (VCE), with the modifications described below.

\subsubsection{Spatial extent of VCE and CPA data}
\begin{figure}
  \centering
  \includegraphics[width=\textwidth]{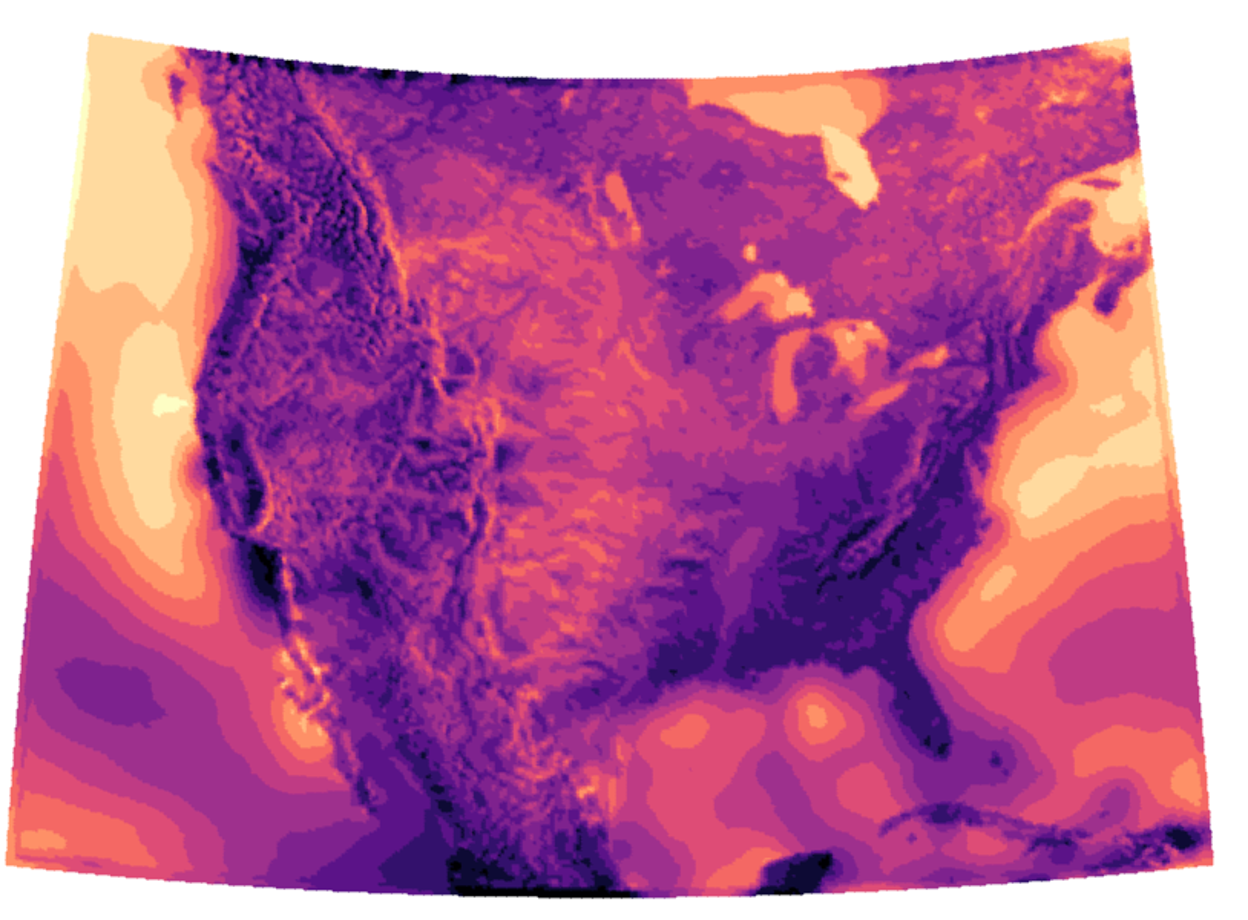}
  \caption{Spatial extent of the VCE generation data, as illustrated by 100 meter hub height wind generation.}
  \label{fig:spatialExtentVCECPA}
\end{figure}

\begin{figure}
  \centering
  \includegraphics[scale=0.5]{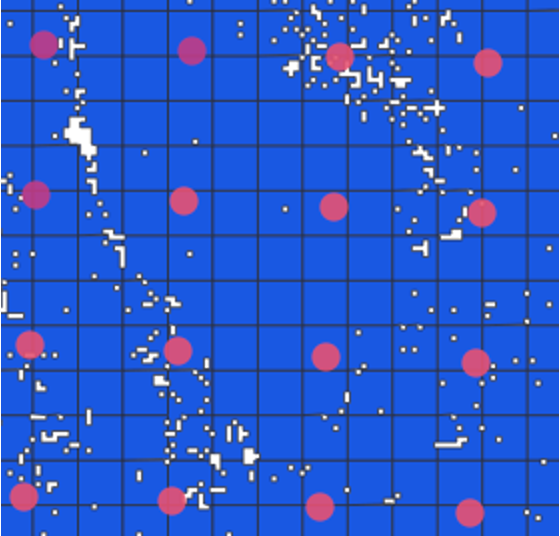}
  \caption{Grid lines represent solar CPA boundaries. Dots represent locations of VCE generation profiles.}
  \label{fig:VCE_CPA_grid}
\end{figure}

The VCE data include profiles for ~152,000 sites that are on an approximately 13$\times$13km\textsuperscript{2} grid, with complete coverage for the continental US and relevant offshore regions (Fig \ref{fig:spatialExtentVCECPA}). Solar PV and offshore wind CPAs are based on a 4$\times$4km\textsuperscript{2} in size. Onshore wind CPAs are based on a larger 8$\times$8km\textsuperscript{2} grid. The VCE grid has lower spatial resolution than the CPA grid, as shown in Fig \ref{fig:VCE_CPA_grid} where VCE sites are shown on top of a grid of solar PV CPAs. Input data sets used in the characterization of raw resource potential are summarized in Table \ref{table:rawResource}.

\begin{table*}
\caption{Raw resource potential inputs}\label{table:rawResource}
\resizebox{\textwidth}{!}{%
\begin{tabular}{llll}
\toprule
Technology & Source & Reference & Spatial Resolution \\
\hline
Solar PV & Vibrant Clean Energy Renewable Generation Dataset & \cite{clack2020renewable} & 13 km centers \\
Onshore wind & reV Model & \cite{draxl2015wind} &  2 km centers  \\
Offshore wind & Vibrant Clean Energy & \cite{clack2016demonstrating} & 13 km centers \\
\bottomrule
\end{tabular}%
}
\end{table*}

\subsubsection{Solar PV}
Annual energy generation profiles were updated to assume single-axis tracking solar configuration, to reflect industry trends indicating increasing market share of tracking versus fixed tilt solar installations \citep{34412}. VCE single-axis tilt solar profiles were developed assuming an inverter loading ratio (ILR) of 1.0 and losses of 3.3\% in the DC:AC conversion. We modified the profiles to use an ILR of 1.34 \citep{nrel2020} and additional average losses of 12\%.\footnote{4.5\% array shading, 1.5\% system degradation, 1\% availability, and 0.5\%/year system degradation losses (over 10 years = 5\%)}

\begin{align}
    & VCEprofileDC = VCEprofileAC / (1 - 0.033) \\
    & profileDC = VCEprofileDC * 1.34 * (1 - 0.12) \\
    & profileAC = profileDC * (1 - 0.033) 
\end{align}

A higher ILR will lead to some hours where the generation profile is greater than 100\% of the site capacity. All profiles are clipped so the hourly generation value cannot be greater than 1. Solar PV CPAs are assigned the generation profile of the VCE site closest to the CPA centroid.

\subsubsection{Onshore and offshore wind}

VCE wind generation profiles are available at hub heights ranging from 80m to 140m. We use 100m hub heights for onshore wind and 140m hub heights for offshore wind. Unlike solar PV, where the nearest VCE site profile is used for each CPA, wind generation profiles for each CPA are calculated using an inverse distance weighting (IDW) method. Hourly generation profiles from the nearest 4 VCE sites are averaged using the inverse square of distance to the CPA centroid as weights. 

Each of the interpolated wind profiles is then scaled to match the 2007-2013 average capacity factor (CF) at 100m hub height from NREL’s Renewable Energy Potential (reV) model \citep{maclaurin2019renewable,maclaurin2020open}. Scaling is done by multiplying the generation potential by the ratio of the reV CF over the generation profile CF. This can lead to hourly generation potential greater than 100\% in cases where the reV CF is larger than the VCE CF. An iterative process is performed, where generation potential is clipped at 1 and all hours are then scaled up again, until the CFs match within a tolerance of 0.005.

\begin{figure}
  \centering
  \includegraphics[scale=0.5]{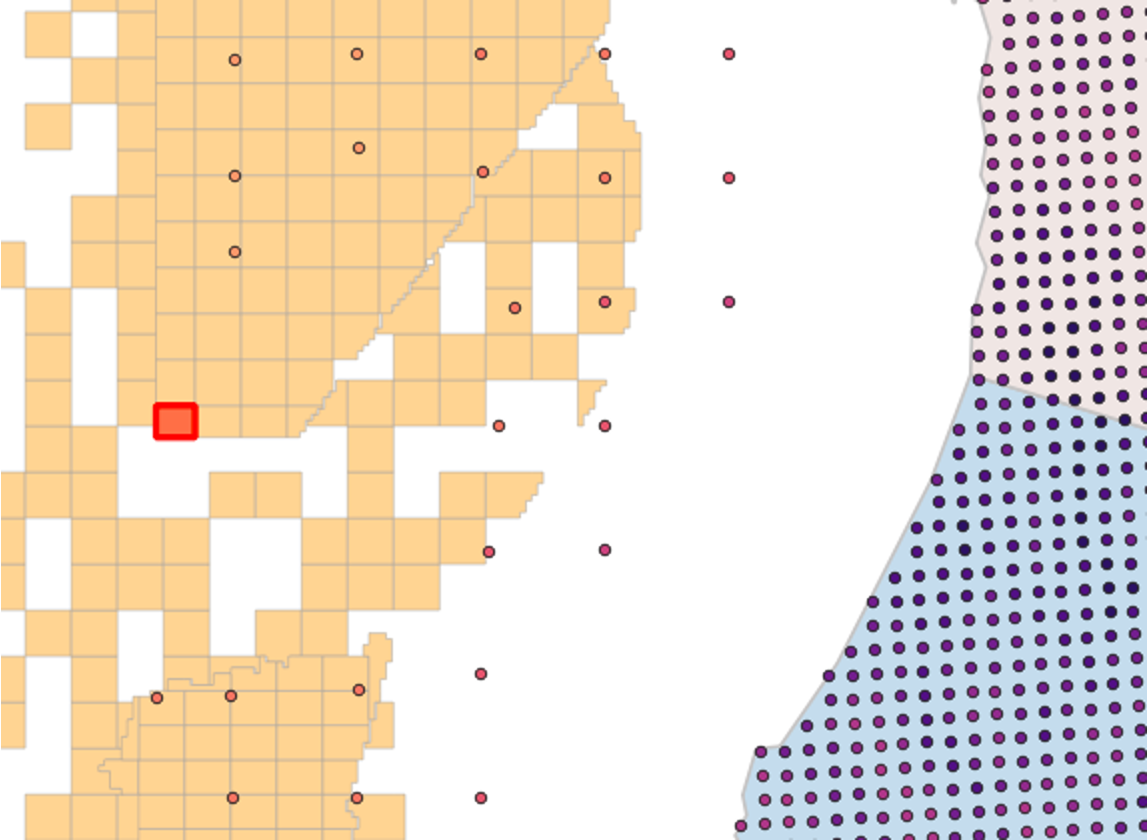}
  \caption{Orange polygons are offshore wind. Polygons with other color are onshore wind. Grid lines represent the boundary of offshore wind CPAs off the coast of California and Oregon. Dots represent location of sites with capacity factors from NREL ReV.  }
  \label{fig:CPA_grid_1}
\end{figure}

\begin{figure}
  \centering
  \includegraphics[scale=0.5]{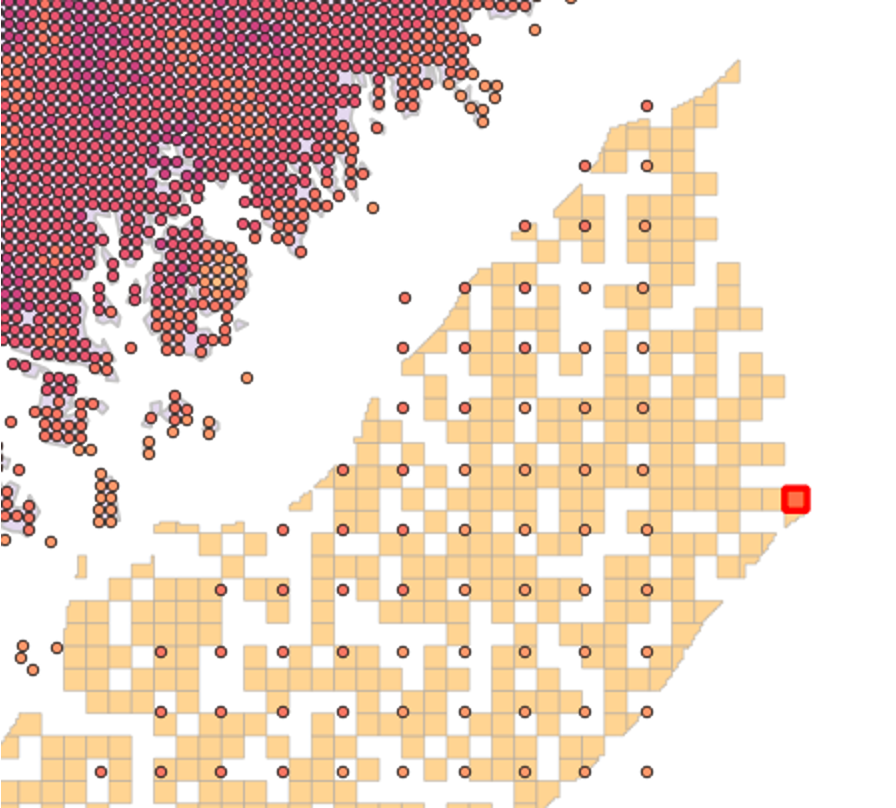}
  \caption{Orange polygons are offshore wind. Polygons with other color are onshore wind. Grid lines represent the boundary of offshore wind CPAs off the coast of Maine. Dots represent location of sites with capacity factors from NREL ReV.}
  \label{fig:CPA_grid_2}
\end{figure}

The reV data points are on a $~$2km grid on land, and the median distance from an onshore wind CPA centroid to a reV data point is 0.8km. There are fewer reV data points in offshore wind areas, and they are not all regularly spaced. The largest distances from CPAs used in this study to reV sites occur near the California/Oregon border (15.5km) and off the coast of Maine (27km), shown in Fig \ref{fig:CPA_grid_1} and Fig \ref{fig:CPA_grid_2}. 

For offshore wind CPAs, shipping lanes were considered ineligible for commercial wind development. Where publicly available GIS data fell short, conical ingress/egress polygons were further identified, defined by 20 deg north and 20 deg south of major ports.  This was based on the assumption that shipping traffic would pose challenges to development, and these areas should be avoided in modeling.

\subsection{Post-processing} \label{SI:CPA_postprocessing}
CPAs with slopes and population densities exceeding cut-offs shown in Table \ref{table:SiteSuitabilityDatasets} are removed as a first step in CPA identification process. In the post processing step, we adjust the CPAs to allow for more realistic development patterns, which account for the relative difficulty of acquiring parcels and siting projects in higher population density areas. As a final step in CPA post-processing before site selection, CPA locations are compared with the locations reported in the EIA data \citep{EIA8602020,EIA2019}, and an attribute is added to CPAs to enable the tracking and inclusion of existing and planned projects during site selection. 

\subsubsection{Population Density} \label{}
Candidate Project Areas have been made broadly eligible for development in  densely populated areas.  For the publicly available CPA dataset \citep{leslieemily20215021146}, population density attributes have been calculated and users can now apply their own population density filters and thresholds, in order to test the predictive power of population density on siting trends.

To reflect the spacing of individual projects for utility-scale solar PV, we derate the capacity of all solar CPAs by 80\%, meaning that only 20\% of the CPA area can be built upon. Similarly, to reflect the spacing and wake effect  for wind farms, we derate the capacity of all wind CPAs by 40\%, meaning that only 60\% of the CPA area can be built upon. We further derate the CPA capacity based on empirically derived data for historical capacity (per \cite{EIA2019}) and population density (per \cite{bright2018landscan}) as summarized in Table \ref{table:PopulationDensityOverview}. 

\begin{table*}
\caption{Population density overview\label{table:PopulationDensityOverview}}
\centering
\begin{tabular}{llll}
\toprule
Technology & Population density &  Derate \\
\hline
Solar & $>=$ 60 persons per km$^2$ & 100\% \\
Solar & $>=$ 40 persons per km$^2$ & 98\% \\
Solar & $>=$ 30 persons per km$^2$ & 97.5\% \\
Solar & $>=$ 10 persons per km$^2$ & 95\% \\
Solar & $>=$ 5 persons per km$^2$ & 90\% \\
Solar & $>=$ 0 persons per km$^2$ & 80\% \\
Onshore Wind & $>=$ 10 persons per km$^2$ & 100\% \\
Onshore Wind & $>=$ 5 persons per km$^2$ &  95\% \\
Onshore Wind & $>=$ 3 persons per km$^2$ &90\% \\
Onshore Wind & $>=$ 1 persons per km$^2$ &  80\% \\
Onshore Wind & $>=$ 0 persons per km$^2$ &  40\% \\
Offshore Wind - Preferred & - & 0\%\\
Offshore Wind - Non-preferred & - & 40\%\\
\bottomrule
\end{tabular}
\end{table*}

\subsubsection{Human Modification Index}
For each CPA, Human Modification Index (HMI) has been calculated as an attribute, in order to enable users to investigate the range of possible development trends with respect to human-modified environments \citep{theobalddavidetaldetailed2020}. Utility-scale energy infrastrcuture development trends tend to favor locations which are not too heavily modified (high-intensity developed areas) but also not completely unmodified (pristine wilderness and intact landscapes). 

\subsubsection{Prime farmland}
For each CPA, the presence of prime farmland is indicated as an attribute, in order to enable users to explore the importance of prime farmland as a siting criteria \citep{USDANRCS2020}. Interactions between energy development and agriculture are complex, and policy treatment varies considerably from region to region. In some jurisdictions (for example the San Joaquin Valley in California), energy development is being promoted and incentivized by local policies to support groundwater conservation and provide an alternate use for degraded agricultural soils.  In other jurisdictions (for example in Oregon and New York), prime farmland protections pose a challenge to utility-scale renewable energy development. In still another emerging trend, agrivoltaics have been promoted as a way to unlock land use  co-benefits \citep{barron2019agrivoltaics}.    

\subsubsection{Land cover type}
Land cover type is indicated as an attribute for each CPA, in order to enable exploring the effect of land cover type on the quality, attractiveness, or probability of energy development. Land cover type has been calculated using zonal statistics, and using the NLCD 2016 dataset \citep{jin2019national} which includes categories such as forest, shrub, grassland, wetland, agriculture, developed low-intensity, and developed high-intensity \citep{MRLC2019}.

\subsubsection{Existing and planned facilities}
The presence of existing and planned facilities is indicated based on USGS data \citep{WindUSGS,SolarUSGS} and EIA 860 data \citep{EIA8602020}. USGS data provide indication of existing facility footprints, with wind turbines grouped into facilities by name, and facility boundaries identified using the convex hull technique. EIA data provide indication of proposed facilities at the time of this writing. Because EIA data are publicly available in the form of points (coordinate pairs), these were converted to polygons, using a buffer radius identified as a function of facility nameplate capacity. For this reason, planned facilities are represented as circular features in this CPA dataset.

\subsubsection{Aspect}
For solar CPAs the following attributes were calculated:
\begin{itemize}
\item Aspect indicates the compass direction toward which a slope faces, measured in degrees from North in a clockwise direction from 0 to 360 \cite{USGSGapAnalysisProject2011}.
\end{itemize}
While annual energy production for solar power can be heavily influenced by aspect, most national scale models fail to account for the locational variation of this siting criteria.

\subsubsection{Airspace and military siting factors}
For wind CPAs the following airspace and military exclusions have been applied based on consultation with wind industry experts: 
\begin{itemize}
\item Airfields: For airfields with arrivals and departures $>=$ 1000 flights per year, a buffer was applied to exclude areas $<$ 5 mi \citep{EZM2021}. Note: industry stakeholders indicate exclusion buffer should be calculated as a function of turbine tip height:  5 miles for 500' TH, and 6 miles for 600' TH
\item Weather radar stations: a buffer was applied to exclude areas $<$ 5 mi
\item NEXRAD radar: a buffer was applied to exclude areas $<$ 3 km \citep{NCEI2021}
\item Military installations, ranges and training areas: a buffer was applied to exclude areas $<$ 5 mi \citep{GISdataset2021}
\end{itemize}

For wind CPAs the following attributes have been calculated, to enable future analyses exploring the relative importance of these siting factors:
\begin{itemize}
\item Visual Flight Route navigation landmarks  (buffered 2 nautical mi) \citep{VFRLandmark2021}
\item Navaid systems (buffered 8 nautical mi) \citep{NAVAIDSystem2021}
\item Special Use Airspace (floor $<=$1000 ft above ground level) \citep{GISdataset2021} Note: military stakeholders recommended this siting criteria should be considered "early consultation recommended," but should not be treated as an outright exclusion
\end{itemize}

While the airspace and military siting features listed above were visually confirmed to be quite numerous and widespread throughout the study area, the overall effect of these exclusions was to reduce wind CPA area by 5-10\%.

\section{Methodology for creating real and routing cost surface} \label{SI:cost surface}
In this section, we build two sets of real and routing cost surfaces - one for onshore CPA locations, and one for offshore locations. We begin the construction of our cost surfaces by dividing the surface area of the continental United States into pixels (we chose 250$\times$250 meter\textsuperscript{2} pixels as that was the minimum resolution selected for all the project GIS work). We then assign a cost value to each pixel. The process of assigning a cost value involves the determination of what cost multipliers should be applied to each pixel. The process is described in the pioneering work of prior studies \citep{larsonnetzero2021, wu_et_al_ecosystem_2022}. The following sections summarize the cost multipliers used for creating the real and routing cost surfaces in this study.

\subsection{Utility-scale solar PV and onshore wind turbines} \label{Section:Solar and wind cost surface}

\begin{figure}
  \centering
  \includegraphics[width=\textwidth]{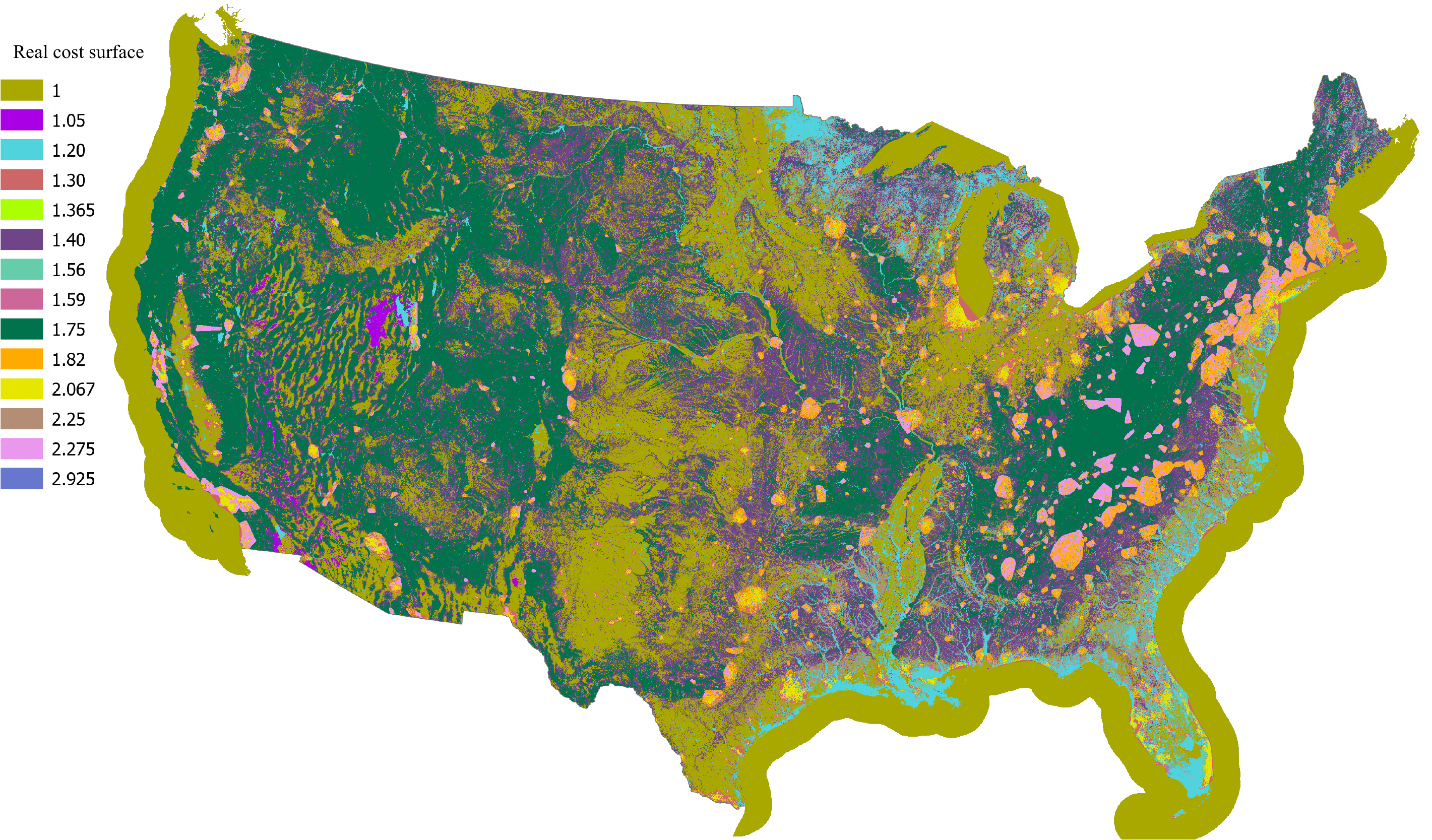}
  \caption{Real cost surface for utility-scale solar PV and onshore wind turbines}
  \label{fig:solar_cost_surface}
\end{figure}

\begin{figure}
  \centering
  \includegraphics[width=\textwidth]{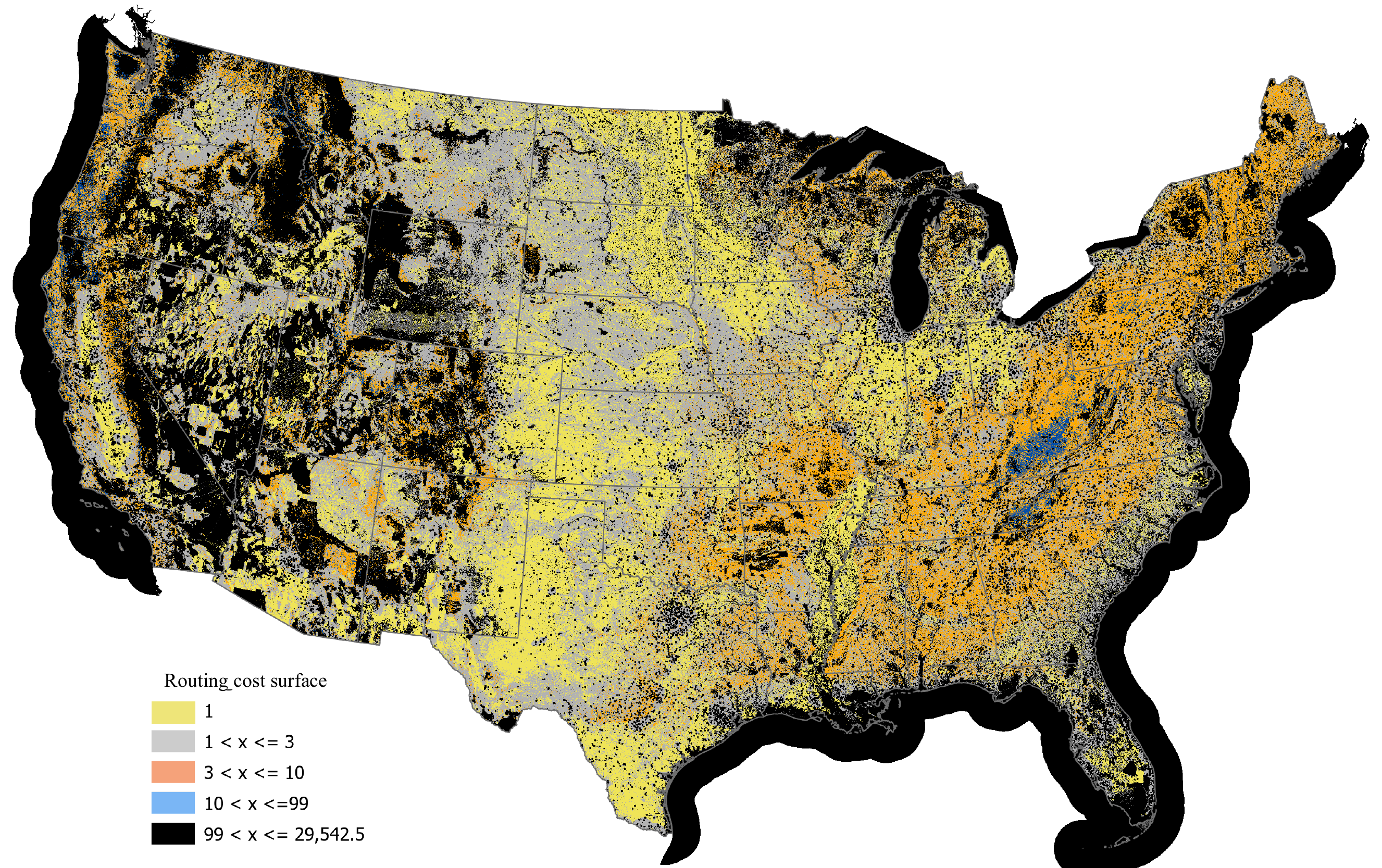}
  \caption{Routing cost surface for utility-scale solar PV and onshore wind turbines}
  \label{fig:solar_transmission_cost_surface}
\end{figure}

\begin{figure}
  \centering
  \includegraphics[width=\textwidth]{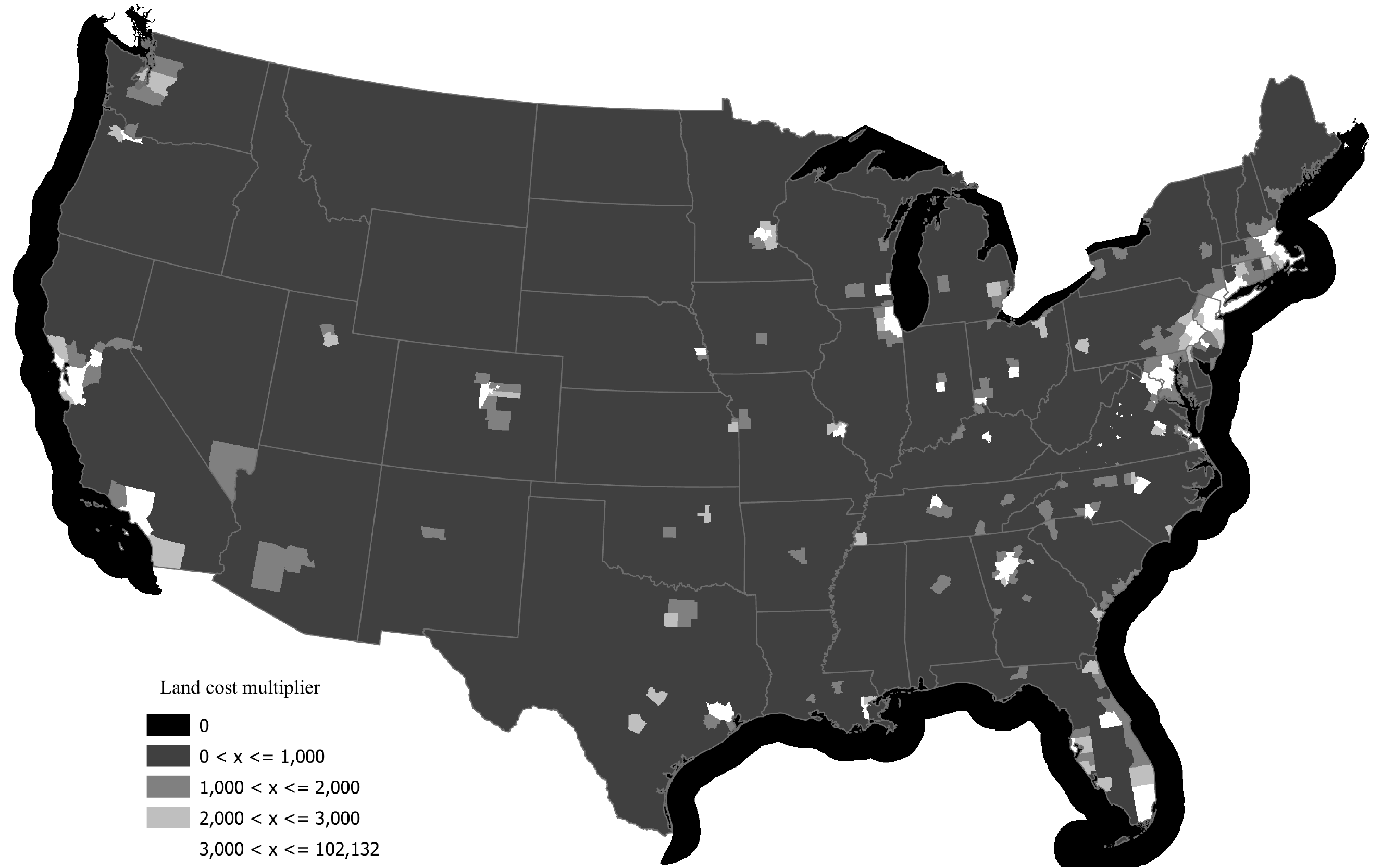}
  \caption{Land cost layer for utility-scale solar PV and onshore wind turbines}
  \label{fig:solar_land_cost_surface}
\end{figure}

The real cost surface for utility scale solar PV and onshore wind turbine is shown in Figure \ref{fig:solar_cost_surface}. The routing cost surface for utility scale solar PV and onshore wind turbine is shown in Figure \ref{fig:solar_transmission_cost_surface}. The land cost surface for utility scale solar PV and onshore wind turbine is shown in Figure \ref{fig:solar_land_cost_surface}. Table \ref{table:CostSurfAssumptions} summarizes the sources and cost multipliers (when applicable) used for creating onshore cost surfaces.


\begin{table*}
\footnotesize
\centering
\caption{Cost surface assumption for utility-scale solar PV and onshore wind turbine} \label{table:CostSurfAssumptions}
\resizebox{\textwidth}{!}{%
\begin{tabular}{p{3cm}|p{8cm}|p{4cm}}
\toprule
Type & Layer (Reference) & Multiplier (\cite{WECCTransmission} unless marked otherwise)\\
\hline
Real cost surface & Slope and Terrain \citep{USGS,LandCoverData} & 1 - 2.25 \\
& Tower structure (MSA convex hull) & 1.3$^a$ \\
Routing cost surface & Slope \citep{USGS} & 1 - 5$^b$ \\
& Terrain \citep{LandCoverData} & 1 - 2.25\\
& Tower structure (MSA convex hull) & 1.3$^a$ \\
& Exclusion layer (airports, runways and onshore wind environmental exclusions in Table \ref{table:SiteSuitabilityDatasets}) & 100$^c$ \\
& Land cost layer \citep{nolte2020high}$^d$  & - \\
\bottomrule
\end{tabular}
}
\begin{tablenotes}
\scriptsize
\item$^a$ B\&V tower cost multipliers for tubular range from 1.11x for 230 kV to 1.5x for 500 kV. In order to avoid having to generate a different cost surface for each voltage class, we used an average multiplier of 1.3 across all voltage classes.
\item$^b$ In addition to the multipliers suggested in the B\&V tool \citep{WECCTransmission}, we have added a 5x multiplier at slopes over $19^{\circ}$. Nineteen degrees was selected as this is the threshold above which we do not allow wind turbines to be sited (Table \ref{table:SiteSuitabilityDatasets}).
\item$^c$ This value is an arbitrarily selected large value to heavily discourage least-cost algorithms from crossing land parcels in this category. In order to allow transmission to cross long continuous exclusion layers (e.g. national trails), we have reduced the multiplier on existing transmission right-of-ways \citep{hifldelectric2020} crossing these layers to 1.
\item$^d$ County level land cost data from Nolte \citep{nolte2020high} was modified a) to inflate costs from 2010 to 2018, b) change units from hectare to square meter, and c) merge with a US Census Bureau country layer \citep{uscensusbureaucartographic2020}.
\end{tablenotes}
\end{table*}

\subsection{Offshore wind turbines} \label{Section:OffWind cost surface}
Table \ref{table:txCostAssumptionsOffWind} summarizes the sources and cost multipliers (when applicable) used for creating offshore cost surfaces. The offshore wind transmission surfaces are more complex than onshore surfaces as we have to allow for the additional costs and considerations of transmission when transmitting electricity using submarine cables. The additional complexity of running the offshore portions of transmission runs using submarine cables and then the onshore portions using overhead lines, leads to the need to consider more criteria than just line length and CPA capacity when selecting and costing transmission voltage and mode options for each CPA. In order to allow the method to engage with the additional complexity, we include base transmission costs in our cost and routing surfaces and generate separate AC and DC versions of the surfaces. This allows both an AC and DC routes for each CPA to be generated and compete on LCOE. 

\begin{table*}
\footnotesize
\centering
\caption{Cost surface assumption for offshore wind turbine} \label{table:txCostAssumptionsOffWind}
\resizebox{\textwidth}{!}{%
\begin{tabular}{p{3cm}|p{8cm}|p{4cm}}
\toprule
Type & Layer (Reference) & Multiplier (\cite{WECCTransmission} unless marked otherwise)\\
\hline
Real cost surface & Slope and Terrain \citep{USGS,LandCoverData} & 1 - 2.25 \\
& Tower structure (MSA convex hull) & 1.3$^a$ \\
& Transmission cost onshore & 1.7$^b$ \\
& Transmission cost offshore (AC layer) & 4.2$^c$  \\
& Transmission cost offshore (DC layer) & 1.5$^c$  \\
Routing cost surface & Slope \citep{USGS} & 1 - 5$^d$ \\
& Terrain \citep{LandCoverData} & 1 - 2.25\\
& Tower structure (MSA convex hull) & 1.3$^a$ \\
& Exclusion layer (airports, runways, onshore wind and offshore wind environmental exclusions in Table \ref{table:SiteSuitabilityDatasets}) & 100$^e$ \\
& Land cost layer \citep{nolte2020high}$^f$ & - \\
& Transmission cost offshore (AC layer) & 4.2$^b$  \\
& Transmission cost offshore (DC layer) & 1.5$^b$  \\
\bottomrule
\end{tabular}
}
\begin{tablenotes}
\scriptsize
\item$^a$ B\&V tower cost multipliers for tubular range from 1.11x for 230 kV to 1.5x for 500 kV. In order to avoid having to generate a different cost surface for each voltage class, we used an average multiplier of 1.3 across all voltage classes.
\item$^b$ For AC lines we make the  simplifying assumption that a substation is installed at landfall and the remainder of the line is run at 230 kV AC. For DC lines we make the additional simplifying assumption that a HVDC/HVAC converter station to 230 kV is also installed at landfall.
\item$^c$ Offshore transmission costs are estimated using Tables 8-2 and 8-3 from a report put out by the New York State Public Service Commission \citep{newyorkdepartmentofpublicservicestaffinitial2021} in 2021. In order to arrive at per kilometer costs using data from those tables we take the per mile CAPEX costs from the table and convert to kilometers and then select a capacity rating from the tables to divide by (we arbitrarily selected a DC rating of 1,300 MW and an AC rating of 400 MW).
\item$^d$ In addition to the multipliers suggested in the B\&V tool \citep{WECCTransmission}, we have added a 5x multiplier at slopes over $19^{\circ}$. Nineteen degrees was selected as this is the threshold above which we do not allow wind turbines to be sited (Table \ref{table:SiteSuitabilityDatasets}).
\item$^e$ This value is an arbitrarily selected large value to heavily discourage least-cost algorithms from crossing land/coean parcels in this category. In order to allow transmission to cross long continuous exclusion layers (e.g. national trails), we have reduced the multiplier on existing transmission right-of-ways \citep{hifldelectric2020} crossing these layers to 1. We have also allowed offshore wind transmission cables to cross exclusion layers at the sites of existing submarine cables, by buffering existing submarine cable routes \citep{noaaofficeforcoastalmanagementsubmarine2020} by 1km and setting the multiplier for the buffered submarine cables to 1.
\item$^f$ County level land cost data from Nolte \citep{nolte2020high} was modified a) to inflate costs from 2010 to 2018, b) change units from hectare to square meter, and c) merge with a US Census Bureau country layer \citep{uscensusbureaucartographic2020}.
\end{tablenotes}
\end{table*}

\begin{figure}
  \centering
  \includegraphics[width=\textwidth]{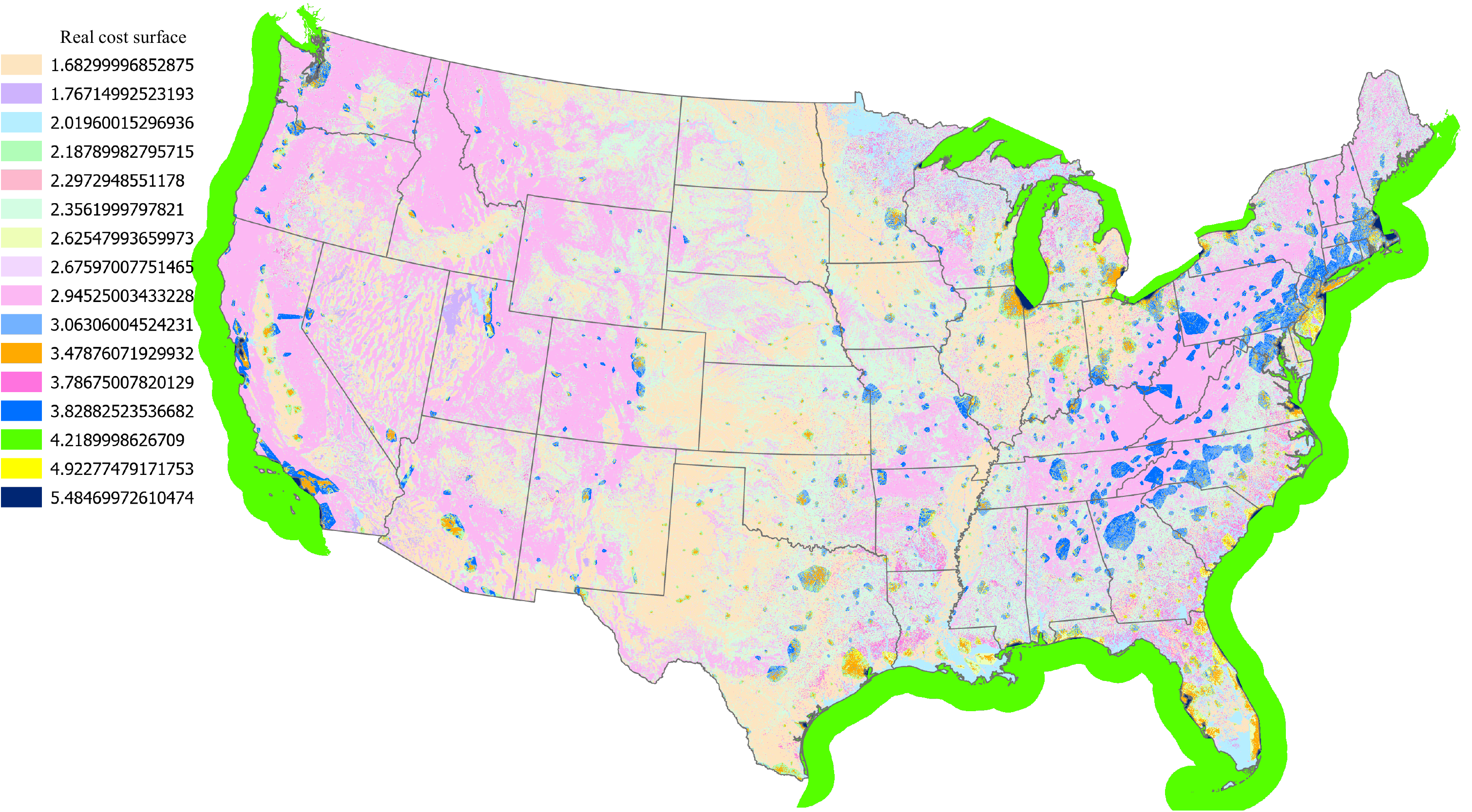}
  \caption{Real cost surface for offshore wind turbines with AC lines}
  \label{fig:offshore_cost_surface_AC}
\end{figure}

\begin{figure}
  \centering
  \includegraphics[width=\textwidth]{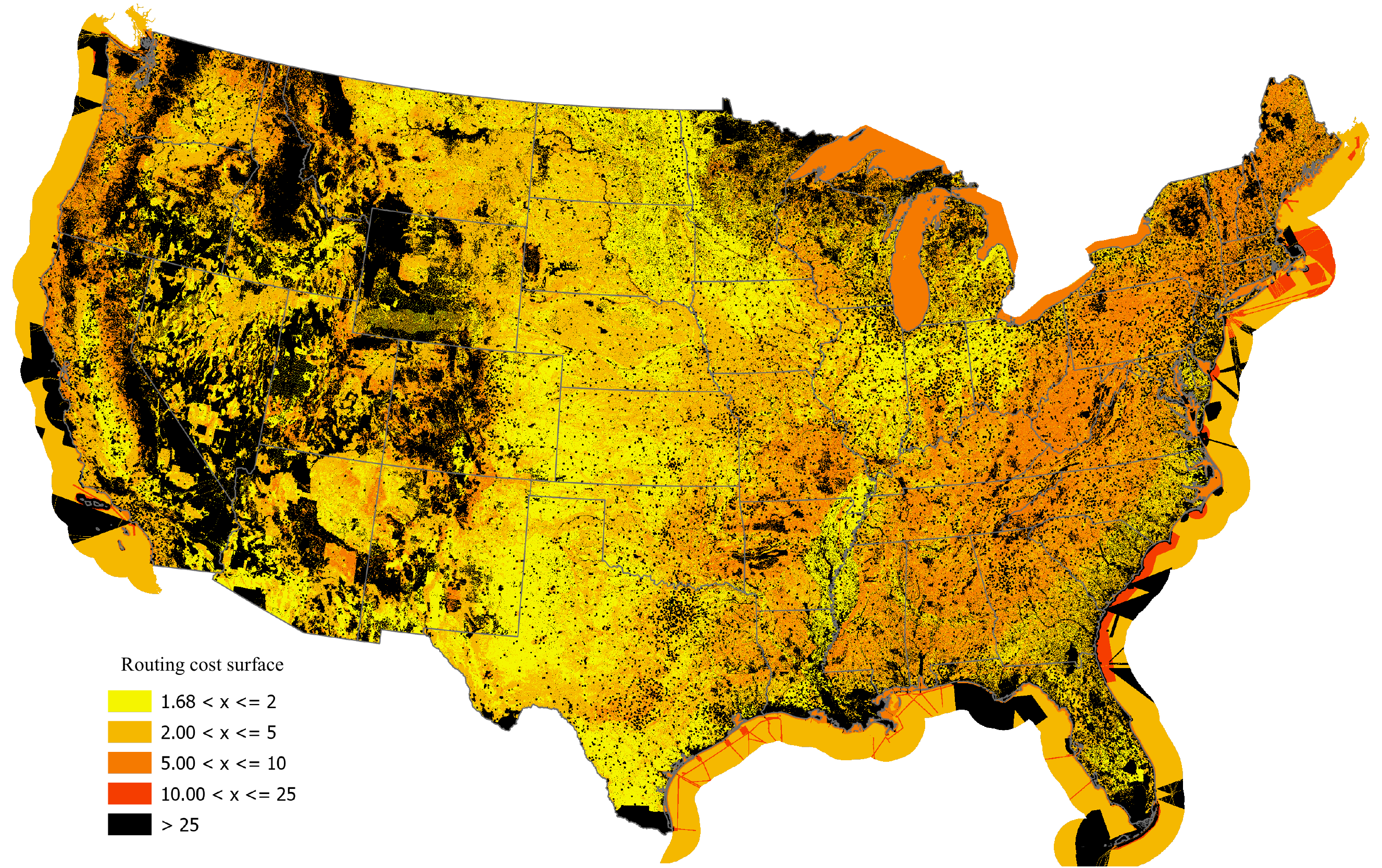}
  \caption{Routing cost surface for offshore wind turbines with AC lines}
  \label{fig:offshore_route_surface_AC}
\end{figure}

\begin{figure}
  \centering
  \includegraphics[width=\textwidth]{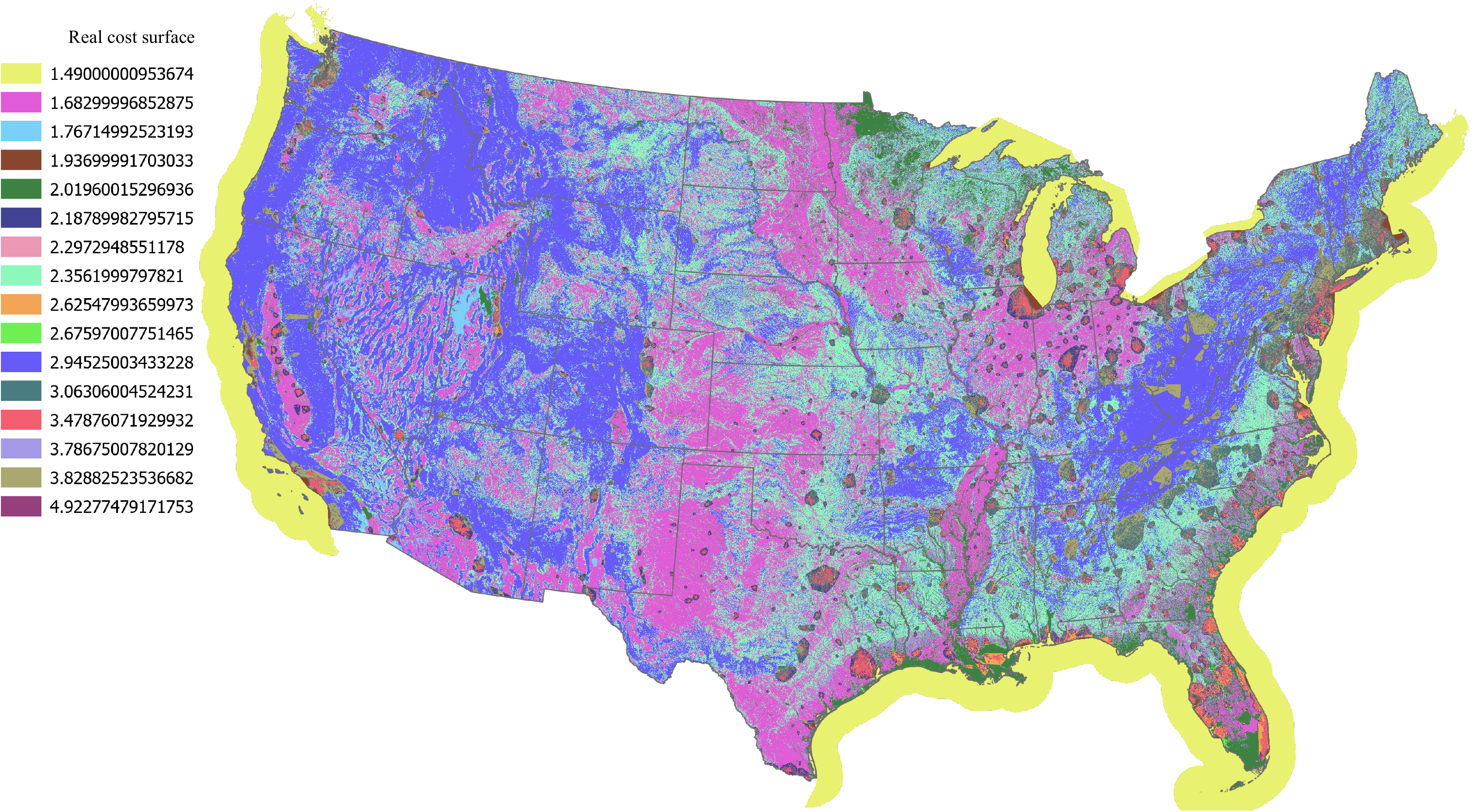}
  \caption{Real cost surface for offshore wind turbines with DC lines}
  \label{fig:offshore_cost_surface_DC}
\end{figure}

\begin{figure}
  \centering
  \includegraphics[width=\textwidth]{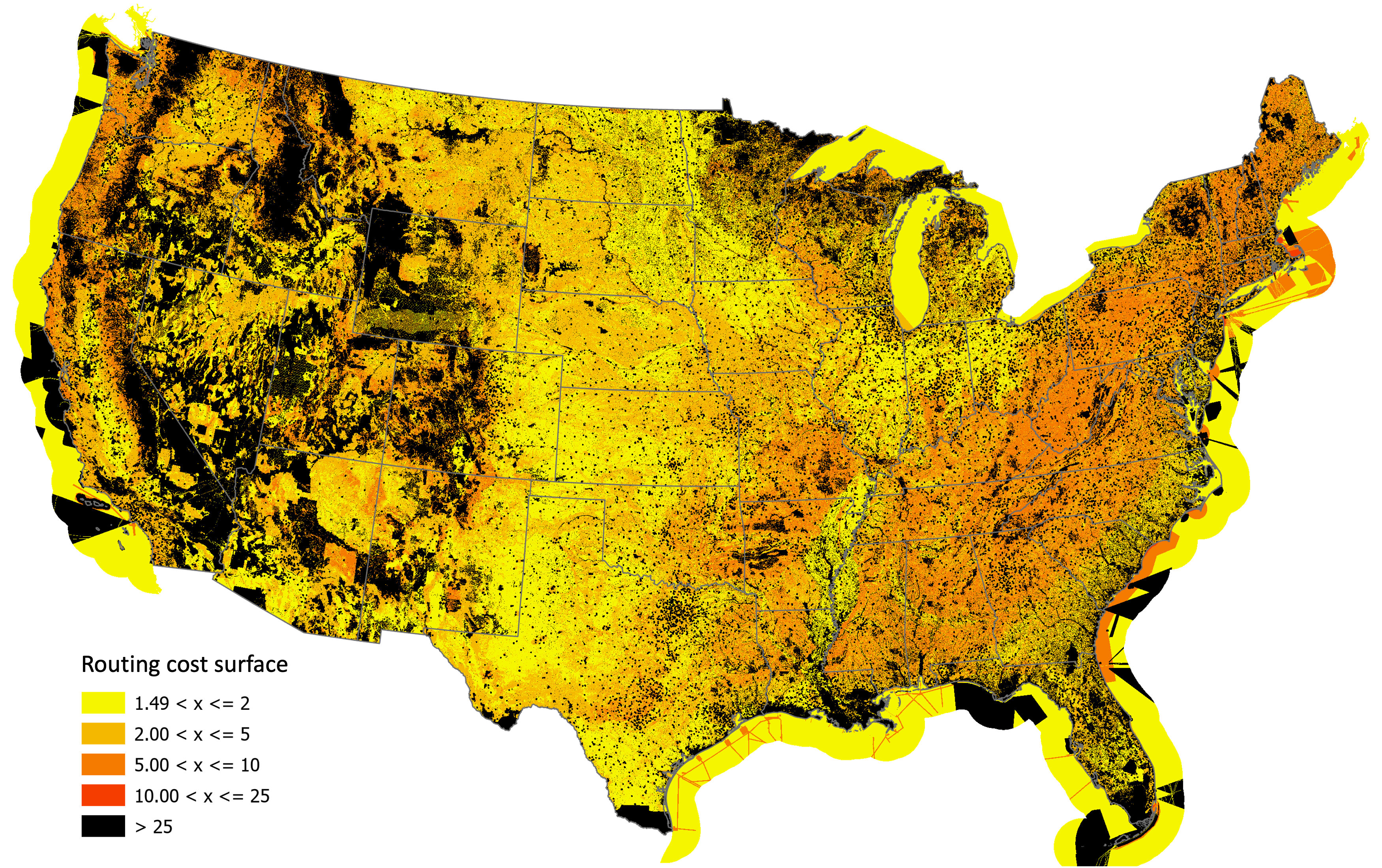}
  \caption{Routing cost surface for offshore wind turbines with DC lines}
  \label{fig:offshore_route_surface_DC}
\end{figure}

The real and routing cost surfaces for offshore wind turbines with AC export lines are shown in Figure \ref{fig:offshore_cost_surface_AC} and Figure \ref{fig:offshore_route_surface_AC} respectively. The real and routing cost surfaces for offshore wind turbines with AC export lines are shown in Figure \ref{fig:offshore_cost_surface_DC} and Figure \ref{fig:offshore_route_surface_DC} respectively. Note that we have not applied a 'land' cost for the right-of-ways for transmission sited offshore, however after a transmission line makes landfall, the land costs used for onshore transmission costing (Figure \ref{fig:solar_land_cost_surface}) is used. 


We do not include a regional multiplier in our cost surfaces to account for regional differences in labor costs, equipment costs, transportation and regulations. Regional differences connected to transmission technologies are partially picked up via transmission voltage caps applied to some regions (e.g. NREL \cite{cohen2019regional} suggests a maximum regional voltage of 345 kV in states in the northeast US). Regional differences connected to land costs will be reflected in our land cost layer \citep{nolte2020high}. We are using the PAD database \citep{GeologicalSurvey2018} GAP status 1 and 2 on the west and gulf coasts, and only 1 on the east coast. This follows the method of prior work done in this space \citep{larsonnetzero2021}. If we were to use Gap status 2 on the east coast, options for transmission siting (and CPAs) would be extremely limited.

\section{Methodology for estimating levelized cost of electricity (LCOE) for each CPA} \label{SI:LCOE}
Calculating per-CPA levelized cost of electricity (LCOE) is done by defining interconnection points for CPAs, calculating routing and the cost of transmission, and combining these costs with the overall capital and operation \& maintenance costs of the CPA. In the methods described below, multiple interconnection routes -- each with their own LCOE -- are considered for every CPA. This process can be enumerated into a six-step methodology; these steps are elucidated below.

In step one (Section \ref{Section:LoadCenters}), we define the parameters of the metropolitan statistical areas (MSA) that constitute the destinations for electricity from CPAs. These parameters include the substations that serve as interconnection points, the model region that each MSA is located in, and an approximation of hourly demand associated with the MSA. In step two, (Section \ref{SI:cost surface}), we create the cost surface layers for utility-scale solar PV, onshore wind, and offshore wind. Cost surface is based on the terrain, slope and tower structure required for transmitting renewable generation to the demand centre. This layer is henceforth called "real cost surface". In this step, we also increase the cost multipliers for some parts of the real cost surface to avoid undesirable transmission routes. In this layer we add a large multiplier to avoid exclusion layers such as military base, existing infrastructure, national/state forests or other reserved landscapes. This layer is henceforth called "routing cost surface".

Step three creates a set of proxy substations used to improve the delivery of renewable electricity from a previously undeveloped area to its final destination. We add the set of proxy substations to the database of existing substations (Section \ref{Section:proxy substation}). Step four enumerates all possible least-cost path route segments between each CPA and potential delivery locations. This step uses the routing cost surfaces, substation layer, and MSA layer (Section \ref{Section:CPA_MSA_routes}).

Step five applies the transmission line costs and characteristics selected in Section \ref{Section:transmission cost}, to the route segments determined in the prior step in order to determine the sited attributes of each transmission line segment (e.g. cost, voltage class, circuits, right-of-way width). This step computes the actual cost of the path using the real cost surface. The cost of CPA to MSA interconnection lines with more than one segment is simply the sum of all line segments included in the path.\footnote{For example, the cost of an interconnection line from CPA "1" to MSA substation "B" that connects through transmission substation "A" would be the sum of the line costs from "1" to "A" and from "A" to "B".} The annualized cost of connecting a CPA to a delivery MSA is henceforth called the interconnect annuity of the CPA.

Finally, we use the interconnect annuity, resource quality, capital cost, and fixed operations and maintenance cost for each CPA-MSA route to compute the levelized cost of electricity for that potential route (Section \ref{Section:LCOE calculation}). Potential CPA-MSA routes are sorted based on their LCOE, and each CPA is assigned to a single MSA as described in Section \ref{Section:CPA MSA assignment}.

\subsection{Defining load centers}
\label{Section:LoadCenters}
In order to calculate the cost of building new transmission lines from new wind and solar facilities, we begin by defining load centers that can use the power. 

\subsubsection{Metropolitan Statistical Area urban boundaries} \label{Section:MSAconvex}
\begin{figure}
  \centering
  \includegraphics[scale=0.5]{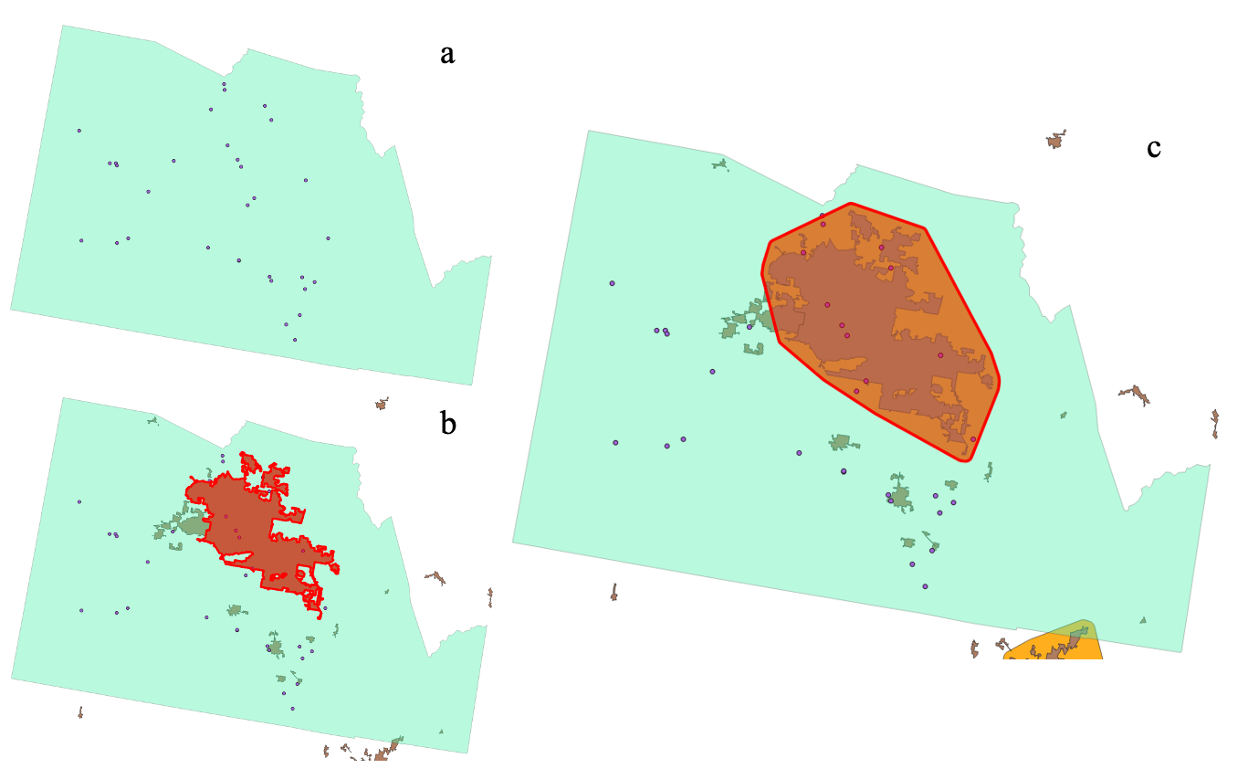}
  \caption{Creation of the Phoenix MSA urban boundary. From the top-left, it shows (a) the MSA boundary, (b) the Phoenix urban area (outlined in red), and (c) the convex hull around Phoenix’s urban area. Dots on the figure are the locations of transmission substations with a connecting line of at least 161kV.}
  \label{fig:MSA_boundary}
\end{figure}

The US Census Bureau \citep{USCensusBureau2019,USCensusBureau2020} defines core based statistical areas (CBSAs, referred to here as MSAs), and provides shapefiles with their extent as defined using “central” or “outlying” counties \citep{u.s.censusbureauMetropolitanMicropolitan2021}. Unfortunately, this geographic extent causes many MSAs in western states to appear much larger than their underlying urban area. Most MSAs are named based on the “principal city”, which is the largest city in the MSA. Some MSAs have longer names that are based on up to three principal cities (e.g. “Atlanta-Sandy Springs-Alpharetta, GA”). In an attempt to define the urban extent within MSAs, we first geospatially join the MSA geodata with a US Census Bureau urban area shapefile \citep{USCensusBureau2018}. All urban areas that intersect with an MSA and form some part of the MSA name (e.g. “Atlanta, GA” is part of the MSA name “Atlanta-Sandy Springs-Alpharetta, GA”) are combined to form a single polygon object. The total extent of the MSA urban area is the convex hull around this urban area polygon with an additional 2.5 km buffer.\footnote{The 2.5 km buffer ensures that every region in EPA's IPM has at least one MSA with a delivery substation.} This process of going from county-based boundaries of MSAs to polygons around their urban centers is illustrated in Figure \ref{fig:MSA_boundary}.

\subsubsection{Power delivery locations}
Electricity generated at a wind or solar CPA must be delivered to a substation that falls within the buffered convex hull of an MSA urban area. Substation data is from the Homeland Infrastructure Foundation-Level Data website \citep{hifldsubstations2020}. We use only substation locations that are in service and have an existing transmission line connection of at least 161kV.

Regions in the GenX model are defined along county borders, although some counties are assigned to more than one region. MSAs -- and the substations within their boundaries -- are assigned to GenX model regions based on the county designation of their centroid. In cases where an MSA centroid is in a county assigned to more than one model region, the MSA is assigned to the region where a larger share of the county households are located.\footnote{In MSAs with a population of at least 250,000, the smallest shares used for assignment are Fort Smith, AR-OK (0.65), Kansas City, MO-KS (0.72) and Des Moines-West Des Moines, IA (0.89).}

There are many MSAs with small populations and limited demand for wind or solar power. To solve this problem, we allocate regional hourly demand from E+ scenario from \cite{larsonnetzero2021} in the year 2050 to MSAs within each region based on their population. Approximate regions for the GenX model and urban MSA boundaries are  The demand at small and moderately-sized MSAs with a population of less than 1 million people is satisfied on an hourly basis as CPAs are assigned to them. The biggest MSA within each region and all MSAs with populations greater than 1 million persons are treated as infinite sinks. These MSAs are thus always available to accept generation; their demand is not diminished by incoming power. This allows a region to build more wind or solar capacity than it can use on an hourly basis. Inter-regional power transfer is subsequently handled by the capacity expansion model.

\begin{figure}
  \centering
  \includegraphics[width=\textwidth]{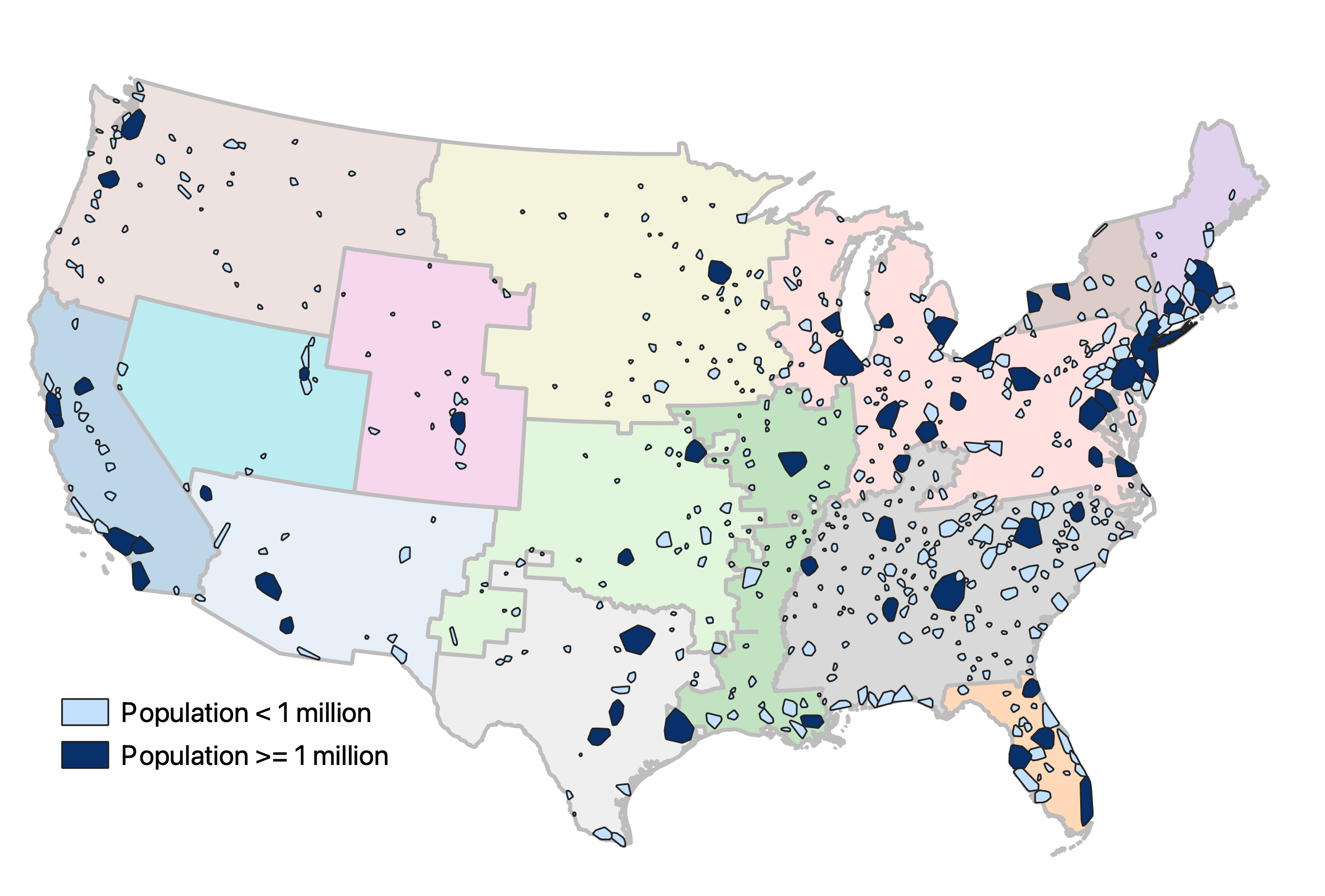}
  \caption{Major (population $>=$ 1 million) and minor (population $<$ 1 million) MSAs. MSAs are shown on a map approximating GenX region boundaries.}
  \label{fig:major_minor_msas}
\end{figure}

\subsection{Development of proxy substations} \label{Section:proxy substation}

Creating proxy substation requires 4 steps: 
\begin{enumerate}
    \item Create a fishnet grid of equally spaced points at 35$\times$35km\textsuperscript{2} resolution.
    \item Select only those points which fall within 500 meter of an existing transmission line (161 kV+).
    \item From that selection, select the subset of points which fall outside a specified radius of existing substations, since we assume CPAs within that radius will just connect to the existing substation.
    \item Assume $~$35 km radius since $~$90\% of existing facilities have interconnection lines $<=$ 35 km. The substations inside the convex hull around the metro area are called "delivery substations" or "MSA substations". The rest of the substations are called "transmission substations".
\end{enumerate}

\begin{figure}
  \centering
  \includegraphics[width=\textwidth]{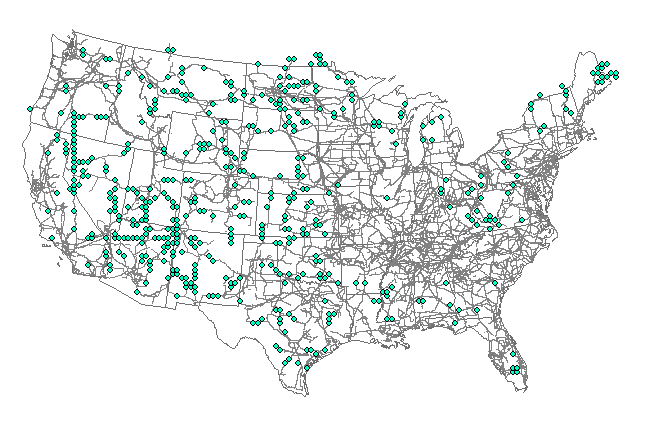}
  \caption{Set of proxy substations along the existing transmission corridors}
  \label{fig:substations}
\end{figure}

\subsection{Algorithm for determining interconnection paths for each CPA} \label{Section:CPA_MSA_routes}

Interconnection costs are the capital expenditures related to the combination of 230kV spur lines, higher voltage transmission lines, and substations needed to bring power from new renewable generation sites to demand centers. In this study we calculated possible interconnection costs using the location of all potential solar PV, onshore wind, and offshore wind candidate project areas (CPAs); raster layers with the cost to build transmission across the United States; metropolitan statistical areas (MSAs) defined by the US Census Bureau; and the location of all existing substations and transmission lines. Each CPA was assigned to a final delivery MSA based on the cost of building new transmission infrastructure and the capacity of the MSA to accept renewable generation.

In this step the routing, costing and land cost surfaces from in Section \ref{SI:cost surface} and substation layers from Section \ref{Section:proxy substation} are used to develop costs for CPA-to-substation and substation-to-substation transmission route segments. These route segments are used to build potential interconnection paths between CPAs and MSAs.

CPAs can connect to an MSA substation in one of four ways:
\begin{enumerate}
    \item Directly connect to a MSA substation. Onshore wind and solar PV use a 230kV single line, offshore wind can either use 230kV single or HVDC.
    \item Directly connect to a MSA substation, then connect to a substation in one or more subsequent MSAs. The voltages of additional lines are determined based on the length of a line.
    \item Connect to a transmission line substation using a 230kV line, then connect to a MSA substation. Offshore wind can only connect directly to an MSA, so this option is for onshore wind or solar PV. The voltages of additional lines are determined based on the length of a line.
    \item Connect to an MSA via a transmission substation, then connect to another MSA substation within the same region. Substations at additional MSAs within the same model region can be reached by building subsequent lines, with voltage levels determined by line length.
\end{enumerate}

All connections from transmission substations or MSA substations to MSA substations are double lines. Based on existing and proposed infrastructure, we limit voltage in New York and New England to 345kV (AC) or 500kV HVDC. In all other regions, lines are either 230kV, 500kV, or 500kV HVDC.

Using the routing raster layer\footnote{The routing layers have relative costs to travel across every 250x250m area of land. The routing costs can be increased in environmentally sensitive areas, reflecting a societal preference to avoid disturbing these areas (e.g. a region with routing costs 100x higher than the surrounding area will only be crossed if it reduces the distance traveled by at least 100x). Actual transmission costs are used for any lines crossing these regions.} described in Section \ref{SI:cost surface}, the following least-cost path (LCP) route segments are calculated using the minimum cost path routing algorithm in the scikit-image package created by \cite{skimage2014}.

\begin{itemize}
    \item Onshore wind and solar PV CPAs and all transmission substations. This path will connect each CPA with a single (lowest-cost) transmission substation.
    \item The substations of a single MSA and all onshore wind/solar PV CPAs within 400 km and substations within 1000 km.\footnote{Straight-line distance. The 1000 km cutoff was used to limit the number of routes calculated. Any lines originating at a CPA that are longer than 252km are discarded.} This provides paths from CPAs and substations (both transmission and MSA) to every delivery MSA.
    \item Offshore wind CPAs and all MSA substations, using both AC and HVDC routing layers.\footnote{The HVDC routing layer reflects higher connection costs and lower offshore transmission line costs. It will prefer to keep a line offshore longer. The AC routing layer has higher costs for offshore transmission, which will bias it towards longer onshore line segments.} The AC and DC least cost routes may connect offshore wind CPAs with different MSA substations.
    \item MSA substations in Boston and Los Angeles and offshore wind CPAs within 500 km of those MSAs, using the HVDC routing layer. These paths are calculated because initial calculations using only the least-cost MSA connection often resulted in long on-shore transmission line segments through smaller MSAs that terminated in Boston or Los Angeles. The option of a longer offshore HVDC line to these MSAs is included, and will only be selected if it is cheaper than routing through a smaller MSA.
\end{itemize}

To calculate the cost of a transmission line, as described in Section \ref{Section:transmission cost}, we need to know the line voltage class and whether it is a single or double-circuit. But to determine this, we would first want to know how much capacity the line is connecting. Since line costs are needed as inputs to a capacity expansion model, and the amount of capacity built is a model output, we must use heuristics to assign line voltage classes. We assume that all transmission lines originating at CPAs are 230kV single-circuit. Transmission lines between substations are assumed to be double-circuit,\footnote{Double-circuit lines can connect more capacity than single-circuit lines and have lower per-km costs. This assumption makes it more likely that multiple CPAs will route through a common substation when connecting to the same MSA.} and the voltage class is determined by the line length as shown in Table \ref{table:txCostAssumptions}. The cost of each route segment -- CPA to transmission/MSA substation and transmission/MSA substation to MSA substation -- are calculated using the appropriate routing cost surface, land cost surface, and methods described in Section \ref{Section:transmission cost}. These route segments can be combined to create interconnection paths from a CPA to many different MSAs. Each interconnection path has a different capital cost, leading to many potential levelized cost of electricity (LCOE) values for a single CPA depending on which MSA it delivers power to.

To calculate the cost of transmission interconnection between a CPA and potential MSA connections, we begin by assigning each MSA to a model region. The least-cost route connecting each MSA within a region to all other MSAs\footnote{Routes are between substations located within each MSA.} is found using Dijkstra’s algorithm \citep{dijkstra1959note} as implemented in NexworkX \citep{hagberg2008exploring}. Each CPA has one or more initial substations it can connect to (a transmission substation and possibly substations at one or more MSAs). The transmission substation can also connect to multiple MSA substations, leading to a set of connections from the CPA to MSAs. These MSAs may be physically located in a different model region -- no limit is placed on what model region a CPA initially connects to. But the following constraints are placed on connections between transmission substations and MSA substations:

\begin{itemize}
    \item A transmission substation can only connect directly with a single MSA per region. Any connections to other MSAs must flow through this initial line.
    \item Transmission substations cannot connect to MSAs in other model regions using HVDC lines. HVDC lines are most cost effective for long distance projects. Using them to transfer power from one model region to another should be up to capacity expansion models, not an algorithm that determines interconnection costs.
\end{itemize}

Connections from the first MSA must be to other MSAs within the same model region, and will follow paths previously calculated using Dijkstra's algorithm. The total cost of each possible interconnection between a CPA and delivery MSA is the sum of all route segments traversed along its path. Because MSAs with a population of at least 1 million people (major MSAs) are represented with unlimited demand, we can discard all connections to a final MSA that are more expensive than the first infinite-sink MSA connection.

\subsection{Calculating final transmission costs for each potential CPA transmission option} \label{Section:transmission cost}
In order to calculate the final transmission cost for each transmission option routed in the prior step, a number of actions are required, depending on whether the originating CPA is onshore or offshore.

\subsubsection{Method for onshore CPA and substation locations} \label{Section:OnshoreCosts}
Calculating the final transmission line costs for portions of transmission routes that stay entirely onshore involves stepping through a process of systematically combining the transmission costs and characteristics found in Table \ref{table:txCostAssumptions} with outputs from the prior step (Section \ref{Section:CPA_MSA_routes}). The process of generating final costs is listed below Table \ref{table:txCostAssumptions}.

\begin{table*}
\caption{Assumptions for onshore transmission lines }\label{table:txCostAssumptions}
\resizebox{\textwidth}{!}{
\begin{tabular}{lllllllllll}
\toprule
Voltage class & Type & Circuits & Type & min MW & Max MW & min km & max km$^{a}$ & ROW width (meters) & \$/MW$^{b}$ & Tie-in cost/MW\\
\hline
230 kV & HVAC & single & spur/bulk & 0 & 400 & 0 & 252 & 38.1 & 1.43211 & 14,759\\
230 kV & HVAC & double & spur/bulk & 400 & 800 & 0 & 252 & 45.72 & 1.14630 & 14,759\\
345 kV & HVAC & single & bulk & 0 & 750 & 0 & 483 & 53.34 & 1.18830 & 14,759\\
345 kV & HVAC & double & bulk & 750 & 1500 & 0 & 483 & 60.96 & 0.95073 & 14,759\\
500 kV & HVAC & single & bulk & 0 & 1500 & 0 & 483 & 60.96 & 0.84866 & 14,759\\
500 kV & HVAC & double & bulk & 1500 & 3000 & 0 & 483 & 76.2 & 0.77227 & 14,759\\
500 kV & HVDC & double & bulk & 500 & 3000 & 483 & 5000 & 60.96 & 0.33965 & 281,290\\

\bottomrule
\end{tabular}
}
\begin{tablenotes}
\scriptsize
    \item$^{a}$ Maximum distances taken from AEP \citep{Gradyfacts}
    \item$^{b}$ Excluding land cost
\end{tablenotes}
\end{table*}

The process of applying selected transmission costs and characteristics involves: 

\begin{itemize}
    \item[A] Multiply the path cost obtained from real cost surface (unitless) by the base cost per MW for the chosen voltage class (the \$/MW column in Table \ref{table:txCostAssumptions})
    \item[B] Multiply the result of (A) by the capacity (in MW) of the candidate project area (CPA)
    \item[C] Multiply the result of (B) by 1.5x if a transmission line’s distance is $<$ 4.83 km, 1x if $>$4 16.1km and 1.2x in all other cases
    \item[D] Multiply the path cost obtained from the land cost surface (in \$/meter) by the ROW width (in meters) for the chosen voltage class (the ROW width (meters) column in Table \ref{table:txCostAssumptions}) in order to get the land related costs for the each line's ROW
    \item[E] Multiply the result of D by the capacity of the CPA (in MW) and divide by the MW max of the voltage class in order to pro rate the land related ROW costs for each CPA
    \item[F] Add the result of C and E and multiply the result times 1.175x (17.5\% base AFUDC in B\&V calculator \citep{WECCTransmission})
    \item[G] Multiply the capacity of the CPA (in MW) by the grid tie-in cost for the chosen voltage class (the Tie-in cost/MW column in Table \ref{table:txCostAssumptions})
    \item[H] Add the result of F and G to get the total cost of a transmission line
 
\end{itemize}

\subsubsection{Method for offshore CPA locations} \label{Section:OffshoreCosts}
Calculating the final transmission line costs for routes that have both offshore and onshore portions involves stepping through a process of systematically combining the transmission costs and characteristics found in Table \ref{table:txOffCostAssumptions} with outputs from the prior step (Section \ref{Section:CPA_MSA_routes}). The process of generating final costs is listed below Table \ref{table:txOffCostAssumptions}.

\begin{table*}
\caption{Assumptions for offshore transmission lines }\label{table:txOffCostAssumptions}
\resizebox{\textwidth}{!}{
\begin{tabular}{p{2cm}|p{2cm}|p{2.3cm}|p{2.3cm}|p{2.3cm}|p{2.3cm}}
\toprule
Turbine type & Seafloor depth (m) & Fixed cost AC (USD2018/kw)$^{a}$  & Fixed cost DC (USD2018/kw)$^{b}$ & Grid tie-in cost (USD2018MW)$^{c}$  & Remainder of grid upgrade costs (\$/MW)$^{d}$\\
\hline
Fixed & 0 to 50 & 707.6 & 853.4 & 14,749 & 398,251\\
Floating & $>$50 & 876.2 & 1135.4 & 14,749 & 398,251 \\
\bottomrule
\end{tabular}
}
\begin{tablenotes}
\scriptsize
    \item$^{a}$ We estimate fixed AC costs for each turbine type as the remainder of the ATB2020's \citep{nrel2020} moderate projection for 2021 in (USD2018/kW), after subtracting off the estimated variable (wrt length) costs in (USD2018/kW) for the turbine type's average distance to shore (53.4 km) \citep{nrel2020} when using the unit costs for 220 kV, 300MW AC submarine cable (in USD2018/kW-km) \citep{newyorkdepartmentofpublicservicestaffinitial2021}.
    \item$^{a}$ We estimate fixed DC costs for each turbine type as the remainder of the ATB2020's \citep{nrel2020} moderate projection for 2021 in (USD2018/kW), after subtracting off the estimated variable (wrt length) costs in (USD2018/kW) for the turbine type's average distance to shore (95 km) \citep{nrel2020} when using the unit costs for 320 kV, 1300MW DC submarine cable (in USD2018/kW-km) \citep{newyorkdepartmentofpublicservicestaffinitial2021}.
    \item$^{c}$ \citep{maclaurinrenewable2019} 
    \item$^{d}$ This is the average offshore wind grid-tie in cost of 413,000 \$/MW estimated by \cite{burkeoffshore2020} (for PJM and MISO) after subtracting off the generic grid-tie in cost used for all other renewable resources in this study. As no dollar year is explicitly given in the report, the dollar year is expected to be the same as the published report USD2020.
\end{tablenotes}
\end{table*}

\begin{itemize}
    \item[A] Multiply the path cost obtained from the AC and/or DC offshore real cost surface by the capacity of the CPA (in MW)
    \item[B] Select the turbine type for each CPA using the average seafloor depth for each CPA and Seafloor depth (m) column in Table \ref{table:txOffCostAssumptions}, and then multiply the capacity of the CPA (in MW) by the AC and/or DC fixed cost for the selected turbine type from the appropriate column in Table \ref{table:txOffCostAssumptions}. 
    \item[C] (See comment re grid-tie in for GREG above - likely remove this step) Multiply the capacity of the CPA (in MW) by the appropriate grid tie-in cost from Table \ref{table:txOffCostAssumptions}.
    \item[D] Multiply the land path cost from step 3 by as many 38.1 meter right-of-ways are needed for 230 kV single circuit lines to carry the capacity of the CPA after coming onshore (see Table \ref{table:CostSurfAssumptions}).
    \item[E] Multiply the result of D by the capacity of the CPA (in MW) and divide by the MW max of the voltage class in order to pro rate the land related ROW costs for each CPA.
    \item[F] (optional) Multiply the capacity of the CPA (in MW) by the Remainder of the grid upgrade costs column in Table \ref{table:txOffCostAssumptions}.
    \item[G] Add the result of A + B + C + E and allocate these to each CPA.
    \item[H] (optional) Allocate the result of F to entity of choice (e.g. project/developer, federal government, etc).
\end{itemize}

\subsection{Computation of LCOE} \label{Section:LCOE calculation}
The LCOE for each possible interconnection route of a CPA is calculated using the method summarized below.  \\
\[LCOE = (Annuity + FOM) / AnnualGeneration\]\\
Where $AnnualGeneration$ is determined as described in Section \ref{SI:resource_profiles}. $FOM$ is the fixed operation and maintenance costs from NREL ATB 2020 \citep{nrel2020}. $Annuity$ is a function of 1) the 2030 resource capital cost and real weighted average cost of capital (WACC) values from NREL ATB 2020 and 2) the transmission interconnection costs described in \ref{Section:transmission cost}. A WACC of 4.9\% and capital recovery period of 60 years \citep{gorman2019improving} are used for transmission projects. ATB capital costs are adjusted using regional cost factors from EIA \citep{eiacostperformance}. Using their centroid locations, each CPA is assigned to a region from EPA's Integrated Planning Model (IPM) \citep{epaipm}. IPM regions are mapped to Electricity Market Module (EMM) regions in EIA's National Energy Modeling System (NEMS) \citep{eiaemm} as shown in Table \ref{table:epa_emm_match}, and the EMM region determines the appropriate regional cost factor. Using an assumed capital recovery period ($P$) of 20 years, $Annuity$ is calculated:

\[ Annuity = CapitalCost \times e^{WACC \times P} \times \frac{e^{WACC} - 1}{e^{WACC \times P} -1} \]

\begin{figure}
  \centering
  \includegraphics[width=\textwidth]{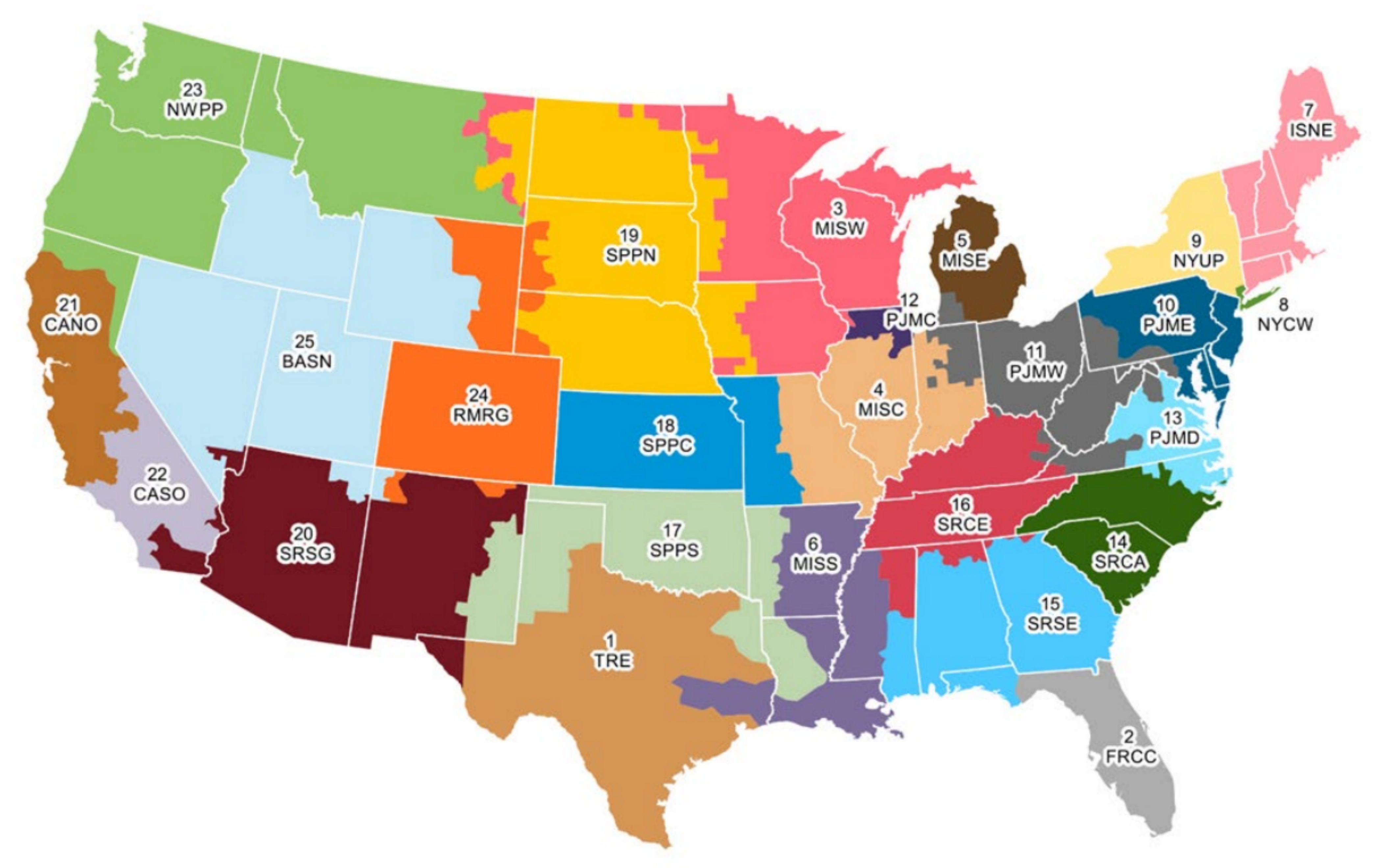}
  \caption{EMM regions in EIA's NEMS \citep{eiaemm}}
  \label{fig:NEMS_EMM_regions}
\end{figure}

\begin{figure}
  \centering
  \includegraphics[width=\textwidth]{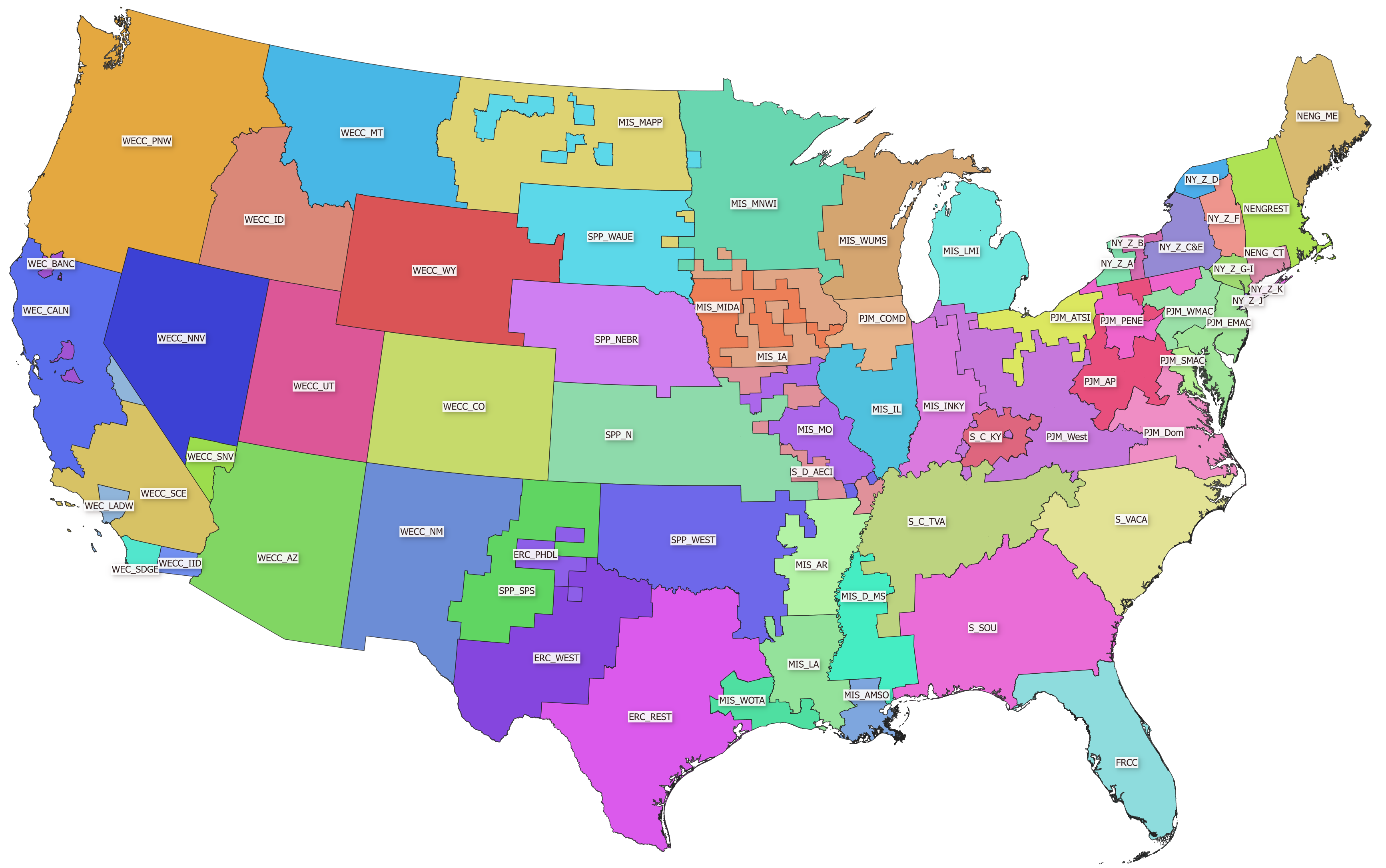}
  \caption{Model regions in EPA's IPM \citep{epaipm}}
  \label{fig:IPM_regions}
\end{figure}

\begin{table*}
\centering
\caption{EPA IPM regions in each NEMS EMM region }\label{table:epa_emm_match}
\resizebox{\textwidth}{!}{
\begin{tabular}{ll}
\toprule
EMM Region & IPM Regions                                                     \\
\hline
TRE        & ERC\_PHDL, ERC\_REST, ERC\_WEST                                 \\
FRCC       & FRCC                                                            \\
MISW       & MIS\_WUMS, MIS\_MNWI, MIS\_IA                                   \\
MISE       & MIS\_LMI                                                        \\
PJMC       & PJM\_COMD                                                       \\
MISC       & MIS\_IL, MIS\_MO, S\_D\_AECI, MIS\_INKY                         \\
SPPN       & MIS\_MAPP, SPP\_WAUE, SPP\_NEBR, MIS\_MIDA                      \\
SPPC       & SPP\_N                                                          \\
SPPS       & SPP\_WEST, SPP\_SPS                                             \\
MISS       & MIS\_AMSO, MIS\_WOTA, MIS\_LA, MIS\_AR, MIS\_D\_MS              \\
SRSE       & S\_SOU                                                          \\
SRCA       & S\_VACA                                                         \\
PJMD       & PJM\_Dom                                                        \\
PJMW       & PJM\_West, PJM\_AP, PJM\_ATSI                                   \\
PJME       & PJM\_WMAC, PJM\_EMAC, PJM\_SMAC, PJM\_PENE                      \\
SRCE       & S\_C\_TVA, S\_C\_KY                                             \\
NYUP       & NY\_Z\_A, NY\_Z\_B, NY\_Z\_C\&E, NY\_Z\_D, NY\_Z\_F, NY\_Z\_G-I \\
NYCW       & NY\_Z\_J, NY\_Z\_K                                              \\
ISNE       & NENG\_ME, NENGREST, NENG\_CT                                    \\
RMRG       & WECC\_CO                                                        \\
BASN       & WECC\_ID, WECC\_WY, WECC\_UT, WECC\_NNV                         \\
NWPP       & WECC\_PNW, WECC\_MT                                             \\
CANO       & WEC\_CALN, WEC\_BANC, CA\_N                                     \\
CASO       & WECC\_IID, WECC\_SCE, WEC\_LADW, WEC\_SDGE, CA\_S               \\
SRSG       & WECC\_AZ, WECC\_NM, WECC\_SNV\\
\bottomrule
\end{tabular}
}
\end{table*}

\section{Creating renewable clusters for the GenX model}
\label{SI:renewable clusters}
The individual CPAs identified in Section \ref{SI:CPA selection} must be grouped into clusters for use in a capacity expansion model. These clusters aggregate the available capacity and generation profiles from multiple CPAs within a model region, and are used in the model as if they are a single generating resource. The capacity expansion model decides how much capacity from each cluster is built, but it does not specify individual CPAs within a cluster. The processes of grouping CPAs into resource clusters and then identifying which CPAs from each cluster are built are described in the following subsections.

\subsection{Assigning CPAs to MSAs}
\label{Section:CPA MSA assignment}
A key element of the method that assigns CPAs to delivery MSAs is that CPAs have interconnection paths to more than one MSA, and most CPAs will not use their lowest-cost interconnection path. Out of approximately 550 MSAs, only 56 can accept unlimited CPA connections. Most CPAs that have their least-cost connection to one of the remaining 498 MSAs will use a more expensive connection path.

The levelized cost of electricity (LCOE) for each CPA to MSA route is calculated using the mid-range 2030 resource cost for each CPA from NREL ATB 2020 as described in Section \ref{Section:LCOE calculation}, the interconnection cost of each route option, and the annual generation of the CPA. CPAs' initial list of potential MSA connections are enumerated using the rules stated in Section \ref{Section:CPA_MSA_routes}. We create a list of all CPA to MSA connections across all CPAs, and sort it by predicted LCOE. Starting with the lowest-LCOE CPA to MSA connection across all source CPAs, CPAs are assigned to target destination-MSAs. As CPAs are assigned to MSAs with a population of less than 1 million, the hourly generation profile of the CPA is subtracted from the MSA's hourly demand profile. If generation from a CPA is greater than demand at the MSA in at least one hour, generation at the CPA is assumed to be curtailed and the LCOE is recalculated.\footnote{This curtailment only affects LCOE calculations for the interconnection algorithm. The full generation profile is used when creating resource clusters that are used in the capacity expansion model, described in Section \ref{Section:upscaling_cpas}} At this point, a connection from the given CPA to a different MSA may have a lower LCOE than the MSA previously thought to be its optimal destination due to demand curtailment at the MSA, despite this secondary MSA having a higher transmission cost.\footnote{If the connection LCOE will increase by less than \$2/MWh and 5\% of the initial LCOE, the CPA is still assigned to the destination MSA. Otherwise the connection is moved down in the queue. After every 10 GW of CPAs assigned the connections are re-sorted based on their LCOE. These steps may allow some CPAs to be assigned out of order (a cheaper connection to the MSA exists), but they reduce calculation times.}

\begin{figure}
  \centering
  \includegraphics[scale=0.5]{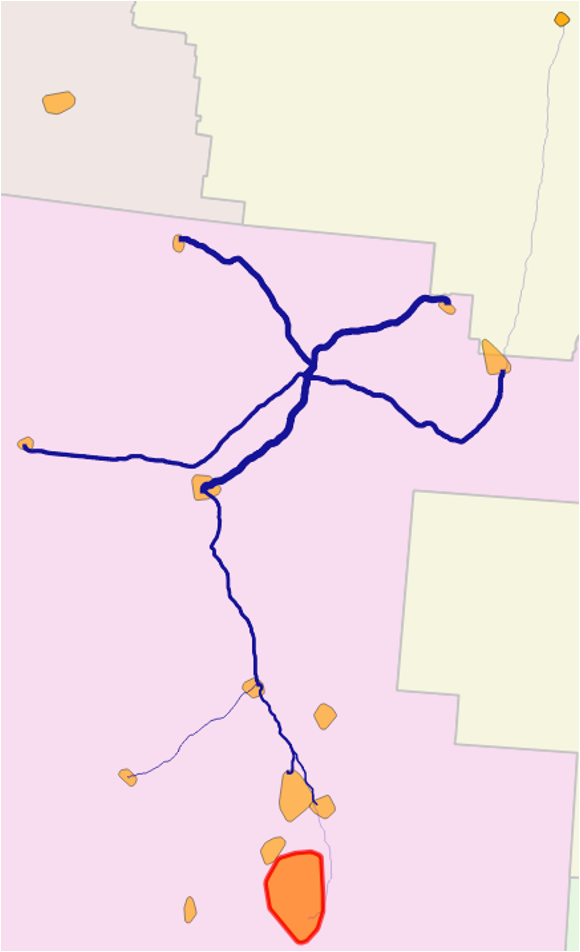}
  \caption{Eleven potential interconnection routes for a solar PV CPA in Wyoming. The width of each line is proportional the inverse of interconnection capital costs. This CPA is ultimately assigned to its most expensive connection -- Denver, CO -- outlined in red.}
  \label{fig:CPA_MSA_routing}
\end{figure}

Figure \ref{fig:CPA_MSA_routing} shows possible connections from a solar PV CPA in Wyoming to eleven MSAs in two different model regions. LCOEs for the different connections range from \$32/MWh (Casper, WY) to \$183/MWh (Denver, CO). The thickness of each line is proportional to the inverse of cost. The low cost connections are to small MSAs, which have their demand satisfied by closer and even lower-cost resources. As each connection for this CPA is considered, enough of the demand during daytime hours is satisfied by other resources that it makes sense to consider building longer transmission lines. This particular CPA is assigned to Denver, CO (outlined in red) for final delivery of electricity at a calculated LCOE of \$183/MWh.

This iterative process involves considering the system-wide lowest-LCOE connections between CPAs and MSAs, recalculating the connection LCOE if the resource will deliver more power than the MSA can use in a given hour, and re-sorting connections as LCOE is updated. This process is repeated until at least 99\% of CPAs are assigned to a destination MSA. These CPAs represent 46 TW of wind and solar PV resources with LCOEs ranging from \$22/MWh to \$500/MWh.

Note that by iterating over a master list of all CPA-connections sorted by connections' LCOE, we ensure an optimal solution across the entire system. Any greedy algorithm which considered connections to CPAs one at a time would be unable to ensure that no MSA was over-supplied, or that few CPAs experience heavy curtailment. By creating a system-level list of CPA connections, we supply MSAs completely and avoid generation surplus, even if such a methodology means that individual CPAs' assigned destination MSAs initially seem suboptimal in terms of transmission cost.

Using an assumed hourly demand to limit connection with all MSAs rather than only connecting to MSAs with a minimum population allows us to to reduce the interconnection costs of some CPAs that are close to small cities but far from large ones. We find that, in MSAs with less than 1 million people, the median interconnected capacity of wind and solar CPAs is 874 MW per 100,000 people (with an interquartile range of 598-1,365 MW per 100,000 people). El Paso, TX has a population of 844 thousand and interconnection capacity of 10 GW. Herber, UT has a population of 76 thousand and interconnection capacity of 759 MW. And Springfield, MA has a population of 697 thousand and interconnection capacity of 3.2 GW. Generally, regions with good wind resources and fewer MSAs can connect more capacity per capita.

\subsection{Upscaling the CPAs}
\label{Section:upscaling_cpas}
\begin{figure}
  \centering
  \includegraphics[width=\textwidth]{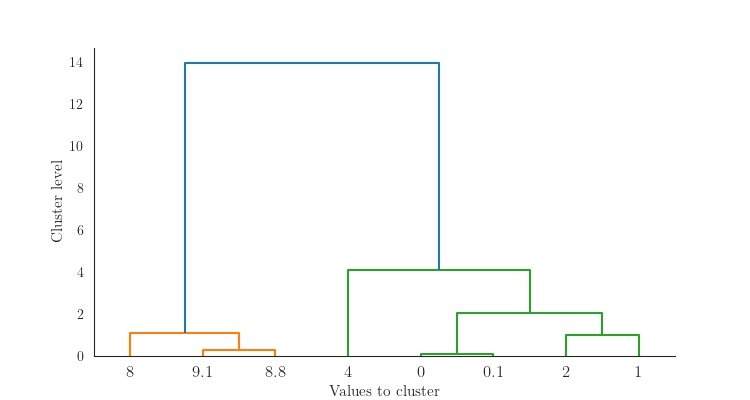}
  \caption{An illustrative example of hierarchical clustering. The value of each point is shown on the x-axis. The tree combines points with the most similar values until all points are part of a single group.}
  \label{fig:agg_clustering}
\end{figure}

Each CPA is assigned a delivery MSA as described in Section \ref{Section:CPA MSA assignment}. Metro regions are assigned to GenX model regions based on their physical location. CPAs of a single resource type are grouped within each MSA using hierarchical (or agglomerative) clustering \citep{bar2001fast, Mullner2011}  -- described in Section \ref{Section:LCOE calculation} -- using methods implemented in SciPy \citep{scipy2020}. The clustering method begins with each CPA as a member of its own resource group, so the number of groups is equal to the number of CPAs connecting to the MSA. Next, the two CPAs nearest in LCOE values are combined into a single group,\footnote{The hourly generation profiles for a group of CPAs is an average of individual CPA generation profiles, weighted by the CPA capacities.} leaving one fewer resource groups than CPAs. This process of aggregation is repeated until all CPAs connecting to the MSA are in a single group. An illustrative example of the clustering method is shown in Fig \ref{fig:agg_clustering}. CPA membership in groups and the group generation profiles are recorded at each level. This deterministic clustering method allows us to pre-calculate and save an entire tree of clusters for each resource type.

To create the final resource clusters for a model, the user specifies the total MW of capacity and the number of clusters for each resource type that should be available within a region. The total capacity might be selected based on how much capacity is available within an LCOE limit. The clusters within a model region with lowest LCOE that satisfy the minimum capacity requirement are selected and aggregated using the saved hierarchical tree structures. Resource groups delivering power to different MSAs are combined using the same clustering method.\footnote{These hierarchical tree structures linking CPA groups across MSAs are not saved because MSA assignments to model regions are a user choice based on the specific study.}

\section{Methodology for creating inter-regional transmission routes and regional backbone network} \label{SI:inter-regional transmission}
The long-distance transmission network is divided into two types: inter-regional transmission and intra-regional transmission. The inter-regional transmission represents the transmission lines that connect the model regions defined in the GenX model. The intra-region transmission network represents the transmission backbone of the region. For each MW increase in the inter-regional transmission capacity, the model needs to build an equivalent capacity of the intra-region network. The following subsections provide details on the assumptions for creating inter and intra-regional transmission network.

\subsection{Development of inter-regional transmission lines}
We assume that all inter-regional transmission lines are 500kV HVDC. For the development of the inter-regional transmission lines, we follow the steps below:
\begin{itemize}
    \item Identify infinite sink MSAs ($>$1,000,000 population) in each of the GenX model regions. If a region doesn't have MSA with $>$1,000,000 population, use the MSA with the highest population in that region as infinite sink MSA. 
    \item Identify substations that fall inside the urban area convex hull around the infinite sink MSAs. These substations are called "MSA-substations".
    \item Create a list of region pairs that are required for inter-regional connections
    \item Use routing cost surface to create least-cost path between all MSA-substations and all infinite sinks in the connecting regions. For example, Pacific Northwest region has 2 infinite sinks: Portland and Seattle. Rocky mountains region has one infinite sink: Denver. Portland has 27 MSA-substations and Denver has 57 MSA-substations. We create a least-cost path between 27 Portland MSA-substations to all 57 Denver MSA-substations using the routing cost surface. We follow the same procedure for Seattle to Denver connection.
    \item Compute the actual cost of all the MSA-substation-to-MSA-substation paths using the real cost surface. 
    \item Identify the path with the least-cost for each pair of regions with inter-regional connection. 
\end{itemize}

\subsection{Development of intra-region network}
We assume that all intra-region network has $<=$500kV and $>=$240kV AC lines. For the development of the intra-regional transmission lines, we follow the steps below:
\begin{itemize}
    \item Identify infinite sink MSAs ($>$1,000,000 population) in each of the GenX model regions. Regions with only one infinite sink MSAs has no intra-region network. 
    \item identify substations that fall inside the convex hull around the infinite sink MSAs. These substations are called "MSA-substations".
    \item Use routing cost surface to create least-cost path between all MSA-substations and all infinite sinks in a given regions. For example, Texas region has 4 infinite sinks: Dallas (49 substations), Austin (7 substations), San Antonio (10 substations), and Houston (38 substations). We create a least-cost paths between MSA-substations of each pair of infinite sinks. 
    \item Compute the actual cost of all the MSA-substation-to-MSA-substation paths using the real cost surface. 
    \item For each pair of infinite sinks, identify the path with the least-cost.
    \item Use a minimum spanning tree algorithm \citep{Naidoo2019} to find a path that connects all infinite sinks in a region at the least cost. For example, least-cost option to connect all infinite sinks in the Texas region is by connecting Dallas to Austin, Austin to San Antonio, and San Antonio to Houston.
\end{itemize}

\subsection{Computation of the cost of inter-regional transmission routes}
We normalize the cost of intra-region transmission lines by MSA population so that the total MW of transmission within a region sums to 1MW for every MW of inter-regional transmission. For example, Desert southwest region has 3 infinite sinks: Las Vegas (2.25M), Phoenix (5M), and Tucson (1M) with total population of 8.25M. The minimum spanning tree of the Desert southwest region connects Las Vegas-Phoenix-Tuscon. If an inter-regional line of 1MW  connects to Las Vegas, we model (2.25+5)/((2.25+5)+(5+1)) = 0.55MW from Las Vegas to Phoenix and 0.45MW from from Phoenix to Tucson.

\begin{table*}
\caption{Inter and intra-regional cost breakdown of the transmission lines }\label{table:txCost}
\resizebox{\textwidth}{!}{
\begin{tabular}{lllll}
\toprule
Region from	&	Region to	&	Inter-regional transmission (\$/MW)	&
Intra-region transmission (\$/MW)	&	Total transmission(\$/MW)	\\
\hline
texas agg.	&	louisiana and ozarks agg.	&	\$565,313	&	\$1,511,017	&	\$2,076,330	\\
florida agg.	&	southeast agg.	&	\$620,524	&	\$675,461	&	\$1,295,985	\\
louisiana and ozarks agg.	&	upper midwest agg.	&	\$681,591	&	\$1,239,299	&	\$1,920,890	\\
mid-atlantic and great lakes agg.	&	new york agg.	&	\$22,305	&	\$853,848	&	\$876,153	\\
louisiana and ozarks agg.	&	lower midwest agg.	&	\$417,214	&	\$1,735,795	&	\$2,153,009	\\
louisiana and ozarks agg.	&	southeast agg.	&	\$533,506	&	\$1,598,218	&	\$2,131,724	\\
new england agg.	&	new york agg.	&	\$1,060,447	&	\$742,583	&	\$1,803,030	\\
mid-atlantic and great lakes agg.	&	upper midwest agg.	&	\$550,119	&	\$240,943	&	\$791,062	\\
lower midwest agg.	&	upper midwest agg.	&	\$623,429	&	\$496,494	&	\$1,119,923	\\
mid-atlantic and great lakes agg.	&	southeast agg.	&	\$97,601	&	\$599,860	&	\$697,461	\\
louisiana and ozarks agg.	&	mid-atlantic and great lakes agg.	&	\$443,101	&	\$1,480,243	&	\$1,923,344	\\
california agg.	&	pacific northwest agg.	&	\$851,548	&	\$835,166	&	\$1,686,714	\\
desert southwest agg.	&	utah/nevada agg.	&	\$625,298	&	\$421,880	&	\$1,047,178	\\
california agg.	&	utah/nevada agg.	&	\$879,441	&	\$380,976	&	\$1,260,417	\\
rocky mountains agg.	&	utah/nevada agg.	&	\$730,080	&	\$0	&	\$730,080	\\
pacific northwest agg.	&	utah/nevada agg.	&	\$989,722	&	\$454,189	&	\$1,443,911	\\
desert southwest agg.	&	rocky mountains agg.	&	\$912,618	&	\$421,881	&	\$1,334,499	\\
pacific northwest agg.	&	rocky mountains agg.	&	\$1,334,896	&	\$454,189	&	\$1,789,085	\\
lower midwest agg.	&	texas agg.	&	\$380,080	&	\$768,213	&	\$1,148,293	\\
california agg.	&	desert southwest agg.	&	\$551,118	&	\$802,857	&	\$1,353,975	\\
rocky mountains agg.	&	texas agg.	&	\$741,886	&	\$271,718	&	\$1,013,604	\\

\bottomrule
\end{tabular}
}
\end{table*}

\section{Additional results}
\subsection{Statistics for existing wind and solar sites}\label{SI:wind_solar_stats}
The following figure demonstrates the frequency of existing solar and wind sites under varying population density, human modification index (HMI), and spur line distance (short site-to-metro lines). For this study, we assume that sites with high population density (i.e., population density in 10\textsuperscript{th} percentile) might be unsuitable for future solar or wind development. Similarly, we assign sites in 10\textsuperscript{th} or 90\textsuperscript{th} percentile of HMI and 90\textsuperscript{th} percentile of spur line distance as unsuitable for further development.

\begin{figure}
  \centering
  \includegraphics[width=\textwidth]{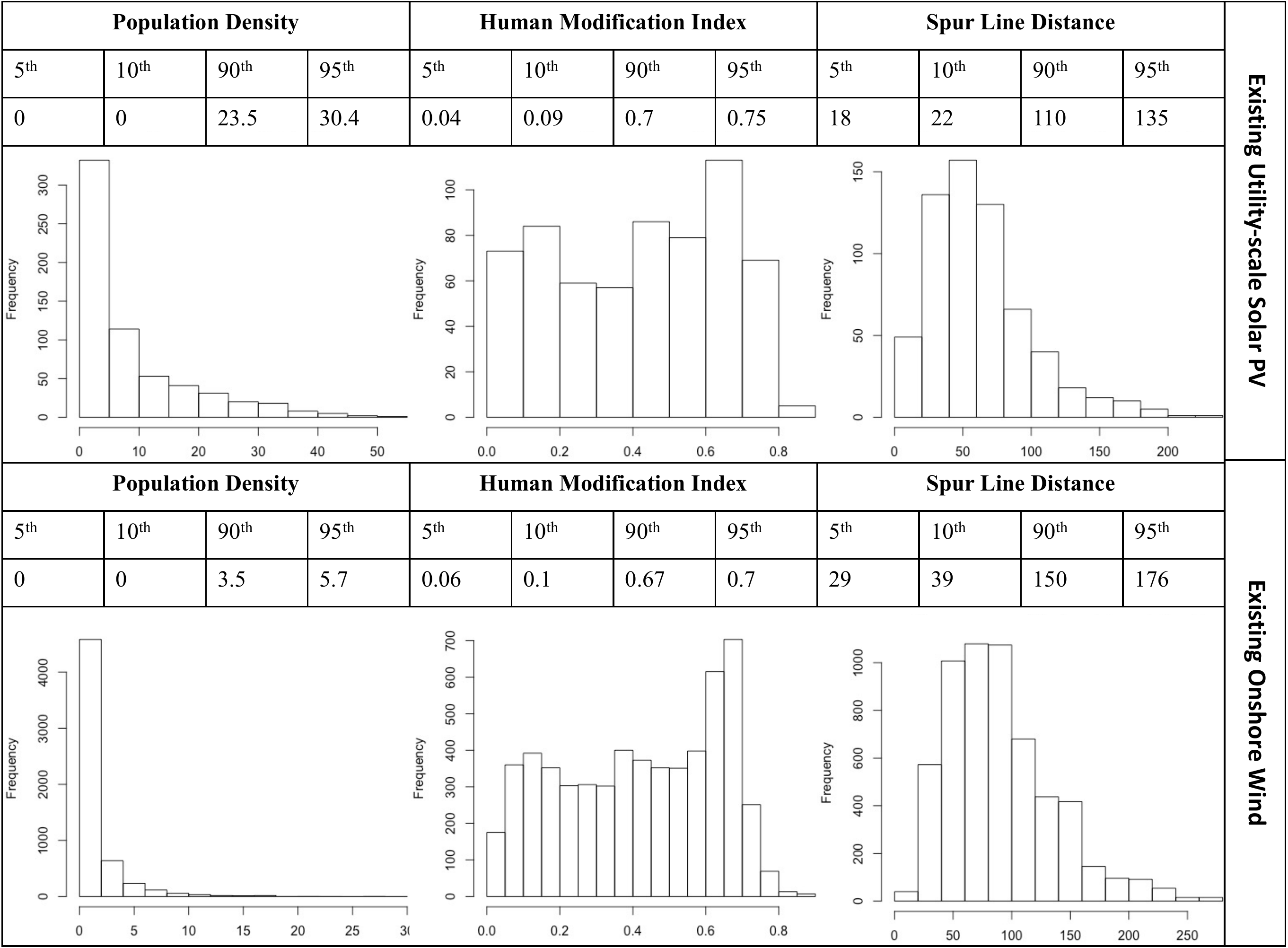}
   \caption{5\textsuperscript{th}, 10\textsuperscript{th}, 90\textsuperscript{th}, and 95\textsuperscript{th} percentile values for spur line distance, population density, and human modification index of existing solar and wind CPAs. Highlighted values are used for the analysis in this paper.}
  \label{fig:solar_wind_stats}
\end{figure}

\subsection{Available land for new solar and wind development} \label{SI:LandAvail}

The following figure demonstrates the Candidate Project Areas (CPAs) available for optimization and the land area/cover type by state. From the MGA iterations performed in this analysis, we learn that the threshold levelized cost of electricity (LCOE) for solar PV is limited to 45 \$/MWh and for onshore wind to 110\$/MWh. In the following figure, we highlight all available solar and wind CPAs that meet this threshold. For context, the total land area available for wind development is 1.48 million km\textsuperscript{2} and solar development is 1.3 million km\textsuperscript{2}. However, the land area available considering the LCOE threshold is less than 10\% of all suitable area (wind = 150,000 km\textsuperscript{2} and solar area = 89,000 km\textsuperscript{2}). The lower figure shows the distribution of suitable and cost-effective land area by land cover type for each state in the WECC power system.

\begin{figure}
  \centering
  \includegraphics[width=\textwidth]{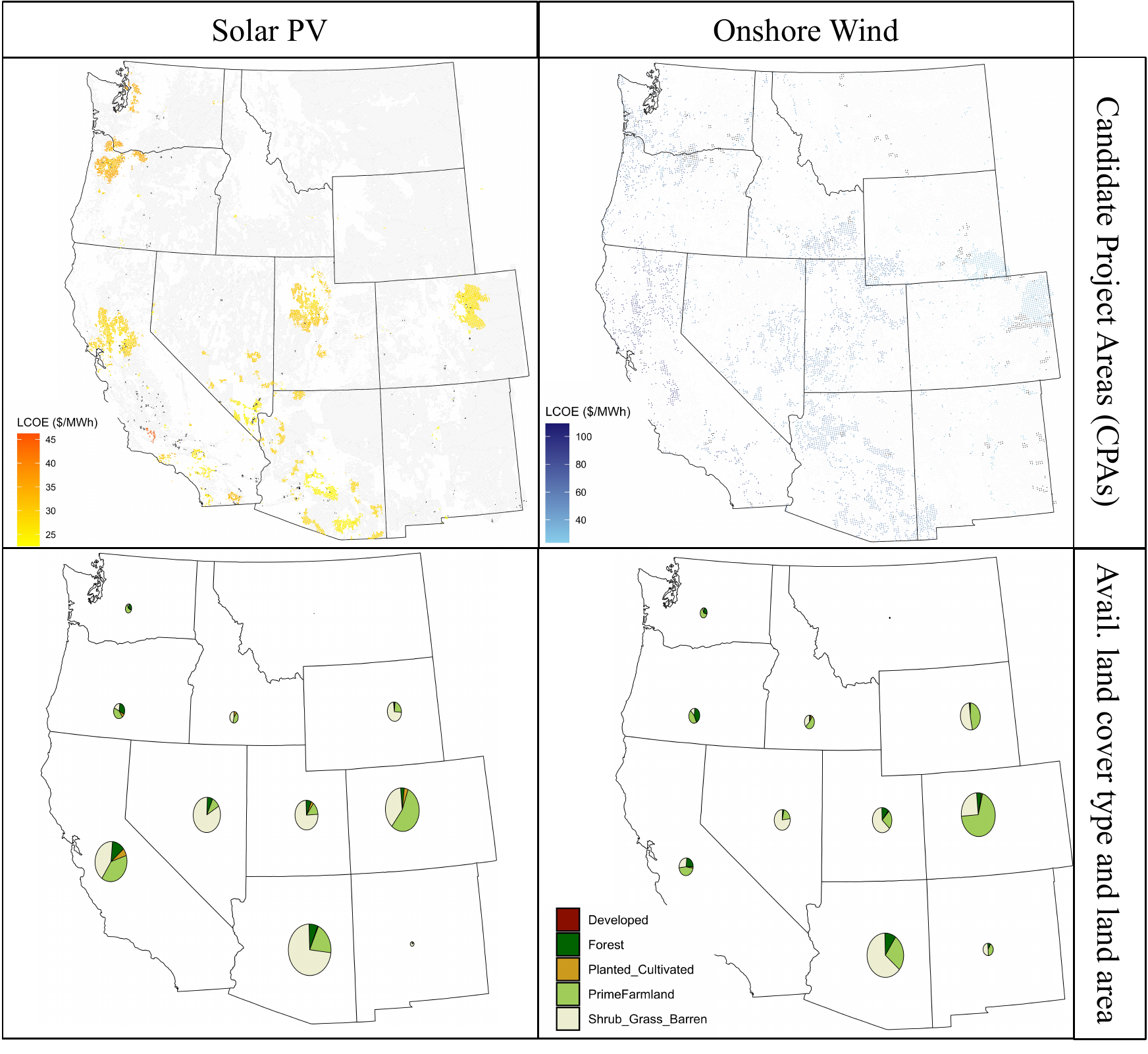}
  \caption{Top row shows the total cost-effective area available for new solar and wind development considering various physical, geographical, and ecological constraints. Bottom row shows the land cover type associated with the cost-effective area in each state. Note that land available at a reasonable cost is approximately ten times lower than the total suitable land available.}
  \label{fig:AvailLandDistribution}
\end{figure}

\subsection{Available, safer, and cost-effective wind and solar sites} \label{SI:Avail_safe_costeffective}
\begin{figure}
  \centering
  \includegraphics[width=\textwidth]{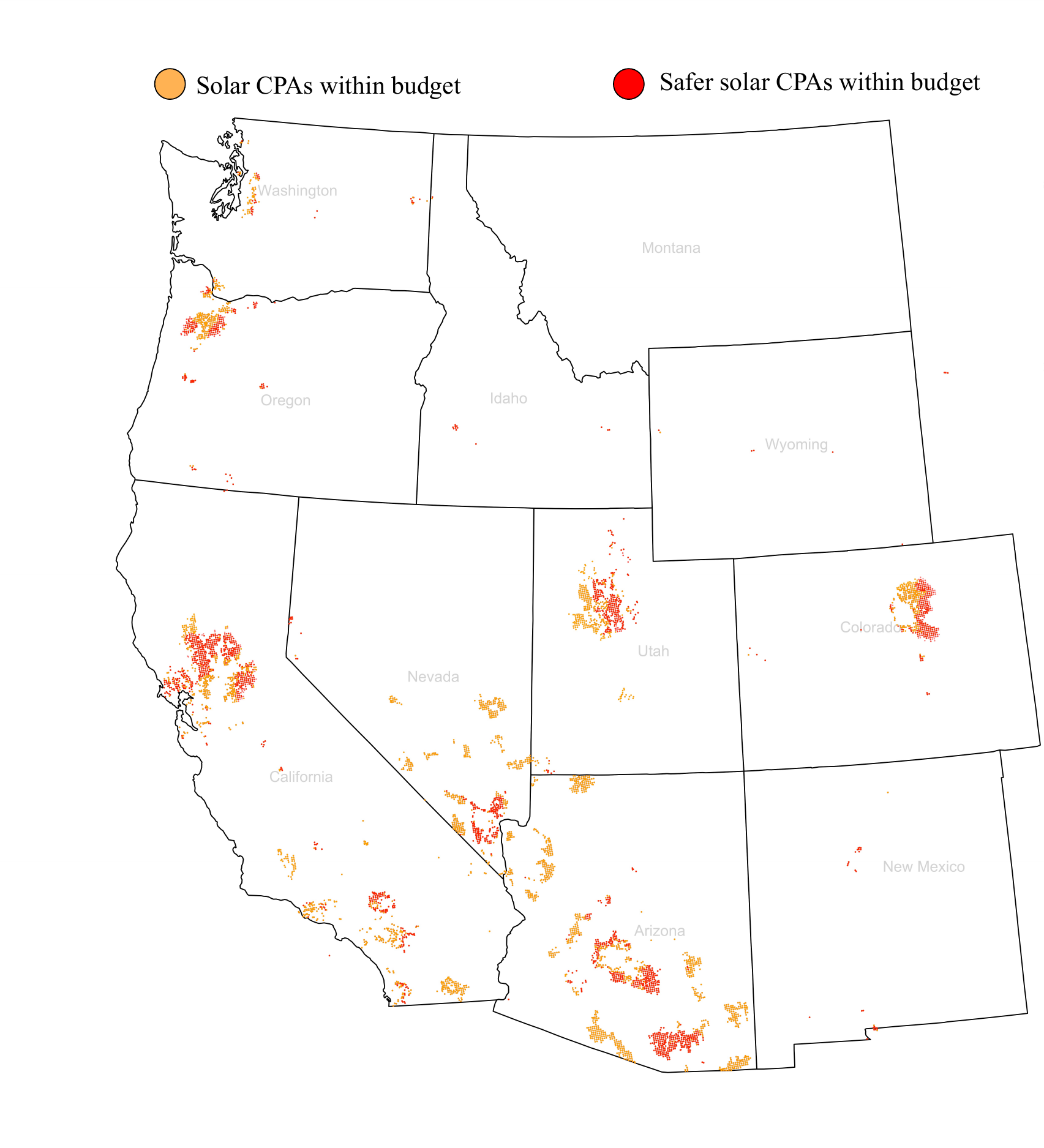}
  \caption{All solar CPAs that can be selected without increasing the total system cost by more than 10\% from the least cost solution. All in-budget wind CPAs that are safer based on the categories described in Section \ref{section:RRisk}}
  \label{fig:safer_S}
\end{figure}

\begin{figure}
  \centering
  \includegraphics[width=\textwidth]{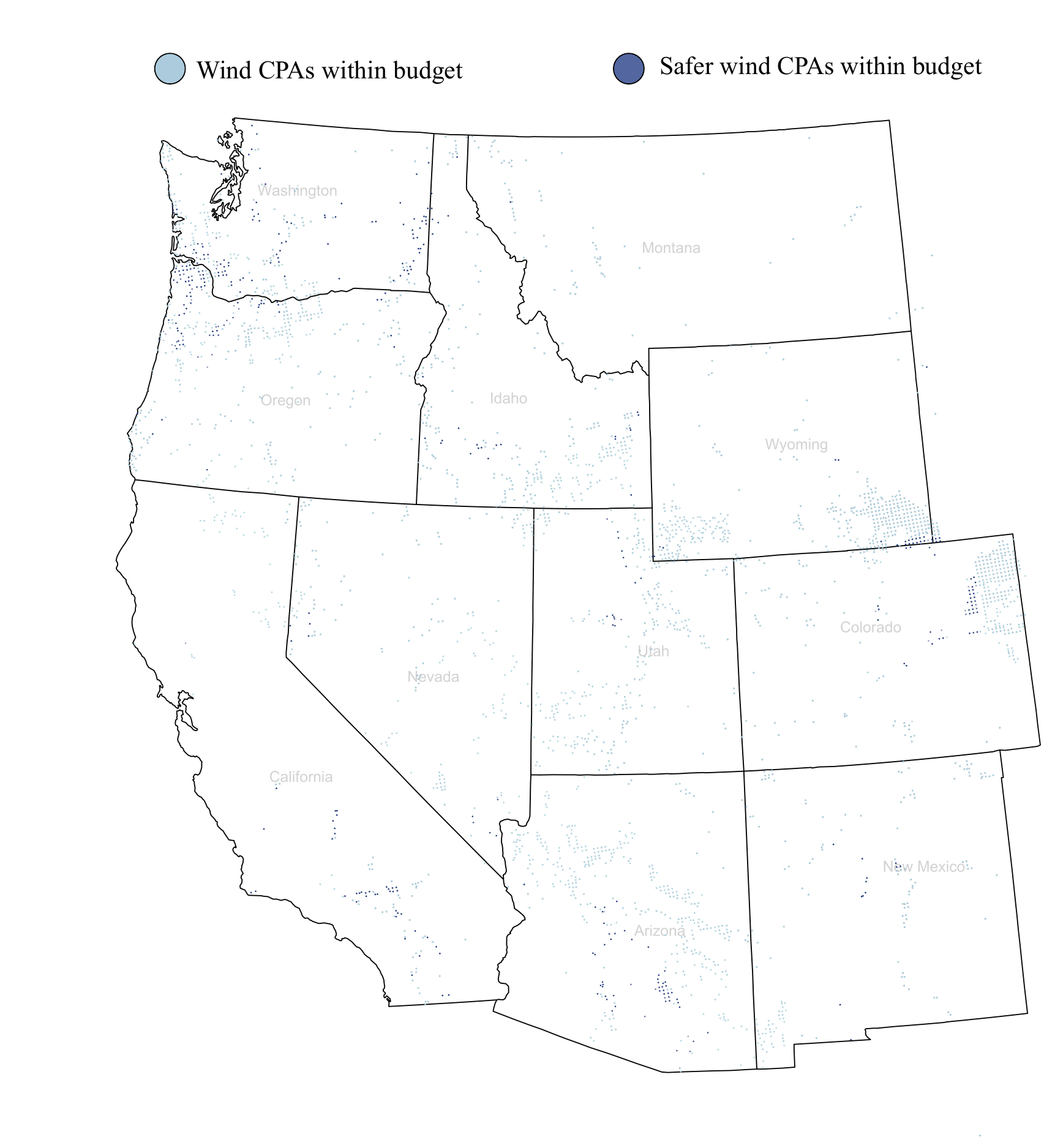}
  \caption{All wind CPAs that can be selected without increasing the total system cost by more than 10\% from the least cost solution. All in-budget wind CPAs that are safer based on the categories described in Section \ref{section:RRisk}}
  \label{fig:safer_W}
\end{figure}

SI Figures \ref{fig:safer_S} and \ref{fig:safer_W} show the cost-effective and safer solar and wind CPAs. Cost-effective CPA is a CPA selected in one or more MGA iteration. The safer CPAs are either located on a prime farmland or satisfy the criteria described in SI \ref{SI:wind_solar_stats}. 

\subsection{Effect of potential objectives on optimal technology capacity} \label{SI:capacity} 

\begin{figure}
  \centering
  \includegraphics[width=\textwidth]{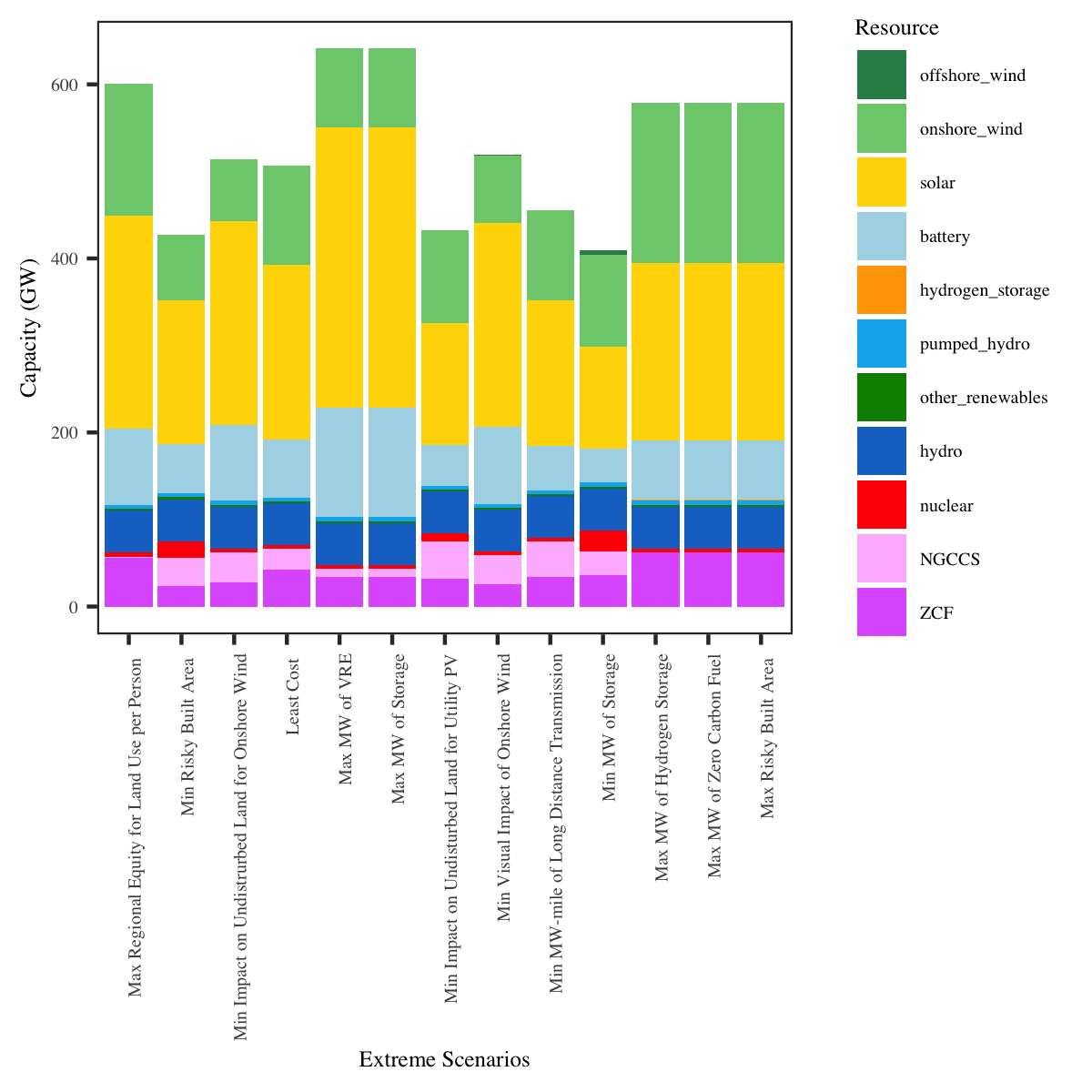}
  \caption{WECC-wide capacity expansion (GW) under ten potential objectives. Criteria are ordered in increasing order of aggregated WECC-wide capacity.}
  \label{fig:Capacity}
\end{figure}

\begin{figure}
  \centering
  \includegraphics[width=\textwidth]{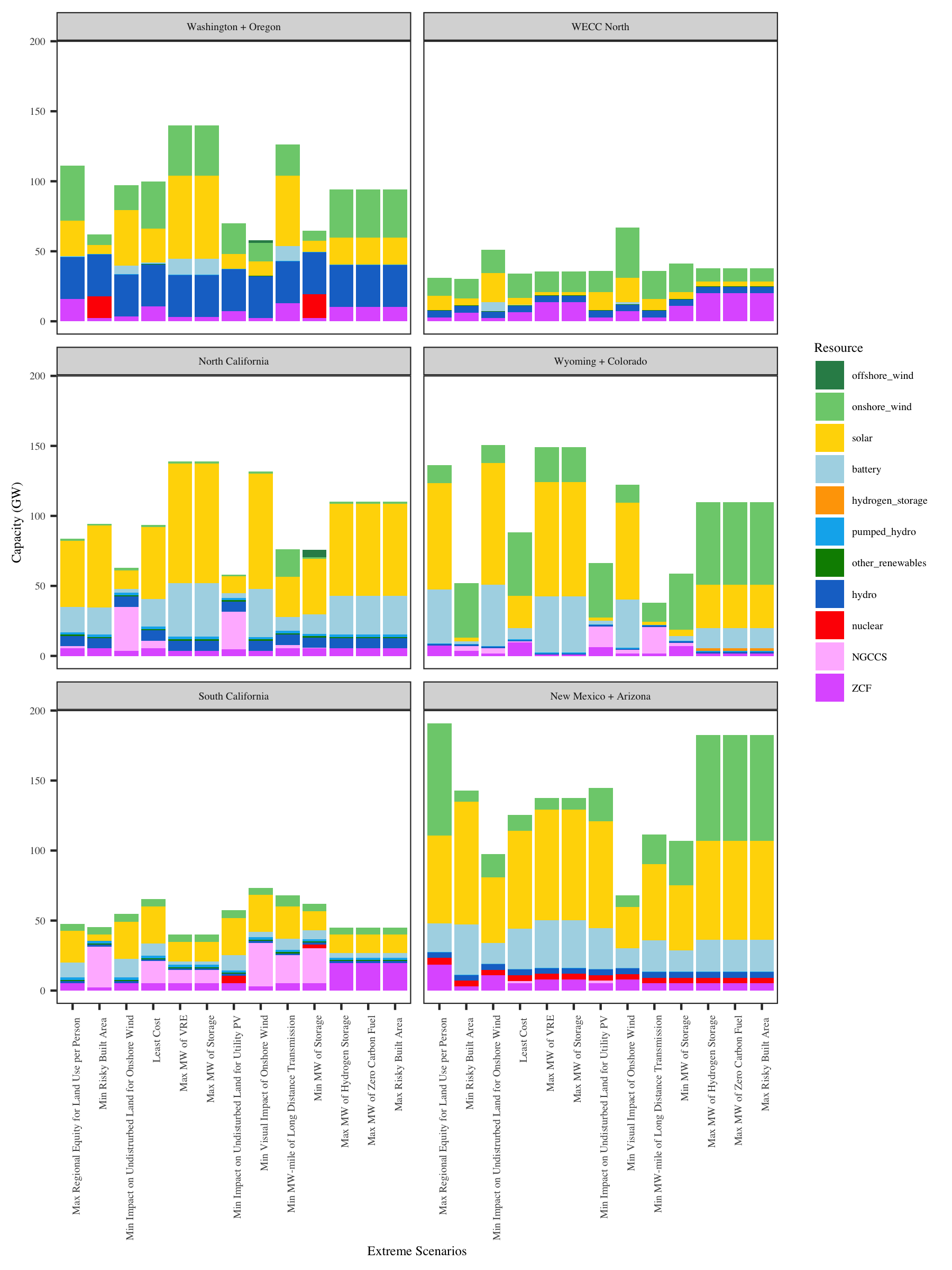}
  \caption{Regional capacity expansion (GW) under ten potential objectives. Criteria are ordered in increasing order of aggregated WECC-wide capacity.}
  \label{fig:Capacity_region}
\end{figure}

We assess the technology breakdown and capacity under ten potential objectives. SI Figure \ref{fig:Capacity} and \ref{fig:Capacity_region} shows the WECC-wide and region specific capacity allocation for each resource under ten potential objectives. Variability at the regional level offers more flexibility than at the systems level, although the range in capacity and technology mix is still significant at the systems level between the lowest capacity (achieved by maximizing clean firm capacity, leading to high penetration of natural gas carbon capture and storage as well as zero-carbon fuels (ZCF)) and the highest capacity (achieved by maximizing storage capacity, leading to almost no natural gas carbon capture and storage and low ZCF, with highest battery and solar). Potential objectives that seem to give fairly homogeneous capacity results at a systems level, such as 'Min impact on Undisturbed Land for Utility PV' compared with 'Min MW of Storage', are shown to have much more recognizable differences in technology capacity at the regional level. This reveals that priorities that seem to barely shift the technology mix of the system as a whole have significant impact on the regions and thus the land area impacts. 
  
\subsection{State-wise total land use trade-off} \label{SI:MinState}

\begin{figure}
  \centering
  \includegraphics[width=\textwidth]{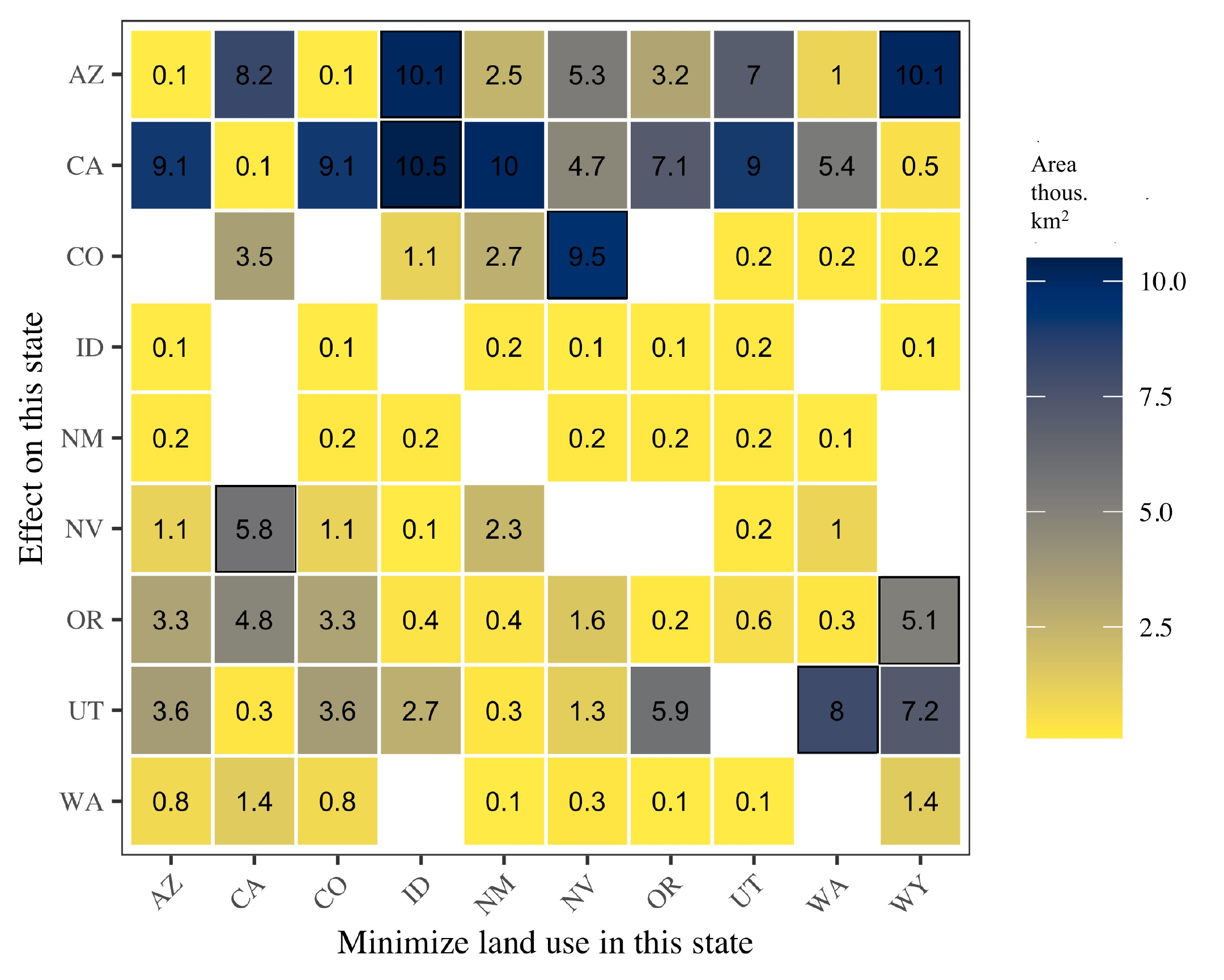}
  \caption{Effects of minimizing solar land footprint in one state on the rest of the states in the WECC. The numbers show the land used in a given state (thousand km\textsuperscript{2}). Diagonal elements ( bottom left - top right) show the minimum required land area in a given state for achieving zero-carbon electricity supply by 2045.}
  \label{fig:MinState_S}
\end{figure}

\begin{figure}
  \centering
  \includegraphics[width=\textwidth]{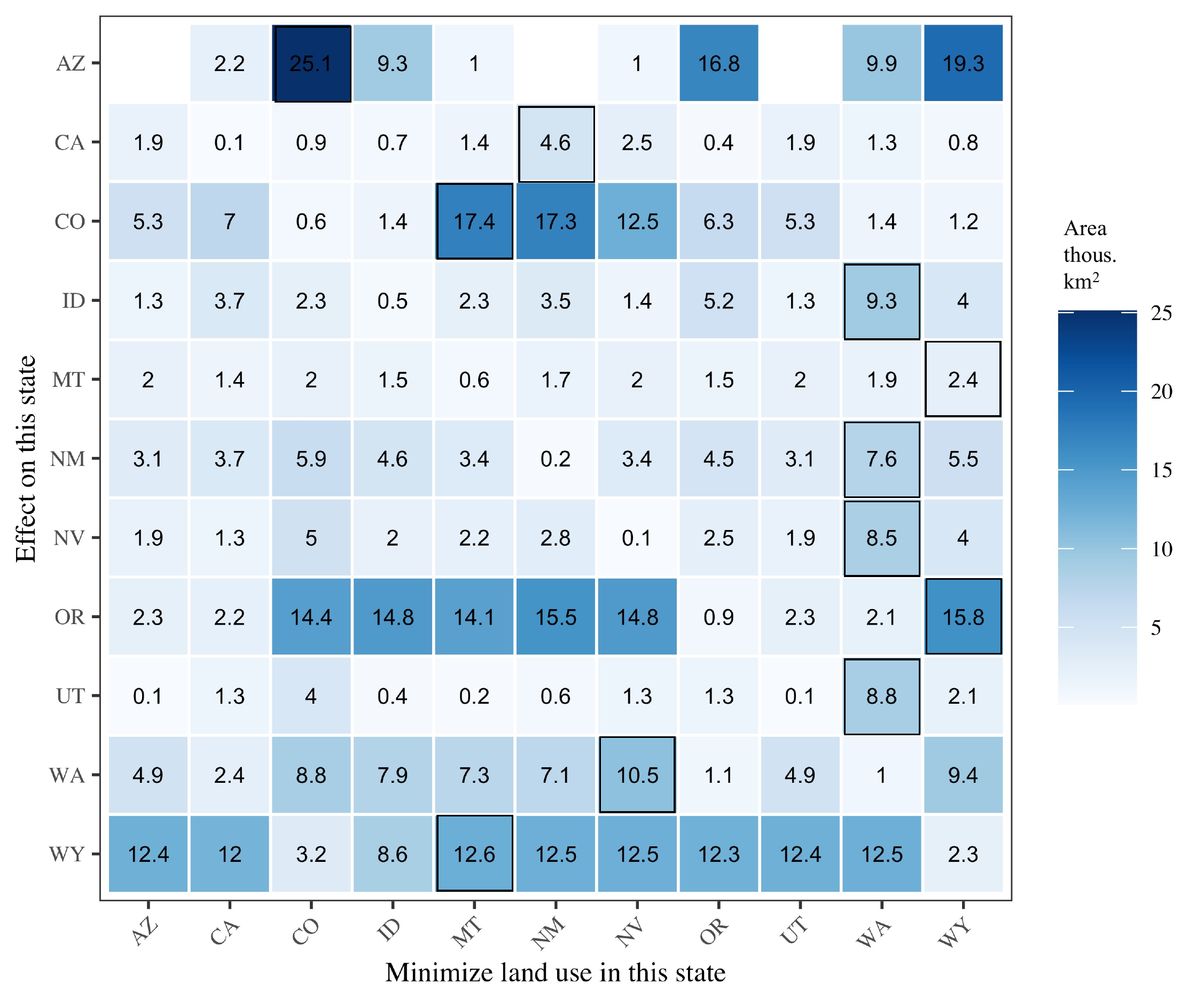}
  \caption{Effects of minimizing wind land footprint in one state on the rest of the states in the WECC. For example, minimizing land impact of wind in Wyoming leads to increase in wind land impact in Montana and vice versa. The numbers show the land used in a given state (thousand km\textsuperscript{2}). Diagonal elements ( bottom left - top right) show the minimum required land area in a given state. Note that these results are valid for 10\% increase in budget from the cost-optimal solution.}
  \label{fig:MinState_W}
\end{figure}

SI Figures \ref{fig:MinState_S} and \ref{fig:MinState_W} demonstrates the effects of minimizing land area impacted by renewable development by state. The x-axis indicates land area in the state that we are minimizing for, and the y-axis is the effect of that minimization on other states. SI Figures \ref{fig:MinState_S} and \ref{fig:MinState_W} show that, under WECC-wide planning for a zero-carbon electricity future, it is feasible in some cases to reduce potential political or ecological conflicts or other bottlenecks by allowing some states to opt out of land use for certain technologies. However, the trade-off of increasing land use in other states can also lead to a decrease in regional equity. It is important to know which states have the capacity to negotiate in this way as well as the effects of such decisions on other states and the region as a whole. The diagonals of both of SI Figures \ref{fig:MinState_S} and \ref{fig:MinState_W} show the minimum amount of land use in each state that can be achieved without increasing the total system cost by more than 10\%. 

\subsection{Land-cover trade-offs} \label{SI:MinCover}

\begin{figure}
  \centering
  \includegraphics[scale=0.6]{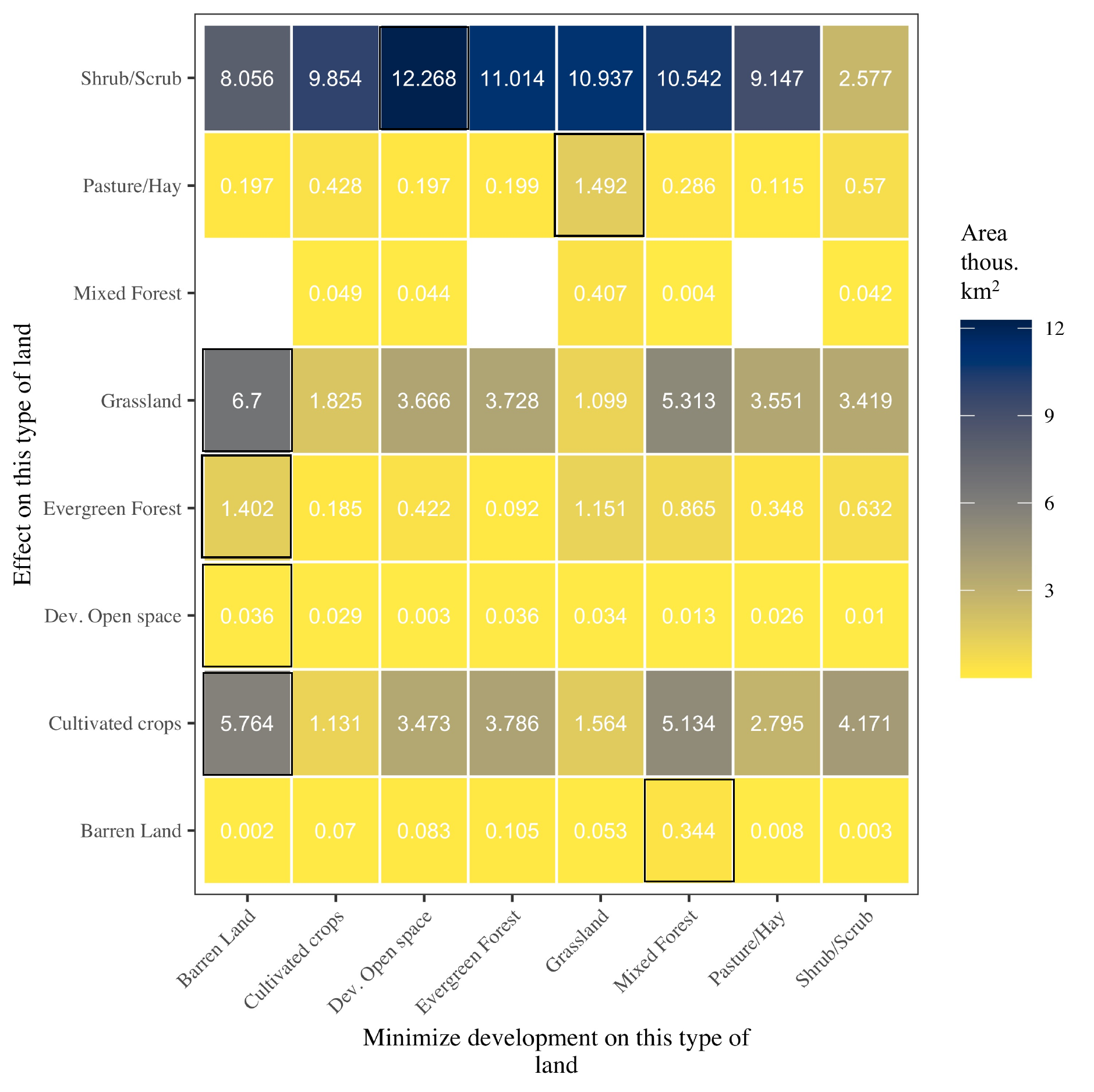}
  \caption{Effects of minimizing a specific land cover type on other land cover types for solar development. The numbers show the area used of a particular land cover type across WECC (thousand km\textsuperscript{2}). Diagonal elements (bottom left - top right) show the minimum required use of a give land cover type across WECC. For example, minimizing shrub land usage for solar development in the WECC increases the grass land use and vice versa. }
  \label{fig:Mincover_S}
\end{figure}

\begin{figure}
  \centering
  \includegraphics[scale=0.6]{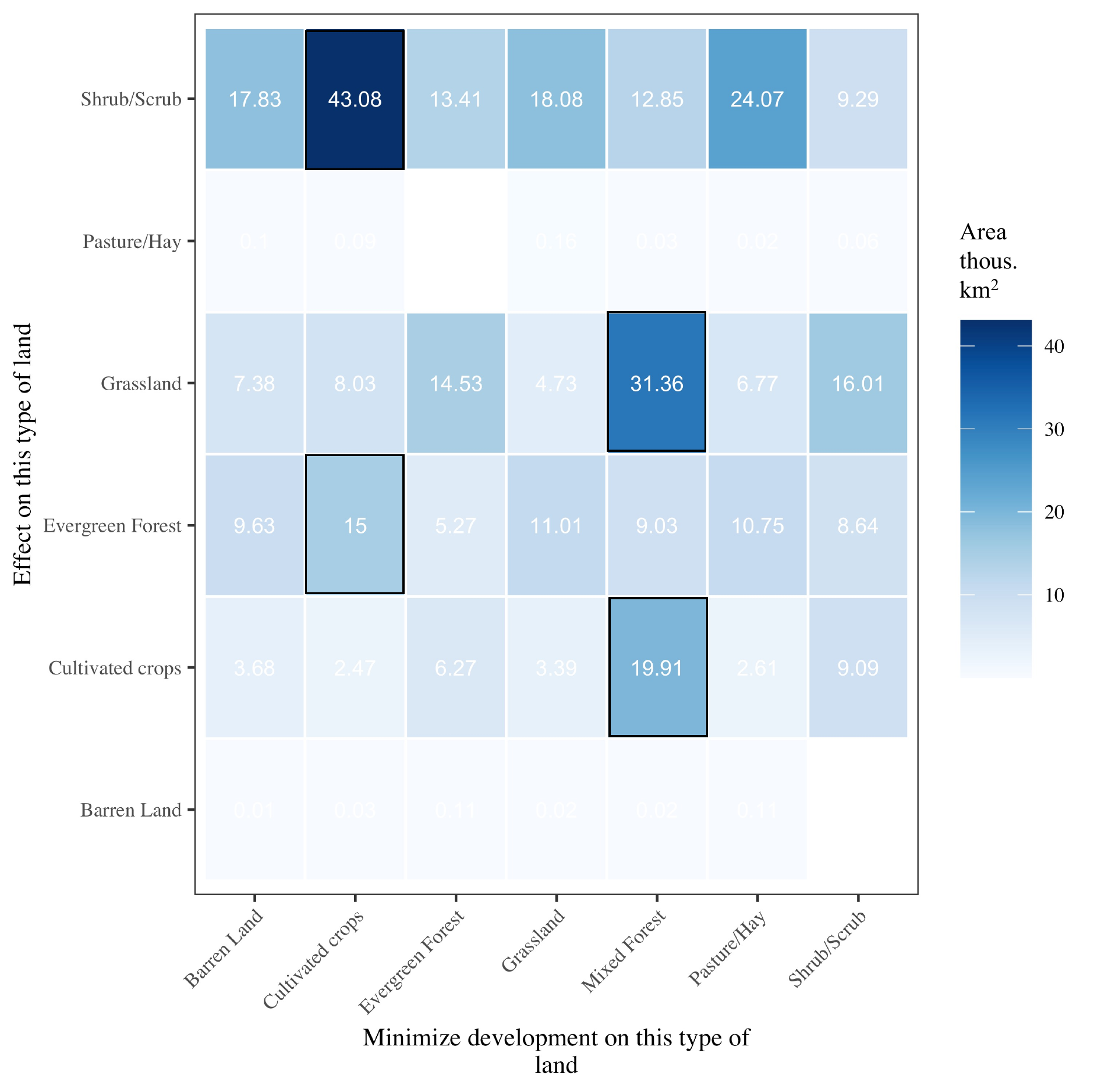}
  \caption{Effects of minimizing a specific land cover type on other land cover types for solar development. The numbers show the area used of a particular land cover type across WECC (thousand km\textsuperscript{2}). Diagonal elements (bottom left - top right) show the minimum required use of a give land cover type across WECC. Note that these results are valid for 10\% increase in budget from the cost-optimal solution.}
  \label{fig:Mincover_W}
\end{figure}

In this section, SI Figures \ref{fig:Mincover_S} and \ref{fig:Mincover_W} show the impact of minimizing land area use according to the land type. In reducing green vs. green conflict, reducing impacts on certain ecologically important lands like forests is a likely priority of stakeholders and accounting for it ahead of time can reduce bottlenecks in the expansion of renewable energy in the American West. Highlighted squares show the maximum value of each row. For example, minimizing the use of evergreen forest for solar development has the worst impact on shrub/grass land. While minimizing the use of shrub/scrub land for solar development has negative impact on cultivated crops land and grass land.
  
\subsection{State-specific land cover trade-offs} \label{SI:state_landcover}
\begin{figure}
  \centering
  \includegraphics[width=\textwidth]{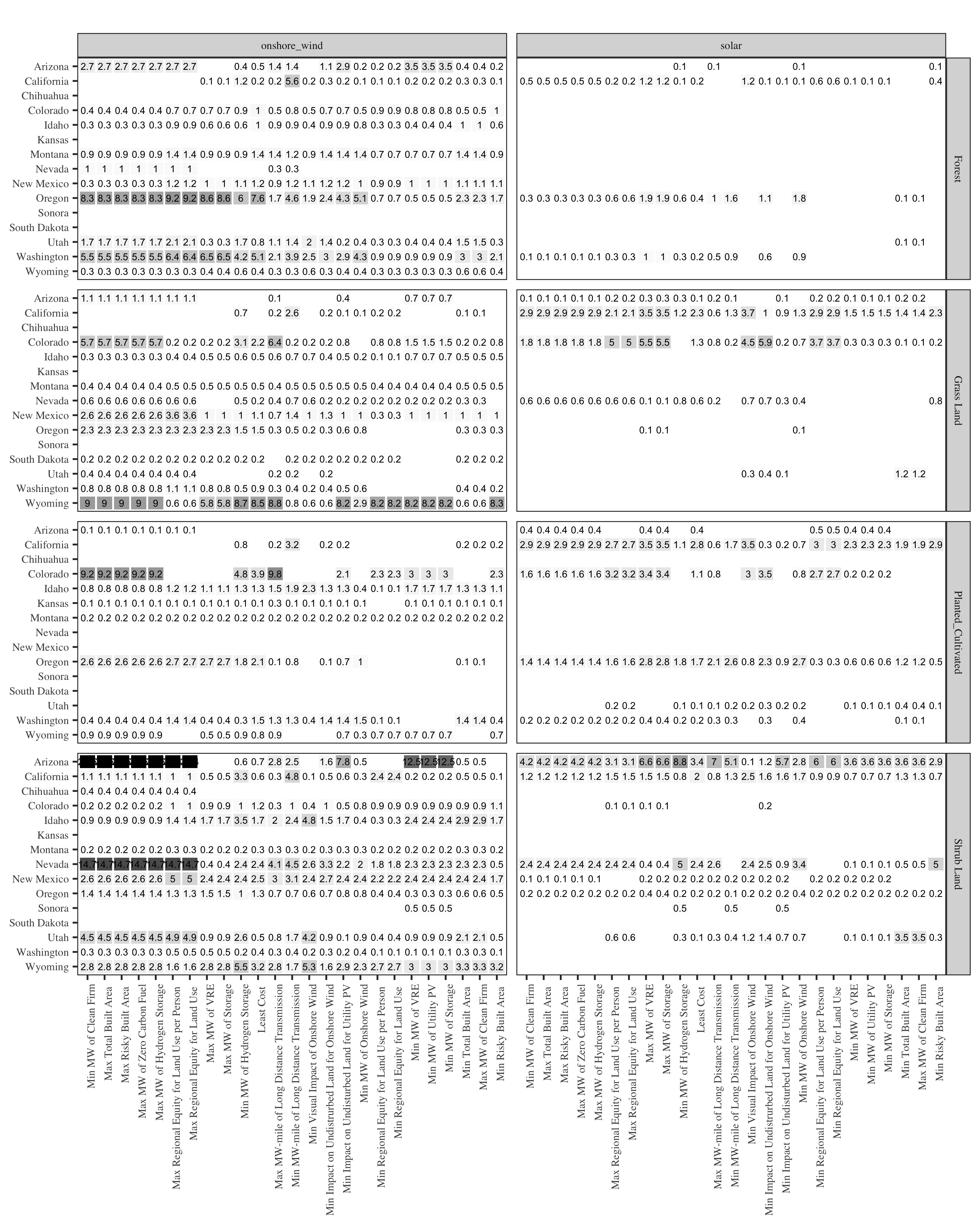}
  \caption{Impact of potential objectives on land cover distribution across the states in the WECC. Numbers show thousand km\textsuperscript{2} land area required to achieve zero-carbon electricity supply in the WECC by 2045.}
  \label{fig:Mincover_state}
\end{figure}

SI Figure \ref{fig:Mincover_state} show the state-wide trade-offs in the usage of different land cover types. Across all potential objectives for utility-scale solar PV, California is the most impacted state for all land types and especially for shrub land. Shrub land shows regional land type trade-offs, offering some relief for California if needed. For example, a policy that maximizes the use of zero-carbon fuel resources minimizes shrub land impacted in Nevada and directly increases land use in California for shrub land, grass land, and planted and cultivated land. Maximizing the use of zero-carbon resources decreases solar capacity overall, but not so significantly in California due to its concentration of low-LCOE solar resources. For solar, a policy that maximizes capacity of clean firm resources evenly distributes shrub land use across land types, decreasing ecological impacts on the states that tend to be most impacted by renewables expansion. Furthermore, maximizing clean firm capacity maintains a medium level of impact across land types (forest, grass land, planted\_cultivated, and shrub land) and is a good option for distributing effects in an equitable way across land types and states. For solar, land type impacts are also smoothed by policies that aim to both maximize and minimize technology site boundary, shifting land use impacts to wind development and which shifts impacts across more states and particularly increases land type impact on Oregon. 

\section{Input data} \label{SI:Data}

SI Table \ref{Tab:datasource} outlines the key system parameters underlying the WECC 6 zone model for all criteria. We use 9000\$/MW to approximate the value of lost load (or the cost incurred by consumers during involuntary emergency demand curtailment during periods of supply scarcity, e.g. “rolling brownouts”). We increase base technology investment costs from NREL’s Annual Technology Baseline 2020 \citep{akar20202020} with the spur line cost to account for the cost of transmission from generators to the metropolitan statistical area, and natural gas plants with carbon capture and storage as well as nuclear (SI Table \ref{Tab:3}). We also use NREL’s regional multipliers to account for regional differences in technology cost due to prevailing costs of labor, land, and other factors \citep{Nrel_regional_multiplier} .  

\begin{table*}
\caption{Sources and assumptions for input data} \label{Tab:datasource}
\resizebox{\textwidth}{!}{%
\begin{tabular}{llllll}
  \toprule
  Data  &	Criteria & Type & Sub-Type & Specifications & Source\\
	\hline
	Capacity & Existing & All power plants &- & - & \cite{PUDL} \\
	& Max/region &Onshore Wind & LCOE cut-off & 150 \$/MWh & \cite{larsonnetzero2021}\\
	& Max/region & Utility PV & LCOE cut-off & 90 \$/MWh & \cite{larsonnetzero2021}\\
	& Max/region & Offshore Wind & LCOE cut-off & 200 \$/MWh & \cite{larsonnetzero2021}\\
	\hline
	VRE Profiles & Existing & Utility PV & Single axis tracking & & Vibrant Clean Energy \\
	& Existing & Onshore Wind & 80m hub height &- & Vibrant Clean Energy \\
	& New & Utility PV & Single axis tracking &- & Vibrant Clean Energy \\
	& New & Onshore Wind & 100m hub height &- & Vibrant Clean Energy \\
	& New & Offshore Wind & Floating turbine &- & Vibrant Clean Energy \\
	& New & Rooftop PV & Single axis tracking &- & \cite{pfenninger2017dealing} \\
	\hline
	Technical Characteristics & Ramp up/down & Thermal generators &- & -& \cite{CPUCIRP}\\
	& Min Power & & -&- & \\
	\hline
	Fixed Cost & Existing & All power plants & Fixed O\&M Cost &- & \cite{PUDL} \\
	& New & Utility PV & Los Angeles &- & \cite{akar20202020} \\
	& New & Onshore Wind & LTRG1 &- &  \cite{akar20202020} \\
	& New & Offshore Wind & OTRG6 & -&  \cite{akar20202020}\\
	& New & NGCT & CTAvgCF & - &  \cite{akar20202020} \\
	& New & NGCC & CCAvgCF & - &  \cite{akar20202020} \\
	& New & NGCCS & CCCSAvgCF & - &  \cite{akar20202020} \\
	& New & Hydrogen Storage & Electrolysis & -& Same as NGCT\\
	& New &  & Storage & & \cite{steward2009lifecycle}\\
	\hline
	Operational Cost & Fuel & ZCF & high/low & 33/15 \$/MMBtu & \cite{baik2021different} \\
	& Fuel & All & -& -& \cite{center2020annual} \\
	& Start up & All fuel types & & & \cite{energy2010western}\\
	\hline
	Load & Profile & Electrification & High/Moderate &- & \cite{mai2018electrification} \\
	& Demand Flexibility & EV load & High/Moderate & 5 hour flexibility & \cite{mai2018electrification} \\
	\hline
	Emission & Rate & -&- & -& \cite{center2020annual} \\
	& Cost & CO\textsubscript{2} pipeline/transport & -& 10 \$/tonne & \\
	\hline
	Transmission & Existing & Capacity &- &- & EPA IPM\\
	& New & Cost & inter-regional & &  \cite{gorman2019improving}\\
	& & Cost & Spur line &- &  \cite{larsonnetzero2021}\\
	\bottomrule
\end{tabular}%
}
\begin{tablenotes}
\scriptsize
\item
Utility PV - Utility scale solar photovoltaic; O\&M - Operations and Maintenance; ZCF - zero-carbon fuel; wacc - weighted average cost of capital; NGCT - Natural gas combustion turbine; NGCC - natural gas combined cycle turbine; NGCCS - natural gas combined cycle with carbon capture and sequestration
\end{tablenotes}
\end{table*}

\subsection{Generation units }

This study solves an optimization problem for a predicted future planning year of 2045, with each year cut into 8760 hourly time intervals. We roughly represent existing generators through a clustering approach where we group individual generators into nine generator types and one storage type based on installed capacity data from \cite{center2020annual}: solar (single axis tracking), onshore wind, hydroelectric reservoir, hydroelectric pumped storage, small hydroelectric (run of the river resources $<$30 MW), biomass, geothermal, and nuclear. As all scenarios run considering fully decarbonized electricity portfolios, we do not model existing coal, natural gas, or petroleum-fired generators. 
For each region in the all technology (REF) case, we allow new capacity in the following generator clusters: solar, onshore wind, battery, hydrogen storage, and hydropower. WECC WYCO, WECC NMAZ, and each CA region are allowed new capacity in natural gas plants with carbon capture and storage (NGCCS), based on limited availability of proximate CO\textsubscript{2} sequestration locations.Offshore wind is allowed in CA N and WECC PNW. New nuclear is allowed in all regions except CA due to an existing state moratorium on new nuclear development. We use NREL ATB \cite{akar20202020} values for predicted capital as well as fixed and variable operations and maintenance costs and convert these to regional values with NREL regional cost multipliers \citep{Nrel_regional_multiplier}. SI Table \ref{Tab:datasource} details cost and performance parameters for new generation technologies. Operational performance characteristics for each resource are from \cite{CPUCIRP}. 

This study’s WECC power system model has 164 eligible resources, 30 thermal units subject to linearized unit commitment constraints, and 91 variable renewable profiles. Regional data for fuels comes from EIA Annual Energy Outlook \citep{center2020annual}. We assume 100\% CCS post-combustion capture rate and add additional ca[ital cost, heat rate, and O\&M costs relative to NREL ATB 2020 assumptions for NGCC w/CCS with 90\% capture rate to reflect higher costs of 100\% capture \citep{feron2019towards}. We assume biomass generation is carbon-neutral. Uranium price for all WECC regions is 0.71 \$/MMBtu. Fuel cost for NGCC w/100\% CCS  is 5.68 \$/MMBtu in CA and in all other regions is 5.92 \$/MMBtu (inclusive of a \$10/ton cost for CO\textsubscript{2} transport and sequestration). CO\textsubscript{2} content for these technologies is 0 tons/MMBtu. 

\begin{table*}
\caption{Summary of techno-economic assumptions for generation technologies \citep{CPUCIRP}} \label{Tab:3}
\resizebox{\textwidth}{!}{%
\begin{tabular}{llllllllll}
\toprule
Technology & Capex [\$]& FOM [\$/MW-a] & VOM [\$/MWh]& LT [yr] & HR & UT/DT [hr]& RU/RD [\%] & SC [\$/MW] & SLD [miles]\\
\hline
Onshore wind & 1202179 & 38186 & 0 & 30 & - & 0 & 100 & - & varies\\
Offshore wind & 2372406 & 64072 & 0 & 30 & -  & 0 & 100 & - & varies\\
Solar & 818823 & 12623 & 0 & 40 & -  & 0 & 100 & - & varies\\
\hline
NGCC ZCF & 769560 & 10463 &	2.1 & 40 & 6.15  & 6 & 64 & 91 & 0\\
NGCT ZCF & 656055 & 7143 &	11.2 & 40 & 8.8  & 1 & 378 & 118 & 0\\
NGCC CCS & 2069884 & 44735 & 8.1 & 40 & 7.86  & 6 & 100 & 91 & 20\\
nuclear & 6166584 & 105225 & 2.4 & 60 & 10.46  & 24 & 25 & 245 & 50\\
\bottomrule
\end{tabular}%
}
\begin{tablenotes}
    \scriptsize
    \item
    Capex - Overnight capital cost; FOM - Fixed Operations and Maintenance; VOM - Variable operations and maintenance; LT - Life time; HR - Heat rate; UT/DT - up time/down time; RU/RD - ramp up/ramp down; SC - Start-up cost; SLD - Spur line distance 
    \end{tablenotes}
\end{table*}

\subsection{Assumptions for variable renewables }

Data for developable wind and solar capacity and cost of grid interconnection (spur line cost) in each zone are based on a detailed, nationwide geospatial analysis of suitable sites for wind and solar development performed for Princeton’s Net Zero America study \citep{larsonnetzero2021}. Thousands of individual candidate project areas suitable for development are then clustered using agglomerative clustering method based on levelized cost electricity accounting for resource quality (capacity factor) and spur line cost and a selection of the best clusters are included as candidate resources in this study.  Hourly capacity factor profiles for all renewable resources come from simulated weather model data sets purchased from Vibrant Clean Energy using NOAA RUC assimilation model data for weather year 2012 to align with load time series \citep{choukulkar2016new, clack2016demonstrating, clack2017modeling}. The wind and solar variability data series is at a 13-km spatial resolution for the United States. The maximum solar development density per candidate project area/cluster is set to 20\% \citep{wu2020low} to reflect the fact that while many areas are suitable for solar deployment, it is unlikely that solar development will proceed at high density across wide geographic areas due to its high impact intensity.

Reservoir hydro input data assumes initial level of water in the reservoir as half full at model start and end, inflow data and minimum reservoir level. We take monthly inflow data for state reservoir hydro from \cite{center2020annual}, and combine state level data into our model regions, distributing monthly inflow equally over each hour.  Existing reservoir capacity data comes from \cite{center2020annual}. Reservoir hydro and small hydro (run-of-the-river) are assumed to have no new capacity additions.

\subsection{Storage technologies}

Pumped hydro storage is assumed to have no new capacity.  We use the data from NREL ATB report \citep{akar20202020} for battery storage costs. Cost assumptions for hydrogen storage consider charging, discharging, and storage, with cost projections for charging capacity  (electrolysis) and fixed O\&M for 2045 projected out from 2030 \citep{saba2018investment, steward2009lifecycle} and 2050 \citep{michalski2017hydrogen, hydrogenIrena19} data. Capital cost projections for hydrogen storage capacity come from \citep{lord2014geologic} and assume underground cavern storage. No fixed or variable O\&M is included for charging or storage capacity. We assume all cost assumptions for hydrogen storage discharging capacity to be the same as natural gas combustion turbines (NGCT) \citep{akar20202020}. SI Table \ref{Tab:stor} lists techno-economic assumptions for hydrogen storage and lithium-ion battery storage. 

\begin{table*}
    \footnotesize
		\centering
		\caption{Summary of techno-economic assumptions for generation technologies} \label{Tab:stor}
		\begin{tabular}{p{4cm}|p{2.5cm}|p{2.5cm}}
    \toprule
    Parameter & Hydrogen storage (DeSolve) & Li-ion Battery (NREL ATB)\\
			\hline
			Capex [\$] (discharging) & 656055 & 342308\\
			IC [\$/MWh] (charging) & 27593 & -\\
			IC [\$/MWh] (storage) & 81 & 11395\\
			FOM [\$/MW-a] & 0 & 8557\\
			VOM [\$/MWh] & 11.2 & 0\\
			LT [yr] & 30 & 40\\
			EU/ED [\%] & 67/40 & 92/92\\
			P2E & 1 & 1\\
			UT/DT [hr]& 1/1 & 0/0\\
			RU/RD [\%] & 100 & 100\\
			min duration [hr] & 48 & 1\\
			max duration [hr] & 2000 & 10\\
			\bottomrule
	  \end{tabular}
\end{table*}

\subsection{Cost and demand profile assumptions} \label{SI:ScenarioDef}

This study uses electrification criteria and their accompanying data from NREL’s 2018 Electrification Futures study scenarios \citep{mai2018electrification}. Reference electrification is a base case with the least incremental change in electrification through 2050, which serves as a baseline of comparison to the other scenarios. High electrification is a result of technological advancement, policy, and consumer priorities such that larger shares of energy service demands in transportation, heating, and industrial sectors are met by electricity. Under the assumptions of the dataset used in NREL and this study, electrification increases for all sectors in all electrification cases such that approximate total electricity share of final energy consumption in the US rises from 20 to 40\%, with the transportation sector increasing from 1\% to 30\% electrified, the commercial sector increasing from 60 to 75\% electrified, the residential sector increasing from 45 to 60\% electrified, and the industrial sector increasing from 20 to 25\% electrified. 

This study assumes low cost projections of variable renewable energy technologies from the NREL ATB new build values \citep{akar20202020}. This study assumes two cases for moderate levels of demand flexibility and draws these values from NREL’s Electrification Futures study.



\end{appendices}

%%===========================================================================================%%

\bibliography{sn-bibliography.bib}% common bib file

\begin{thebibliography}{126}
\providecommand{\natexlab}[1]{#1}
\providecommand{\url}[1]{{#1}}
\providecommand{\urlprefix}{URL }
\providecommand{\doi}[1]{\url{https://doi.org/#1}}
\providecommand{\eprint}[2][]{\url{#2}}
 \bibcommenthead

\bibitem[{USG(2011)}]{USGSGapAnalysisProject2011}
 (2011) Usgs gap analysis project. \doi{https://DOI.org/10.5066/F7FT8JXP},
  \urlprefix\url{https://www.sciencebase.gov/catalog/item/5540ec40e4b0a658d7939628}

\bibitem[{Geo(2018)}]{GeologicalSurvey2018}
 (2018) Protected areas database of the united states ({PAD}-{US}): U.s.
  geological survey data release.
  \urlprefix\url{https://DOI.org/10.5066/P955KPLE}

\bibitem[{MRL(2019)}]{MRLC2019}
 (2019) National land cover database 2019 (nlcd2019).
  \urlprefix\url{https://www.mrlc.gov/data/legends/national-land-cover-database-2019-nlcd2019-legend}

\bibitem[{usc(2020)}]{uscensusbureaucartographic2020}
 (2020) Cartographic boundary shapefiles - county.
  \urlprefix\url{https://www2.census.gov/geo/tiger/GENZ2018/shp/cb_2018_us_county_500k.zip}

\bibitem[{EIA(2020)}]{EIA8602020}
 (2020) {EIA} (last) (2019). {Preliminary} {Monthly} {Electric} {Generator}
  {Inventory} ({Based} on {Form} {EIA}-{860M} as a {Supplement} to {Form}
  {EIA}-860). \urlprefix\url{https://www.eia.gov/electricity/data/eia860m/}

\bibitem[{Lan(2020)}]{LandCoverData}
 (2020) Multi-resolution land characteristics (mrlc) consortium. \url{).
  https://www.mrlc.gov/data?f\%5B0\%5D=category\%3ALand\%20Cover}, Accessed:
  1-6-2020

\bibitem[{USD(2020)}]{USDANRCS2020}
 (2020) Usa ssurgo - farmland class.
  \urlprefix\url{https://www.arcgis.com/home/item.html?id=9708ede640c640aca1de362589e60f46}

\bibitem[{EZM(2021)}]{EZM2021}
 (2021) Energy zone mapping tool. \urlprefix\url{https://ezmt.anl.gov/}

\bibitem[{GIS(2021)}]{GISdataset2021}
 (2021) Geographic information systems datasets.
  \urlprefix\url{https://www.acq.osd.mil/dodsc/fast41_gisdatasets.html}

\bibitem[{NCE(2021)}]{NCEI2021}
 (2021) National center for environmental information: Radar data.
  \urlprefix\url{https://www.ncei.noaa.gov/maps/radar/}

\bibitem[{NAV(2021)}]{NAVAIDSystem2021}
 (2021) Navaid system.
  \urlprefix\url{https://services6.arcgis.com/ssFJjBXIUyZDrSYZ/ArcGIS/rest/services/NAVAIDSystem/FeatureServer}

\bibitem[{USG(2021)}]{USGS}
 (2021) Usgs - core science analytics, synthesis, and library.
  \url{https://www.sciencebase.gov/catalog/file/get/5540ebe2e4b0a658d7939626?f=__disk__9c\%2F24\%2Fd5\%2F9c24d5062c98ecf82988b4e6c827d07c374e9776&transform=1&allowOpen=true},
  Accessed: 1-6-2020

\bibitem[{VFR(2021)}]{VFRLandmark2021}
 (2021) Vfr navigation landmarks.
  \urlprefix\url{https://services6.arcgis.com/ssFJjBXIUyZDrSYZ/ArcGIS/rest/services/VFR_Navigation_Landmarks/FeatureServer}

\bibitem[{Akar et~al(2020)Akar, Beiter, Cole, Feldman, Kurup, Lantz, Margolis,
  Oladosu, Stehly, Rhodes et~al}]{akar20202020}
Akar S, Beiter P, Cole W, et~al (2020) 2020 annual technology baseline (atb)
  cost and performance data for electricity generation technologies. Tech.
  rep., National Renewable Energy Laboratory-Data (NREL-DATA), Golden, CO
  (United~…

\bibitem[{{Anthony Lopez, Billy Roberts, Donna Heimiller,} and {Nate Blair, and
  Gian Porro}(2012)}]{lopez2012}
{Anthony Lopez, Billy Roberts, Donna Heimiller,}, {Nate Blair, and Gian Porro}
  (2012) U.{S}. {Renewable} {Energy} {Technical} {Potentials}: {A}
  {GIS}-{Based} {Analysis}.
  \urlprefix\url{https://www.nrel.gov/docs/fy12osti/51946.pdf}

\bibitem[{Baik et~al(2021)Baik, Chawla, Jenkins, Kolster, Patankar, Olson,
  Benson, and Long}]{baik2021different}
Baik E, Chawla KP, Jenkins JD, et~al (2021) What is different about different
  net-zero carbon electricity systems? Energy and Climate Change 2:100,046

\bibitem[{Bar-Joseph et~al(2001)Bar-Joseph, Gifford, and
  Jaakkola}]{bar2001fast}
Bar-Joseph Z, Gifford DK, Jaakkola TS (2001) Fast optimal leaf ordering for
  hierarchical clustering. Bioinformatics 17(suppl\_1):S22--S29

\bibitem[{Barron-Gafford et~al(2019)Barron-Gafford, Pavao-Zuckerman, Minor,
  Sutter, Barnett-Moreno, Blackett, Thompson, Dimond, Gerlak, Nabhan
  et~al}]{barron2019agrivoltaics}
Barron-Gafford GA, Pavao-Zuckerman MA, Minor RL, et~al (2019) Agrivoltaics
  provide mutual benefits across the food--energy--water nexus in drylands.
  Nature Sustainability 2(9):848--855

\bibitem[{Beiter et~al(2016)Beiter, Musial, Smith, Kilcher, Damiani, Maness,
  Sirnivas, Stehly, Gevorgian, Mooney et~al}]{beiter2016spatial}
Beiter P, Musial W, Smith A, et~al (2016) A spatial-economic cost-reduction
  pathway analysis for us offshore wind energy development from 2015--2030.
  Tech. rep., National Renewable Energy Lab.(NREL), Golden, CO (United States)

\bibitem[{Berntsen and Trutnevyte(2017)}]{berntsen2017ensuring}
Berntsen PB, Trutnevyte E (2017) Ensuring diversity of national energy
  scenarios: Bottom-up energy system model with modeling to generate
  alternatives. Energy 126:886--898

\bibitem[{Bolinger et~al(2020{\natexlab{a}})Bolinger, Lantz, Wiser, Hoen, Rand,
  and Hammond}]{bolinger2020}
Bolinger M, Lantz E, Wiser RH, et~al (2020{\natexlab{a}}) Opportunities for and
  {Challenges} to {Further} {Reductions} in the “{Specific} {Power}”
  {Rating} of {Wind} {Turbines} {Installed} in the {United} {States}. Wind
  Engineering \doi{10.1177/0309524X19901012},
  \urlprefix\url{https://eta-publications.lbl.gov/sites/default/files/wind_engineering_accepted_manuscript_w_disclaimer_copyright.pdf}

\bibitem[{Bolinger et~al(2020{\natexlab{b}})Bolinger, Seel, Robson, and
  Warner}]{34412}
Bolinger M, Seel J, Robson D, et~al (2020{\natexlab{b}}) Utility-scale solar
  data update: 2020 edition. Tech. rep.

\bibitem[{Borrmann et~al(2018)Borrmann, Rehfeldt, Wallasch, and
  Lüers}]{rasmus2018}
Borrmann R, Rehfeldt K, Wallasch A, et~al (2018) Capacity densities of european
  offshore wind farms

\bibitem[{Bright et~al(2018)Bright, Rose, Urban, and
  McKee}]{bright2018landscan}
Bright EA, Rose AN, Urban ML, et~al (2018) Landscan 2017 high-resolution global
  population data set. Tech. rep., Oak Ridge National Lab.(ORNL), Oak Ridge, TN
  (United States)

\bibitem[{Brill et~al(1990)Brill, Flach, Hopkins, and Ranjithan}]{brill1990mga}
Brill ED, Flach JM, Hopkins LD, et~al (1990) Mga: A decision support system for
  complex, incompletely defined problems. IEEE Transactions on systems, man,
  and cybernetics 20(4):745--757

\bibitem[{Brill~Jr et~al(1982)Brill~Jr, Chang, and Hopkins}]{brill1982modeling}
Brill~Jr ED, Chang SY, Hopkins LD (1982) Modeling to generate alternatives: The
  hsj approach and an illustration using a problem in land use planning.
  Management Science 28(3):221--235

\bibitem[{Burke et~al(2020)Burke, Goggin, and Gramlich}]{burkeoffshore2020}
Burke B, Goggin M, Gramlich R (2020) Offshore wind transmission white paper.
  \urlprefix\url{https://www.ferc.gov/sites/default/files/2020-10/Panel-4-Michael-Goggin-Business-Network-OSW.pdf}

\bibitem[{Carr et~al(2016)Carr, Fancher, Freeman, and
  Battles~Manley}]{SolarUSGS}
Carr N, Fancher T, Freeman A, et~al (2016) Surface area of solar arrays in the
  conterminous united states. \doi{http://dx.doi.org/10.5066/F79S1P57},
  \url{https://www.sciencebase.gov/catalog/item/57a25271e4b006cb45553efa},
  Accessed: 10-30-2021

\bibitem[{{Carr, N.B., Fancher, T.S., Freeman, A.T., Battles Manley,
  H.M.}(2016)}]{carr2016}
{Carr, N.B., Fancher, T.S., Freeman, A.T., Battles Manley, H.M.} (2016) Surface
  area of solar arrays in the conterminous {United} {States}. US Geological
  Survey \doi{http://dx.DOI.org/10.5066/F79S1P57},
  \urlprefix\url{https://www.sciencebase.gov/catalog/item/57a25271e4b006cb45553efa}

\bibitem[{Catalyst-Cooperative(2020)}]{PUDL}
Catalyst-Cooperative (2020) Public utility data liberation (pudl).
  \doi{10.5281/zenodo.3672068},
  \urlprefix\url{https://doi.org/10.5281/zenodo.3404014}

\bibitem[{Center(2020)}]{center2020annual}
Center BP (2020) Annual energy outlook 2020. Energy Information Administration,
  Washington, DC

\bibitem[{Choukulkar et~al(2016)Choukulkar, Pichugina, Clack, Calhoun, Banta,
  Brewer, and Hardesty}]{choukulkar2016new}
Choukulkar A, Pichugina Y, Clack CT, et~al (2016) A new formulation for rotor
  equivalent wind speed for wind resource assessment and wind power
  forecasting. Wind Energy 19(8):1439--1452

\bibitem[{Clack(2017)}]{clack2017modeling}
Clack CT (2017) Modeling solar irradiance and solar pv power output to create a
  resource assessment using linear multiple multivariate regression. Journal of
  Applied Meteorology and Climatology 56(1):109--125

\bibitem[{Clack et~al(2016)Clack, Alexander, Choukulkar, and
  MacDonald}]{clack2016demonstrating}
Clack CT, Alexander A, Choukulkar A, et~al (2016) Demonstrating the effect of
  vertical and directional shear for resource mapping of wind power. Wind
  Energy 19(9):1687--1697

\bibitem[{Clack et~al(2020)Clack, Choukulkar, Cote, and
  McKee}]{clack2020renewable}
Clack CT, Choukulkar A, Cote B, et~al (2020) Renewable Generation, Electric
  Demand, Transmission Line Ratings \& Losses, and Climate Change: Dataset
  Overview. Vibrant Clean Energy, LLC

\bibitem[{for
  Coastal~Management(2020)}]{noaaofficeforcoastalmanagementsubmarine2020}
for Coastal~Management NO (2020) Submarine cable. \url =
  {https://ezmt.anl.gov/}, Accessed: 2020-11-13

\bibitem[{Cohen et~al(2014)Cohen, Reichl, and Schmidthaler}]{cohen2014re}
Cohen JJ, Reichl J, Schmidthaler M (2014) Re-focussing research efforts on the
  public acceptance of energy infrastructure: A critical review. Energy 76:4--9

\bibitem[{Cohen et~al(2019)Cohen, Becker, Bielen, Brown, Cole, Eurek
  et~al}]{cohen2019regional}
Cohen S, Becker J, Bielen D, et~al (2019) Regional energy deployment system
  (reeds) model documentation: Version 2018. national renewable energy
  laboratory (nrel). golden, co (united states)

\bibitem[{DeCarolis et~al(2017)DeCarolis, Daly, Dodds, Keppo, Li, McDowall,
  Pye, Strachan, Trutnevyte, Usher et~al}]{decarolis2017formalizing}
DeCarolis J, Daly H, Dodds P, et~al (2017) Formalizing best practice for energy
  system optimization modelling. Applied energy 194:184--198

\bibitem[{DeCarolis(2011)}]{decarolis2011using}
DeCarolis JF (2011) Using modeling to generate alternatives (mga) to expand our
  thinking on energy futures. Energy Economics 33(2):145--152

\bibitem[{DeCarolis et~al(2016)DeCarolis, Babaee, Li, and
  Kanungo}]{decarolis2016modelling}
DeCarolis JF, Babaee S, Li B, et~al (2016) Modelling to generate alternatives
  with an energy system optimization model. Environmental Modelling \& Software
  79:300--310

\bibitem[{Dijkstra et~al(1959)}]{dijkstra1959note}
Dijkstra EW, et~al (1959) A note on two problems in connexion with graphs.
  Numerische mathematik 1(1):269--271

\bibitem[{Dolf~Gielen and Miranda(2019)}]{hydrogenIrena19}
Dolf~Gielen ET, Miranda R (2019) Hydrogen: A renewable energy perspective.
  \url{https://www.irena.org/-/media/Files/IRENA/Agency/Publication/2019/Sep/IRENA_Hydrogen_2019.pdf},
  Accessed: 6-25-2020

\bibitem[{Draxl et~al(2015)Draxl, Clifton, Hodge, and McCaa}]{draxl2015wind}
Draxl C, Clifton A, Hodge BM, et~al (2015) The wind integration national
  dataset (wind) toolkit. Applied Energy 151:355--366

\bibitem[{EIA(2019)}]{EIA2019}
EIA (2019) Form eia-860 detailed data with previous form data (eia-860a/860b).
  \urlprefix\url{https://www.eia.gov/electricity/data/eia860/}

\bibitem[{Energiewende et~al(2020)Energiewende, Verkehrswende
  et~al}]{energiewende2020making}
Energiewende A, Verkehrswende A, et~al (2020) Making the most of offshore wind:
  Re-evaluating the potential of offshore wind in the german north sea. Agora
  Energiewende pp 1--81

\bibitem[{Energy(2010)}]{energy2010western}
Energy G (2010) Western wind and solar integration study. Tech. rep., National
  Renewable Energy Lab.(NREL), Golden, CO (United States)

\bibitem[{Feron et~al(2019)Feron, Cousins, Jiang, Zhai, Thiruvenkatachari,
  Burnard et~al}]{feron2019towards}
Feron P, Cousins A, Jiang K, et~al (2019) Towards zero emissions from fossil
  fuel power stations. International Journal of Greenhouse Gas Control
  87:188--202

\bibitem[{Gabrielli et~al(2018)Gabrielli, Gazzani, Martelli, and
  Mazzotti}]{gabrielli2018optimal}
Gabrielli P, Gazzani M, Martelli E, et~al (2018) Optimal design of multi-energy
  systems with seasonal storage. Applied Energy 219:408--424

\bibitem[{Gini et~al(1912)Gini, Pizetti, and Salvemini}]{gini1912memorie}
Gini C, Pizetti E, Salvemini T (1912) Memorie di metodologica statistica rome:
  Libreria eredi virgilio veschi

\bibitem[{Gorman et~al(2019)Gorman, Mills, and Wiser}]{gorman2019improving}
Gorman W, Mills A, Wiser R (2019) Improving estimates of transmission capital
  costs for utility-scale wind and solar projects to inform renewable energy
  policy. Energy Policy 135:110,994

\bibitem[{Grady(2011)}]{Gradyfacts}
Grady W (2011) American electric power transmission facts.
  \url{https://web.ecs.baylor.edu/faculty/grady/_13_EE392J_2_Spring11_AEP_Transmission_Facts.pdf},
  Accessed: 10-25-2021

\bibitem[{Hagberg et~al(2008)Hagberg, Swart, and
  S~Chult}]{hagberg2008exploring}
Hagberg A, Swart P, S~Chult D (2008) Exploring network structure, dynamics, and
  function using networkx. Tech. rep., Los Alamos National Lab.(LANL), Los
  Alamos, NM (United States)

\bibitem[{Hernandez et~al(2015)Hernandez, Hoffacker, Murphy-Mariscal, Wu, and
  Allen}]{hernandez2015solar}
Hernandez RR, Hoffacker MK, Murphy-Mariscal ML, et~al (2015) Solar energy
  development impacts on land cover change and protected areas. Proceedings of
  the National Academy of Sciences 112(44):13,579--13,584

\bibitem[{HIFLD(2019)}]{hifldelectric2020}
HIFLD (2019) Electric power transmission lines.
  \url{https://hifld-geoplatform.opendata.arcgis.com/datasets/electric-power-transmission-lines},
  Accessed: 2020-11-13

\bibitem[{Hobbs(1995)}]{hobbs1995optimization}
Hobbs BF (1995) Optimization methods for electric utility resource planning.
  European Journal of Operational Research 83(1):1--20

\bibitem[{Hoen et~al(2018)Hoen, Diffendorfer, Rand, Kramer, Garrity, and
  Hunt}]{WindUSGS}
Hoen B, Diffendorfer J, Rand J, et~al (2018) The u.s. wind turbine database.
  \doi{https://doi.org/10.5066/F7TX3DN0},
  \url{https://eerscmap.usgs.gov/uswtdb/}, Accessed: 10-30-2021

\bibitem[{{Hoen, B.D., Diffendorfer, J.E., Rand, J.T., Kramer, L.A., Garrity,
  C.P., and Hunt, H.E.}(2018)}]{hoen2018}
{Hoen, B.D., Diffendorfer, J.E., Rand, J.T., Kramer, L.A., Garrity, C.P., and
  Hunt, H.E.} (2018) United states wind turbine database (ver. 3.1, july 2020).
  \doi{https://DOI.org/10.5066/F7TX3DN0}

\bibitem[{{H.R.5376 - 117th Congress}(2021-2022)}]{IRA}
{H.R.5376 - 117th Congress} (2021-2022) Inflation reduction act of 2022.
  \url{http://www.congress.gov/}

\bibitem[{Jenkins et~al(2022)Jenkins, Mayfield, Farbes, Jones, Patankar, Xu,
  and Schivley}]{REPEAT}
Jenkins J, Mayfield E, Farbes J, et~al (2022) Preliminary report: The climate
  and energy impacts of the inflation reduction act of 2022.
  \url{https://repeatproject.org/}, Accessed: 10-25-2022

\bibitem[{Jenkins(2018)}]{jenkins2018electricity}
Jenkins JD (2018) Electricity system planning with distributed energy
  resources: new methods and insights for economics, regulation, and policy.
  PhD thesis, Massachusetts Institute of Technology

\bibitem[{Jenkins and Sepulveda(2017)}]{jenkins2017enhanced}
Jenkins JD, Sepulveda NA (2017) Enhanced decision support for a changing
  electricity landscape: the genx configurable electricity resource capacity
  expansion model. An MIT Energy Initiative Working Paper https://energy mit
  edu/wpcontent/uploads/2017/10/Enhanced-Decision-Support-for-a-Changing-Electricity-Landscape
  pdf

\bibitem[{Jenkins et~al(2021)Jenkins, Mayfield, Larson, Pacala, and
  Greig}]{jenkins2021mission}
Jenkins JD, Mayfield EN, Larson ED, et~al (2021) Mission net-zero america: The
  nation-building path to a prosperous, net-zero emissions economy. Joule
  5(11):2755--2761

\bibitem[{Jin et~al(2019)Jin, Homer, Dewitz, Danielson, and
  Howard}]{jin2019national}
Jin S, Homer C, Dewitz J, et~al (2019) National land cover database (nlcd) 2016
  science research products

\bibitem[{Jing et~al(2019)Jing, Kuriyan, Kong, Zhang, Shah, Li, and
  Zhao}]{jing2019exploring}
Jing R, Kuriyan K, Kong Q, et~al (2019) Exploring the impact space of different
  technologies using a portfolio constraint based approach for multi-objective
  optimization of integrated urban energy systems. Renewable and Sustainable
  Energy Reviews 113:109,249

\bibitem[{{John van Zalk, Paul Behrens}(2018)}]{johnvanzalk2018}
{John van Zalk, Paul Behrens} (2018) The spatial extent of renewable and
  non-renewable power generation: {A} review and meta-analysis of power
  densities and their application in the {U}.{S}. Energy Policy 123:83--91.
  \doi{https://DOI.org/10.1016/j.enpol.2018.08.023}

\bibitem[{Klein and Whalley(2015)}]{klein2015comparing}
Klein SJ, Whalley S (2015) Comparing the sustainability of us electricity
  options through multi-criteria decision analysis. Energy Policy 79:127--149

\bibitem[{Kotzur et~al(2018)Kotzur, Markewitz, Robinius, and
  Stolten}]{kotzur2018impact}
Kotzur L, Markewitz P, Robinius M, et~al (2018) Impact of different time series
  aggregation methods on optimal energy system design. Renewable Energy
  117:474--487

\bibitem[{Larson et~al(2020)Larson, Greig, Jenkins, Mayfield, Pascale, Zhang,
  Drossman, Williams, Pacala, Socolow, Baik, Birdsey, Duke, Jones, Haley,
  Leslie, Paustian, and Swan}]{larsonnetzero2021}
Larson E, Greig C, Jenkins J, et~al (2020) Net-zero america: Potential
  pathways, infrastructure, and impacts.
  \url{https://netzeroamerica.princeton.edu/}, Accessed: 10-25-2021

\bibitem[{Leslie et~al(2021)Leslie, Pascale, and
  Jenkins}]{leslieemily20215021146}
Leslie E, Pascale A, Jenkins J (2021) {Wind and Solar Candidate Project Areas
  for Princeton REPEAT}. \doi{10.5281/zenodo.5021146},
  \urlprefix\url{https://DOI.org/10.5281/zenodo.5021146}

\bibitem[{Li and Trutnevyte(2017)}]{li2017investment}
Li FG, Trutnevyte E (2017) Investment appraisal of cost-optimal and
  near-optimal pathways for the uk electricity sector transition to 2050.
  Applied energy 189:89--109

\bibitem[{Likas et~al(2003)Likas, Vlassis, and Verbeek}]{likas2003global}
Likas A, Vlassis N, Verbeek JJ (2003) The global k-means clustering algorithm.
  Pattern recognition 36(2):451--461

\bibitem[{Lombardi et~al(2022)Lombardi, Pickering, and
  Pfenninger}]{lombardi2022redundant}
Lombardi F, Pickering B, Pfenninger S (2022) What is redundant and what is not?
  computational trade-offs in modelling to generate alternatives for energy
  infrastructure deployment. arXiv preprint arXiv:220608637

\bibitem[{Lord et~al(2014)Lord, Kobos, and Borns}]{lord2014geologic}
Lord AS, Kobos PH, Borns DJ (2014) Geologic storage of hydrogen: Scaling up to
  meet city transportation demands. International journal of hydrogen energy
  39(28):15,570--15,582

\bibitem[{Maclaurin et~al(2019{\natexlab{a}})Maclaurin, Grue, Lopez, and
  Heimiller}]{maclaurinrenewable2019}
Maclaurin G, Grue N, Lopez A, et~al (2019{\natexlab{a}}) The renewable energy
  potential ({reV}) model: A geospatial platform for technical potential and
  supply curve modeling.
  \urlprefix\url{https://www.nrel.gov/docs/fy19osti/73067.pdf}

\bibitem[{Maclaurin et~al(2020)Maclaurin, Lopez, Grue, Buster, Rossol, Spencer
  et~al}]{maclaurin2020open}
Maclaurin G, Lopez A, Grue N, et~al (2020) Open source rev (the renewable
  energy potential model). Tech. rep., National Renewable Energy Lab.(NREL),
  Golden, CO (United States)

\bibitem[{Maclaurin et~al(2019{\natexlab{b}})Maclaurin, Grue, Lopez, and
  Heimiller}]{maclaurin2019renewable}
Maclaurin GJ, Grue NW, Lopez AJ, et~al (2019{\natexlab{b}}) The renewable
  energy potential (rev) model: a geospatial platform for technical potential
  and supply curve modeling. Tech. rep., National Renewable Energy Lab.(NREL),
  Golden, CO (United States)

\bibitem[{Mai et~al(2021)Mai, Lopez, Mowers, and Lantz}]{mai2021interactions}
Mai T, Lopez A, Mowers M, et~al (2021) Interactions of wind energy project
  siting, wind resource potential, and the evolution of the us power system.
  Energy p 119998

\bibitem[{Mai et~al(2018)Mai, Jadun, Logan, McMillan, Muratori, Steinberg,
  Vimmerstedt, Haley, Jones, and Nelson}]{mai2018electrification}
Mai TT, Jadun P, Logan JS, et~al (2018) Electrification futures study:
  Scenarios of electric technology adoption and power consumption for the
  united states. Tech. rep., National Renewable Energy Lab.(NREL), Golden, CO
  (United States)

\bibitem[{Mallapragada et~al(2018)Mallapragada, Papageorgiou, Venkatesh, Lara,
  and Grossmann}]{mallapragada2018impact}
Mallapragada DS, Papageorgiou DJ, Venkatesh A, et~al (2018) Impact of model
  resolution on scenario outcomes for electricity sector system expansion.
  Energy 163:1231--1244

\bibitem[{{Mark Bolinger} et~al(2020){Mark Bolinger}, {Joachim Seel}, {Dana
  Robson}, and {Cody Warner}}]{markbolingerutilityscale2020}
{Mark Bolinger}, {Joachim Seel}, {Dana Robson}, et~al (2020) Utility-{Scale}
  {Solar} {Data} {Update}: 2020 {Edition}.
  \urlprefix\url{https://emp.lbl.gov/sites/default/files/2020_utility-scale_solar_data_update.pdf}

\bibitem[{Mason et~al(2020)Mason, Curry, and Wilson}]{WECCTransmission}
Mason T, Curry T, Wilson D (2020) Wecc transmission cost report.
  \url{https://www.wecc.org/reliability/1210_bv_wecc_transcostreport_final.pdf},
  Accessed: 1-6-2020

\bibitem[{Michalski et~al(2017)Michalski, B{\"u}nger, Crotogino, Donadei,
  Schneider, Pregger, Cao, and Heide}]{michalski2017hydrogen}
Michalski J, B{\"u}nger U, Crotogino F, et~al (2017) Hydrogen generation by
  electrolysis and storage in salt caverns: Potentials, economics and systems
  aspects with regard to the german energy transition. International Journal of
  Hydrogen Energy 42(19):13,427--13,443

\bibitem[{Müllner(2011)}]{Mullner2011}
Müllner D (2011) Modern hierarchical, agglomerative clustering algorithms.
  arXiv:11092378 [cs, stat] \urlprefix\url{http://arxiv.org/abs/1109.2378},
  arXiv: 1109.2378

\bibitem[{{Naidoo}(2019)}]{Naidoo2019}
{Naidoo} K (2019) {MiSTree: a Python package for constructing and analysing
  Minimum Spanning Trees}. The Journal of Open Source Software 4(42):1721.
  \doi{10.21105/joss.01721}

\bibitem[{NCSL(2021)}]{CES}
NCSL (2021) Energy state bill tracking database.
  \url{https://www.ncsl.org/research/energy/energy-legislation-tracking-database.aspx},
  Accessed: 9-30-2021

\bibitem[{{New York Department of Public Service Staff} et~al(2021){New York
  Department of Public Service Staff}, {New York State Energy Research and
  Development Authority Staff}, Group, and
  Consulting}]{newyorkdepartmentofpublicservicestaffinitial2021}
{New York Department of Public Service Staff}, {New York State Energy Research
  and Development Authority Staff}, Group TB, et~al (2021) Initial report on
  the new york power grid study.
  \urlprefix\url{https://brattlefiles.blob.core.windows.net/files/20842_initial_report_on_the_new_york_power_grid_study.pdf}

\bibitem[{Nock and Baker(2019)}]{nock2019holistic}
Nock D, Baker E (2019) Holistic multi-criteria decision analysis evaluation of
  sustainable electric generation portfolios: New england case study. Applied
  Energy 242:655--673

\bibitem[{Nolte(2020)}]{nolte2020high}
Nolte C (2020) High-resolution land value maps reveal underestimation of
  conservation costs in the united states. Proceedings of the National Academy
  of Sciences 117(47):29,577--29,583

\bibitem[{{NREL}(2020)}]{nrel2020}
{NREL} (2020) 2020 annual technology baseline.
  \urlprefix\url{https://atb.nrel.gov/electricity/2020/data.php}

\bibitem[{NREL(2020)}]{Nrel_regional_multiplier}
NREL (2020) Nrel: Regional multipliers.
  \url{https://atb.nrel.gov/electricity/2019/regional-capex.html}, Accessed:
  6-25-2020

\bibitem[{{Oak Ridge National Laboratory (ORNL)} et~al(2020){Oak Ridge National
  Laboratory (ORNL)}, {Los Alamos National Laboratory (LANL)}, {Idaho National
  Laboratory (INL)}, and {National Geospatial-Intelligence Agency (NGA)
  Homeland Security Infrastructure Program (HSIP) Team}}]{hifldsubstations2020}
{Oak Ridge National Laboratory (ORNL)}, {Los Alamos National Laboratory
  (LANL)}, {Idaho National Laboratory (INL)}, et~al (2020) Electric
  substations.
  \urlprefix\url{https://hifld-geoplatform.opendata.arcgis.com/datasets/geoplatform::electric-substations/explore}

\bibitem[{{P. Denholm, M. Hand, M. Jackson, and S. Ong,}(2009)}]{denholm2009}
{P. Denholm, M. Hand, M. Jackson, and S. Ong,} (2009) Land-{Use} {Requirements}
  of {Modern} {Wind} {Power} {Plants} in the {United} {States}.
  \urlprefix\url{https://www.nrel.gov/docs/fy09osti/45834.pdf}

\bibitem[{Patankar et~al(2022)Patankar, Basset, Schivley, Leslie, and
  Jenkins}]{patankar_neha_2022_6897346}
Patankar N, Basset X, Schivley G, et~al (2022) {Land Use Trade-offs in
  Decarbonization of Electricity Generation in the American West}.
  \doi{10.5281/zenodo.6897346},
  \urlprefix\url{https://doi.org/10.5281/zenodo.6897346}

\bibitem[{Pfenninger(2017)}]{pfenninger2017dealing}
Pfenninger S (2017) Dealing with multiple decades of hourly wind and pv time
  series in energy models: A comparison of methods to reduce time resolution
  and the planning implications of inter-annual variability. Applied energy
  197:1--13

\bibitem[{Pfenninger and Staffell(2016)}]{pfenninger2016long}
Pfenninger S, Staffell I (2016) Long-term patterns of european pv output using
  30 years of validated hourly reanalysis and satellite data. Energy
  114:1251--1265

\bibitem[{Price and Keppo(2017)}]{price2017modelling}
Price J, Keppo I (2017) Modelling to generate alternatives: A technique to
  explore uncertainty in energy-environment-economy models. Applied energy
  195:356--369

\bibitem[{Rand and Hoen(2017)}]{rand2017thirty}
Rand J, Hoen B (2017) Thirty years of north american wind energy acceptance
  research: What have we learned? Energy research \& social science 29:135--148

\bibitem[{{Ranjit Deshmukh, Grace
  Wu}(2019{\natexlab{a}})}]{ranjit_deshmukh_grace_wu_multi-criteria_2019}
{Ranjit Deshmukh, Grace Wu} (2019{\natexlab{a}}) Multi-criteria {Analysis} for
  {Planning} {Renewable} {Energy}. \urlprefix\url{https://mapre.lbl.gov/}

\bibitem[{{Ranjit Deshmukh, Grace
  Wu}(2019{\natexlab{b}})}]{ranjitdeshmukhgracewumulticriteria2019}
{Ranjit Deshmukh, Grace Wu} (2019{\natexlab{b}}) Multi-criteria {Analysis} for
  {Planning} {Renewable} {Energy}. \urlprefix\url{https://mapre.lbl.gov/}

\bibitem[{RESOLVE(2019)}]{CPUCIRP}
RESOLVE (2019) Resolve.
  \url{https://www.cpuc.ca.gov/General.aspx?id=6442462824}, Accessed: 6-27-2020

\bibitem[{{S. Ong, C. Campbell, P. Denholm, R. Margolis, and G.
  Heath}(2013)}]{Ong2013}
{S. Ong, C. Campbell, P. Denholm, R. Margolis, and G. Heath} (2013) Land-{Use}
  {Requirements} for {Solar} {Power} {Plants} in the {United} {States}.
  \urlprefix\url{https://www.nrel.gov/docs/fy13osti/56290.pdf}

\bibitem[{Saba et~al(2018)Saba, M{\"u}ller, Robinius, and
  Stolten}]{saba2018investment}
Saba SM, M{\"u}ller M, Robinius M, et~al (2018) The investment costs of
  electrolysis--a comparison of cost studies from the past 30 years.
  International journal of hydrogen energy 43(3):1209--1223

\bibitem[{Sasse and Trutnevyte(2019)}]{sasse2019distributional}
Sasse JP, Trutnevyte E (2019) Distributional trade-offs between regionally
  equitable and cost-efficient allocation of renewable electricity generation.
  Applied Energy 254:113,724

\bibitem[{Schivley et~al(2021)Schivley, Welty, and Patankar}]{PowerGenome}
Schivley G, Welty E, Patankar N (2021) Powergenome.
  \doi{10.5281/zenodo.4552835},
  \urlprefix\url{https://doi.org/10.5281/zenodo.4426096}

\bibitem[{Sch{\"u}tz et~al(2018)Sch{\"u}tz, Schraven, Fuchs, Remmen, and
  M{\"u}ller}]{schutz2018comparison}
Sch{\"u}tz T, Schraven MH, Fuchs M, et~al (2018) Comparison of clustering
  algorithms for the selection of typical demand days for energy system
  synthesis. Renewable energy 129:570--582

\bibitem[{Sepulveda et~al(2021)Sepulveda, Jenkins, Mallapragada, Schwartz,
  Patankar, Xu, Morris, and Chakrabarti}]{GenX}
Sepulveda N, Jenkins J, Mallapragada D, et~al (2021) Genx: power system
  capacity expansion model. \url{https://github.com/GenXProject/GenX},
  Accessed: 9-30-2021

\bibitem[{Sepulveda et~al(2018)Sepulveda, Jenkins, de~Sisternes, and
  Lester}]{sepulveda2018role}
Sepulveda NA, Jenkins JD, de~Sisternes FJ, et~al (2018) The role of firm
  low-carbon electricity resources in deep decarbonization of power generation.
  Joule 2(11):2403--2420

\bibitem[{Steward et~al(2009)Steward, Saur, Penev, and
  Ramsden}]{steward2009lifecycle}
Steward D, Saur G, Penev M, et~al (2009) Lifecycle cost analysis of hydrogen
  versus other technologies for electrical energy storage. Tech. rep., National
  Renewable Energy Lab.(NREL), Golden, CO (United States)

\bibitem[{Swisher et~al(1997)Swisher, Martino~Jannuzzi, and
  Redlinger}]{swisher1997tools}
Swisher JN, Martino~Jannuzzi Gd, Redlinger RY (1997) Tools and methods for
  integrated resource planning. improving energy efficiency and protecting the
  environment. Tech. rep., United Nations Environmental Programme

\bibitem[{{Theobald, David et
  al.}(2020{\natexlab{a}})}]{theobald_david_et_al_detailed_2020}
{Theobald, David et al.} (2020{\natexlab{a}}) Detailed temporal mapping of
  global human modification from 1990 to 2017.
  \urlprefix\url{https://doi.org/10.5061/dryad.n5tb2rbs1}

\bibitem[{{Theobald, David et
  al.}(2020{\natexlab{b}})}]{theobalddavidetaldetailed2020}
{Theobald, David et al.} (2020{\natexlab{b}}) Detailed temporal mapping of
  global human modification from 1990 to 2017.
  \urlprefix\url{https://DOI.org/10.5061/dryad.n5tb2rbs1}

\bibitem[{Trutnevyte et~al(2012)Trutnevyte, Stauffacher, Schlegel, and
  Scholz}]{trutnevyte2012context}
Trutnevyte E, Stauffacher M, Schlegel M, et~al (2012) Context-specific energy
  strategies: coupling energy system visions with feasible implementation
  scenarios. Environmental science \& technology 46(17):9240--9248

\bibitem[{{U.S. Census Bureau}(2018)}]{USCensusBureau2018}
{U.S. Census Bureau} (2018) Urban areas.
  \urlprefix\url{https://www2.census.gov/geo/tiger/GENZ2018/shp/cb_2018_us_ua10_500k.zip}

\bibitem[{{U.S. Census Bureau}(2019)}]{USCensusBureau2019}
{U.S. Census Bureau} (2019) Metropolitan and micropolitan statistical areas and
  related statistical areas: Core based statistical areas (cbsas).
  \urlprefix\url{https://www2.census.gov/geo/tiger/GENZ2019/shp/cb_2019_us_cbsa_5m.zip}

\bibitem[{{U.S. Census Bureau}(2020)}]{USCensusBureau2020}
{U.S. Census Bureau} (2020) Metropolitan and micropolitan statistical area
  population estimates and estimated components of change: April 1, 2010 to
  july 1, 2019 (cbsa-est2019-alldata).
  \urlprefix\url{https://www2.census.gov/programs-surveys/popest/datasets/2010-2019/metro/totals/cbsa-est2019-alldata.csv}

\bibitem[{{U.S. Census
  Bureau}(2021)}]{u.s.censusbureauMetropolitanMicropolitan2021}
{U.S. Census Bureau} (2021) Metropolitan and {{Micropolitan}}: About.
  \urlprefix\url{https://web.archive.org/web/20211031192642/https://www.census.gov/programs-surveys/metro-micro/about.html}

\bibitem[{{U.S. Energy Information
  Agency}(2020{\natexlab{a}})}]{eiacostperformance}
{U.S. Energy Information Agency} (2020{\natexlab{a}}) Cost and performance
  characteristics of new generating technologies, annual energy outlook 2020.
  \urlprefix\url{https://www.eia.gov/outlooks/archive/aeo20/assumptions/pdf/table_8.2.pdf}

\bibitem[{{U.S. Energy Information Agency}(2020{\natexlab{b}})}]{eiaemm}
{U.S. Energy Information Agency} (2020{\natexlab{b}}) Electricity market
  module.
  \urlprefix\url{https://www.eia.gov/outlooks/archive/aeo20/assumptions/pdf/electricity.pdf}

\bibitem[{{U.S. Environmental Protection Agency}(2021)}]{epaipm}
{U.S. Environmental Protection Agency} (2021) EPA’s Power Sector Modeling
  Platform v6 using IPM Summer 2021 Reference Case. Environmental Protection
  Agency,
  \urlprefix\url{https://www.epa.gov/airmarkets/epas-power-sector-modeling-platform-v6-using-ipm-summer-2021-reference-case}

\bibitem[{Virtanen et~al(2020)Virtanen, Gommers, Oliphant, Haberland, Reddy,
  Cournapeau, Burovski, Peterson, Weckesser, Bright, and et~al.}]{scipy2020}
Virtanen P, Gommers R, Oliphant TE, et~al (2020) Scipy 1.0: Fundamental
  algorithms for scientific computing in python. Nature Methods 17:261–272.
  \doi{10.1038/s41592-019-0686-2}

\bibitem[{Walt et~al(2014)Walt, Schönberger, Nunez-Iglesias, Boulogne, Warner,
  Yager, Gouillart, and Yu}]{skimage2014}
Walt Svd, Schönberger JL, Nunez-Iglesias J, et~al (2014) scikit-image: image
  processing in python. PeerJ 2:e453. \doi{10.7717/peerj.453}

\bibitem[{Ward~Jr and Hook(1963)}]{ward1963application}
Ward~Jr JH, Hook ME (1963) Application of an hierarchical grouping procedure to
  a problem of grouping profiles. Educational and Psychological Measurement
  23(1):69--81

\bibitem[{Wilson et~al(2013)Wilson, Biewald, and Economics}]{wilson2013best}
Wilson R, Biewald B, Economics SE (2013) Best practices in electric utility
  integrated resource planning: Examples of state regulations and recent
  utility plans. Regulatory Assistance Project

\bibitem[{Wu(2022)}]{wu_et_al_ecosystem_2022}
Wu G (2022) Ecosystem protection and net zero energy systems: renewable
  infrastructure siting and land and ocean use in the western united states

\bibitem[{Wu et~al(2020)Wu, Leslie, Sawyerr, Cameron, Brand, Cohen, Allen,
  Ochoa, and Olson}]{wu2020low}
Wu GC, Leslie E, Sawyerr O, et~al (2020) Low-impact land use pathways to deep
  decarbonization of electricity. Environmental Research Letters 15(7):074,044

\end{thebibliography}


\end{document}